%% file: TOP-17-023_temp.tex
\pdfoutput=1

\documentclass[11pt,twoside,a4paper,cmspaper,final,collab]{cms-tdr}
\def\svnVersion{d0ab7b7-D}\def\svnDate{2020/04/18}\def\cmsCernNoTag{CERN-EP-2019-138}\def\cmsCernDate{\today}\def\cmsMessage{"Published in the European Physical Journal C as \href{http://dx.doi.org/10.1140/epjc/s10052-020-7858-1}{\doi{10.1140/epjc/s10052-020-7858-1}.}"}

\begin{document}\cmsNoteHeader{TOP-17-023}

\hyphenation{had-ron-i-za-tion}
\hyphenation{cal-or-i-me-ter}
\hyphenation{de-vices}
\RCS$HeadURL: svn+ssh://mkomm@svn.cern.ch/reps/tdr2/papers/TOP-17-023/trunk/TOP-17-023.tex $
\RCS$Id: TOP-17-023.tex 494045 2019-04-22 17:58:40Z mkomm $

\newlength\cmsFigWidth
\ifthenelse{\boolean{cms@external}}{\setlength\cmsFigWidth{0.85\columnwidth}}{\setlength\cmsFigWidth{0.4\textwidth}}
\ifthenelse{\boolean{cms@external}}{\providecommand{\cmsLeft}{upper\xspace}}{\providecommand{\cmsLeft}{left\xspace}}
\ifthenelse{\boolean{cms@external}}{\providecommand{\cmsRight}{lower\xspace}}{\providecommand{\cmsRight}{right\xspace}}

\providecommand{\Pell}{{\HepParticle{\ell}{}{}}\Xspace}
\providecommand{\expnote}{\ensuremath{\,\text{(exp)}}\xspace}
\newcommand{\wjets}{\ensuremath{\PW\text{+jets}}\xspace}
\newcommand{\zjets}{\ensuremath{\PZ/\cPgg^{*}\text{+jets}}\xspace}
\newcommand{\wzjets}{\ensuremath{\PW/\PZ/\cPgg^{*}\text{+jets}}\xspace}
\newcommand{\tw}{\ensuremath{\cPqt\PW}\xspace}

\newcommand{\tchannel}{\ensuremath{t\text{-channel}}\xspace}
\newcommand{\tch}{\ensuremath{t\text{-ch}}\xspace}

\newcommand{\sigmat}{\ensuremath{\sigma_{\PQt}}\xspace}
\newcommand{\sigmatbar}{\ensuremath{\sigma_{\PAQt}}\xspace}
\newcommand{\sigmatsum}{\ensuremath{\sigma_{\PQt\text{+}\PAQt}}\xspace}

\newcommand{\muiso}{\ensuremath{I_\text{rel}^{\PGm}}\xspace}
\newcommand{\eiso}{\ensuremath{I_\text{rel}^{\Pe}}\xspace}
\newcommand{\mtop}{\ensuremath{m_{\Pell\PGn{\PQb}}}\xspace}
\newcommand{\met}{\ptmiss}
\newcommand{\mtw}{\ensuremath{\mT(\PW)}\xspace}
\newcommand{\pvmiss}{\ptvecmiss}
\newcommand{\jprime}{\ensuremath{j^{\prime}}\xspace}
\newcommand{\qprime}{\ensuremath{{\cPq}^{\prime}}\xspace}

\newcommand{\Rbj}{\ensuremath{\Delta R(\PQb,\jprime)}\xspace}
\newcommand{\etabl}{\ensuremath{|\Delta \eta(\PQb,\Pell)|}\xspace}

\newcommand{\bdttch}{\ensuremath{\mathrm{BDT}_{\tch}}\xspace}
\newcommand{\bdttw}{\ensuremath{\mathrm{BDT}_{\ttbar/\PW}}\xspace}
\newcommand{\cosw}{\ensuremath{\cos\theta_{\PW}^\star}\xspace}
\newcommand{\anglepol}{\ensuremath{\theta_{\text{pol}}^\star}\xspace}
\newcommand{\cospol}{\ensuremath{\cos\anglepol}\xspace}

\newcommand{\MADSPIN}{\textsc{madspin}\xspace}

\newlength\cmsTabSkip\setlength{\cmsTabSkip}{1ex}

\cmsNoteHeader{TOP-17-023}

\title{Measurement of differential cross sections and charge ratios for $t$-channel single top quark production in proton-proton collisions at $\sqrt{s}=13$\TeV}
\titlerunning{Differential cross sections and charge ratios for $t$-channel single top quark production at $\sqrt{s} = 13\TeV$}

\date{\today}

\abstract{
A measurement is presented of differential cross sections for $t$-channel single top quark and antiquark production in proton-proton collisions at a centre-of-mass energy of $13$\TeV by the CMS experiment at the LHC. From a data set corresponding to an integrated luminosity of 35.9\fbinv, events containing one muon or electron and two or three jets are analysed. The cross section is measured as a function of the top quark transverse momentum ($\pt$), rapidity, and polarisation angle, the charged lepton $\pt$ and rapidity, and the $\pt$ of the \PW~boson from the top quark decay. In addition, the charge ratio is measured differentially as a function of the top quark, charged lepton, and \PW~boson kinematic observables. The results are found to be in agreement with standard model predictions using various next-to-leading-order event generators and sets of parton distribution functions. Additionally, the spin asymmetry, sensitive to the top quark polarisation, is determined from the differential distribution of the polarisation angle at parton level to be $0.440 \pm 0.070$, in agreement with the standard model prediction.
}

\hypersetup{
pdfauthor={CMS Collaboration},
pdftitle={Measurement of differential cross sections and charge ratios for t-channel single top quark production in proton-proton collisions at sqrt(s) = 13 TeV},
pdfsubject={CMS},
pdfkeywords={CMS, physics, top quark, single top}}

\maketitle

\section{Introduction}

\begin{figure*}[htb!]
\centering
\includegraphics[width=0.28\textwidth]{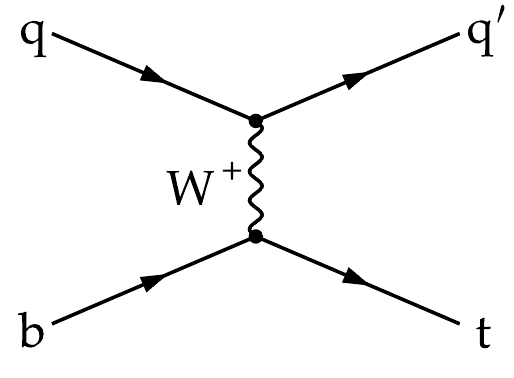}\hspace{0.1\textwidth}
\includegraphics[width=0.28\textwidth]{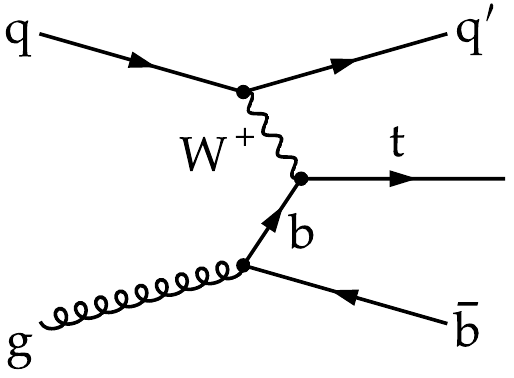}
\caption{\label{fig:intro-tch} Born-level Feynman diagrams for single top quark production in the $t$~channel. Corresponding diagrams also exist for single top antiquark production.}
\end{figure*}

{\tolerance=800
The three main production modes of single top quarks and antiquarks in proton-proton ($\Pp\Pp$) collisions occur via electroweak interactions and are commonly categorised through the virtuality of the exchanged \PW~boson four-momentum. They are called $t$~channel ($t$~ch) when the four-momentum is space-like, $s$ channel when it is time-like, and \PW-associated (\tw) when the four-momentum is on shell. At the CERN LHC, the production via the $t$~channel has the largest cross section of the three modes whose most-relevant Born-level Feynman diagrams are shown in Fig.~\ref{fig:intro-tch}. In the rest of this paper, ``quark'' is used to generically denote a quark or an antiquark, unless otherwise specified.\par
}

The \tchannel production process was first observed by the D0 and CDF experiments at the Tevatron~\cite{Abazov:2009ii,Aaltonen:2009jj}. Its inclusive cross section has been measured with high precision at the CERN LHC by the ATLAS and CMS Collaborations at $\sqrt{s}=7$, 8, and 13\TeV~\cite{Aad:2014fwa,Chatrchyan:2012ep,Aaboud:2017pdi,Khachatryan:2014iya,Aaboud:2016ymp,Sirunyan:2018rlu}. Differential cross sections have been determined as well at 7 and 8\TeV~\cite{Aad:2014fwa,Aaboud:2017pdi,Khachatryan:2015dzz}.

Differential cross section measurements can contribute to constraining the effective field theory operators~\cite{Hartland:2019bjb}, the top quark mass, the renormalisation and factorisation scales, and the parton distribution functions (PDFs) of the proton~\cite{Kant:2014oha}. In particular, the ratio of the \tchannel top quark to antiquark production is sensitive to the ratio of the up to down quark content of the proton~\cite{Berger:2016oht,Alekhin:2017kpj}. Furthermore, differential angular distributions can be used to assess the electroweak coupling structure at the {\PW}{\cPqt}{\cPqb} vertex. A ``vector$-$axial-vector'' (V$-$A) coupling is predicted in the standard model (SM), leading to the production of highly polarised top quarks~\cite{Mahlon:1999gz,Boos:2002xw,AguilarSaavedra:2010nx}. A powerful observable to investigate the coupling structure in \tchannel production is given by the top quark polarisation angle \anglepol, defined via
\begin{linenomath}
\begin{equation}
\cospol = \frac{\vec{p}_{\qprime}^{\star}\cdot \vec{p}_{\Pell}^{\star}}{\abs{\vec{p}_{\qprime}^{\star}} \abs{\vec{p}_{\Pell}^{\star}}},
\end{equation}
\end{linenomath}
where the superscript signifies that the momenta of the charged lepton, $\Pell$ (muon or electron), from the top quark decay, and the spectator quark, $\qprime$, are calculated in the top quark rest frame. The normalised differential cross section as a function of \cospol at the parton level is related to the top quark polarisation, $P$, as
\begin{linenomath}
\begin{equation}
\frac{1}{\sigma}\frac{\rd\sigma}{\rd\cospol}=\frac{1}{2}\left(1+2 A_{\Pell}\cospol\right), \enskip A_{\Pell}=\frac{1}{2}P\alpha_{\Pell},
\label{eq:intro-pol}
\end{equation}
\end{linenomath}
where $A_{\Pell}$ denotes the spin asymmetry and $\alpha_{\Pell}$ is the so-called spin-analysing power of the charged lepton~\cite{AguilarSaavedra:2010nx}. The spin asymmetry and/or polarisation have been measured in $\Pp\Pp$ collision data by the ATLAS and CMS Collaborations at $\sqrt{s}=8$\TeV using various analysis techniques~\cite{Khachatryan:2015dzz,Aaboud:2017aqp,Aaboud:2017yqf}.

In this paper, the differential cross section of combined single top quark and antiquark production in the $t$~channel is measured by the CMS experiment at $\sqrt{s}=13$\TeV as a function of the top quark transverse momentum ($\pt$), rapidity, and polarisation angle, the $\pt$ and rapidity of the charged lepton that originates from the top quark decay, and the $\pt$ of the \PW~boson from the top quark decay. The spin asymmetry is further determined from the measured differential cross section with respect to the polarisation angle. Additionally, a measurement of the differential charge ratio is performed as a function of the $\pt$ and rapidities of the top quark and charged lepton, and the $\pt$ of the \PW~boson. Differential cross sections are measured at both the parton and particle levels using an unfolding procedure.

The analysis strategy and the structure of the paper are outlined in the following. A brief description of the CMS detector is given in Section~\ref{sec:cms}, followed by a summary of the analysed data and simulated event samples in Section~\ref{sec:samples}. The reconstruction of physics objects and the event selection are detailed in Section~\ref{sec:selection}. To determine the contributions from signal and backgrounds a maximum-likelihood fit (ML) is performed separately in each bin of the measurement. In the fit, shape distributions, referred to in the following as templates, are fitted to the data. For the signal and all background processes, samples of simulated events are used to determine the shape distributions, except for the templates of events containing only jets produced through the strong interaction, which are referred to as ``multijet'' events in this paper. The procedure to estimate the templates of multijet events based on data in a sideband region is provided in Section~\ref{sec:multijet}. Section~\ref{sec:fit} describes the measurement of the number of \tchannel single top quark events from data through an ML fit. In the fit, statistical and experimental systematic uncertainties are profiled, where the latter encompasses uncertainties related to the reconstruction, identification, and calibration of the selected events and physics objects. The resulting distributions of the observables are validated in control and signal regions in Section~\ref{sec:validation}. The fit results are input to an unfolding procedure to determine the differential cross sections and charge ratios at the parton and particle levels, as detailed in Section~\ref{sec:unfolding}. The sources of experimental and theoretical systematic uncertainties are described in Section~\ref{sec:systematics}. The results are presented in Section~\ref{sec:results} and the paper is summarised in Section~\ref{sec:summary}.

\section{The CMS detector and event reconstruction}
\label{sec:cms}

The central feature of the CMS apparatus is a superconducting solenoid of 6\unit{m} internal diameter, providing a magnetic field of 3.8\unit{T}. Within the solenoid volume are a silicon pixel and strip tracker, a lead tungstate crystal electromagnetic calorimeter~(ECAL), and a brass and scintillator hadron calorimeter~(HCAL), each composed of a barrel and two endcap sections. Forward calorimeters~(HF) extend the pseudorapidity ($\eta$) coverage provided by the barrel and endcap detectors. Muons are detected in gas-ionisation chambers embedded in the steel flux-return yoke outside the solenoid. A more detailed description of the CMS detector, together with a definition of the coordinate system used and the relevant kinematic variables, can be found in Ref.~\cite{Chatrchyan:2008zzk}.

The particle-flow~(PF) algorithm~\cite{Sirunyan:2017ulk} aims to reconstruct and identify each particle in an event with an optimised combination of information from various elements of the CMS detector. The energy of electrons is estimated from a combination of the electron momentum at the primary interaction vertex, as determined by the tracker, the energy of the corresponding ECAL cluster, and the energy sum of all bremsstrahlung photons spatially compatible with originating from the electron track. The energy of muons is obtained from the curvature of a global track estimated from reconstructed hits in the inner tracker and muon systems. The energy of charged hadrons is determined from a combination of their momentum measured in the tracker and the matching ECAL and HCAL energy deposits. Finally, the energy of neutral hadrons is obtained from the corresponding ECAL and HCAL energy deposits. In the regions $\abs{\eta}>3$, electromagnetic and hadronic shower components are identified in the HF.

Events of interest are selected using a two-tiered trigger system~\cite{Khachatryan:2016bia}. The first level, composed of custom hardware processors, uses information from the calorimeters and muon detectors whereas a version of the full event reconstruction software optimised for fast processing is performed at the second level, which runs on a farm of processors.

The missing transverse momentum vector, \pvmiss, is defined as the projection onto the plane perpendicular to the beams of the negative vector momentum sum of all PF candidates in an event. Its magnitude is referred to as \met.

\section{Data set and simulated samples}
\label{sec:samples}

The analysed $\Pp\Pp$ collision data set was recorded in 2016 by the CMS detector and corresponds to an integrated luminosity of 35.9\fbinv~\cite{CMS-PAS-LUM-17-001}. Events were triggered by requiring at least one isolated muon candidate with $\pt>24$\GeV and $\abs{\eta} < 2.4$ or one electron candidate with $\pt>32$\GeV and $\abs{\eta} < 2.1$, with additional requirements~\cite{Khachatryan:2015hwa} that select genuine electrons with an efficiency of about 80\%.

Various samples of simulated events are used in this measurement to evaluate the detector resolution, efficiency, and acceptance, estimate the contributions from background processes, and determine the differential cross sections at the parton and particle levels.

{\tolerance=800
Single top quark events in the $t$ channel are simulated at next-to-leading order (NLO) in the four-flavour scheme~(4FS) with \POWHEG\,v2~\cite{Alioli:2010xd,Frederix:2012dh} interfaced with \PYTHIA v8.212~\cite{Sjostrand:2014zea} for the parton shower simulation, using the CUETP8M1~\cite{Khachatryan:2015pea} tune interfaced with \MADSPIN~\cite{Artoisenet:2012st} for simulating the top quark decay. For comparison, alternative NLO \tchannel samples have been generated in the 4FS and five-flavour scheme~(5FS), using \MGvATNLO v2.2.2~\cite{Alwall:2014hca} interfaced with \PYTHIA.\par
}

{\tolerance=800
The \POWHEG\,v2 generator is also used to simulate events from top quark pair production (\ttbar) at NLO. Parton showering is simulated with \PYTHIA using the CUETP8M2T4 tune~\cite{CMS-PAS-TOP-16-021}. The production of single top quark events via the tW~channel is simulated at NLO using \POWHEG\,v1~\cite{Re:2010bp} in the 5FS interfaced with \PYTHIA using the CUETP8M1 tune for the parton shower simulation. The overlap with top quark pair production is removed by applying the diagram removal scheme~\cite{Frixione:2008yi}. Samples of \wjets events are generated with \MGvATNLO v2.3.3 at NLO, and interfaced with \PYTHIA using the CUETP8M1 tune. The production of leptonically decaying \PW~bosons in association with jets is simulated with up to two additional partons at the matrix element level, and the FxFx scheme~\cite{Frederix:2012ps} is used for jet merging. Lastly, \zjets events are generated with \MGvATNLO v2.2.2 at leading order (LO), interfaced with \PYTHIA using the MLM jet matching scheme~\cite{Alwall:2007fs}.\par
}

{\tolerance=800
In these simulated samples, the NNPDF3.0~\cite{Ball:2014uwa} NLO set is used as the default PDF, and a nominal top quark mass of $172.5$\GeV is chosen where applicable. The simulated events are overlaid with additional collision interactions (``pileup'') according to the distribution inferred from the data. All generated events undergo a full \GEANTfour~\cite{geant} simulation of the detector response.\par
}

The \tchannel cross section in $\Pp\Pp$ collisions at $\sqrt{s}=13$\TeV is predicted to be $\sigmat=136.0{\,}^{+5.4}_{-4.6}$\unit{pb} for the top quark and $\sigmatbar=81.0{\,}^{+4.1}_{-3.6}$\unit{pb} for the top antiquark, calculated for a top quark mass of 172.5\GeV at NLO in quantum chromodynamics (QCD) using the \textsc{hathor}\,v2.1~\cite{Aliev:2010zk,Kant:2014oha} program. The PDF and the strong coupling constant ($\alpS$) uncertainties are calculated using the PDF4LHC prescription~\cite{Alekhin:2011sk,Botje:2011sn} with the MSTW2008 NLO 68\% confidence level~\cite{Martin:2009iq,Martin:2009bu}, CT10~\cite{Lai:2010vv} NLO, and NNPDF2.3~\cite{Ball:2012cx} NLO PDF sets, and are added in quadrature with the renormalisation and factorisation scale uncertainty. The simulated samples of single top quark and antiquark events employed in this measurement---generated with similar settings---were normalised using the predicted cross sections above. Predictions at next-to-next-to-leading order are available as well~\cite{Berger:2016oht} and are 3\% smaller than the corresponding cross sections at NLO. However, these are not utilised since they have been calculated using a different PDF set and top quark mass value.

\section{Event selection}
\label{sec:selection}

{\tolerance=1000
Proton-proton collision events containing one isolated muon or electron and two or three jets are analysed. This signature selects events where the \PW~boson from a single top quark decays into a charged lepton and a neutrino. One of the selected jets is expected to stem from the hadronisation of a bottom quark that originates from the top quark decay. Another jet (\jprime) from a light-flavoured quark (up, down, or strange) is expected from the spectator quark (labelled \qprime in Fig.~\ref{fig:intro-tch}) that is produced in association with the top quark. The jet from the spectator quark is characteristically found at relatively low angles with respect to the beam axis.\par
}

{\tolerance=800
The reconstructed vertex with the largest value of summed physics-object $\pt^2$ is taken to be the primary $\Pp\Pp$ interaction vertex. The physics objects are the jets, clustered using the jet finding algorithm described in Refs.~\cite{Cacciari:2008gp,Cacciari:2011ma} with the tracks assigned to the vertex as inputs, and the negative vector $\vec{p}_\mathrm{T}$ sum of those jets.\par
}

{\tolerance=800
Muon candidates are accepted if they have $\pt>26$\GeV, $\abs{\eta}<2.4$, and pass the following identification requirements optimised for the selection of genuine muons. A global muon track must have a track fit with a $\chi^2$ per degree of freedom $<$10, have hits in the silicon tracker and muon systems, including at least six in the tracker, of which at least one must be in the pixel detector. Additionally, track segments are required in at least two muon stations to suppress signals from hadronic showers spilling into the muon system. Muon candidates are required to be isolated with a relative isolation parameter $\muiso <6$\%, which is defined as the scalar sum of the transverse energies \ET deposited in the ECAL and HCAL within a cone of radius $\Delta R = \sqrt{\smash[b]{(\Delta\eta)^2+(\Delta\phi)^2}} < 0.4$, divided by the muon $\pt$. The transverse energy is defined as $\ET=E\,\sin(\theta)$ with $E$ and $\theta$ being the energy and polar angle, respectively, of photons and charged and neutral hadrons. Here, $\Delta\eta$ and $\Delta\phi$ are the pseudorapidity and azimuthal angle, respectively, measured relative to the muon direction. The isolation parameter is corrected by subtracting the energy deposited by pileup, which is estimated from the energy deposited by charged hadrons within the isolation cone that are associated with pileup vertices~\cite{Sirunyan:2018fpa}.\par
}

Electron candidates are required to have $\pt>35$\GeV, $\abs{\eta}<1.48$, and fulfil a set of additional quality requirements as follows: the distance between the matched ECAL cluster position and the extrapolated electron track has to be within $\abs{\Delta\eta}<3.08\times10^{-3}$ and $\abs{\Delta\phi}<8.16\times10^{-2}$; the absolute difference between the inverse of the energy estimated from the ECAL cluster and the inverse of the electron track momentum must be less than $12.9\MeV^{-1}$; the ratio of the HCAL to the ECAL energy associated with the electron is required to be less than 4.14\%; the energy-weighted lateral width  of the electron shower in the ECAL along the $\eta$ direction is restricted to ${<}9.98\times10^{-3}$. Electrons from photon conversions are suppressed by requiring that the corresponding track has no missing hits in the inner layers of the tracker and that they do not stem from a photon conversion vertex. Electron candidates have to be isolated using the so-called effective-area-corrected relative isolation parameter~\cite{Cacciari:2007fd} by requiring $\eiso<5.88$\%. This parameter is defined similarly to the muon isolation parameter as the sum of the charged and neutral particle energies within a cone of $\Delta R<0.3$ around the electron candidate, divided by the electron $\pt$. The relative contribution from pileup is estimated as $A_\text{eff}\,\rho$ and subtracted from the isolation parameter, where $A_\text{eff}$ denotes an $\eta$-dependent effective area, and $\rho$ is the median of the \ET density in a $\delta\eta{\times}\delta\phi$ region calculated using the charged particle tracks associated with the pileup vertices.

The selected muon (electron) candidate has to be within 2.0 (0.5)\unit{mm} in the transverse plane and 5.0 (1.0)\unit{mm} along the beam direction of the primary vertex.

Electron candidates with showers in the ECAL endcap ($1.48<\abs{\eta}<2.5$) are not used in the measurement because of the higher background consisting of hadrons misidentified as electrons and of electrons originating from decays of heavy-flavour hadrons, which is found to be about four times larger compared to the ECAL barrel region.

Events are rejected if additional muon or electron candidates passing looser selection criteria are present. The selection requirements for these additional muons/electrons are as follows: looser identification and isolation criteria, $\pt>10$\,(15)\GeV for muons\,(electrons), and $\abs{\eta}<2.5$.

The transverse \PW~boson mass is calculated from the formula
\begin{linenomath}
\begin{equation}
\mtw=\sqrt{\smash[b]{2\pt^{\ell}\met\left[1-\cos(\phi^\ell-\phi^\text{miss})\right]}}
\end{equation}
\end{linenomath}
using the $\pt$ and the $\phi$ of the charged lepton and \pvmiss.

Jets are reconstructed from PF candidates and clustered by applying the anti-\kt algorithm~\cite{Cacciari:2008gp} with a distance parameter of $0.4$ using the \FASTJET package~\cite{Cacciari:2011ma}. The influence of pileup is mitigated using the charged hadron subtraction technique~\cite{CMS-PAS-JME-14-001}. The jet momentum is determined as the vectorial sum of all particle momenta in the jet. An offset correction is applied to the jet $\pt$ to account for contributions from pileup. Further corrections are applied to account for the nonuniform detector response in $\eta$ and $\pt$ of the jets. The corrected jet momentum is found from simulation to be within 2 to 10\% of the true momentum over the whole \pt spectrum and detector acceptance. The corrections are propagated to the measured \pvmiss. A potential overlap of a jet with the selected lepton is removed by ignoring jets that are found within a cone of $\Delta R<0.4$ around a selected lepton candidate. The analysis considers jets within $\abs{\eta}<4.7$ whose calibrated $\pt$ is greater than 40\GeV, with the exception of the HCAL--HF transition region ($2.7<\abs{\eta}<3$) in which jets must have a $\pt$ of at least 50\GeV to reduce the contribution from detector noise. The event is accepted for further analysis if two or three jets are present.

To reduce the large background from \wjets events, a \PQb~tagging algorithm based on a multivariate analysis (MVA) called ``combined MVA''~\cite{Sirunyan:2017ezt}, which combines the results from various other \PQb~tagging algorithms, is used for identifying jets produced from the hadronisation of \PQb~quarks within the acceptance of the silicon tracker ($\abs{\eta}<2.4$). A tight selection is applied on the discriminant of the algorithm, which gives an efficiency of ${\approx}$50\% for jets originating from true \PQb~quarks and misidentification rates of ${\approx}$0.1\% for light jets from \PQu, \PQd, or \PQs quarks or gluons and ${\approx}$3\% for jets from \PQc quarks, as determined from simulation.

Corrections are applied to the simulated events to account for known differences with respect to data. Lepton trigger, reconstruction, and identification efficiencies are estimated with a ``tag-and-probe'' method~\cite{Chatrchyan:2012xi} from \zjets events for data and simulation from which corrections are derived in bins of lepton $\eta$ and \pt. The \PQb~tagging performance in simulation is corrected to match the tagging efficiency observed in data, using scale factors that depend on the $\pt$ and $\eta$ of the selected jets. The scale factors are estimated by dedicated analyses performed with independent data samples~\cite{Sirunyan:2017ezt}. In particular, the mistagging rate of non-\PQb jets in data is determined using the ``negative-tag'' method~\cite{Chatrchyan:2012jua}. A smearing of the jet momenta is applied to account for the known difference in jet energy resolution in simulation compared to data. The profile of pileup interactions is reweighted in simulation to match the one in data derived from the measured instantaneous luminosity.

To classify signal and control samples of events, different event categories are defined, denoted ``$N$j$M$\PQb'', where $N$ is the total number of selected jets (2 or 3) and $M$ is the number of those jets passing the \PQb~tagging requirement (0, 1, or 2). The 2j1\PQb category has the highest sensitivity to the signal yield, whereas the 2j0\PQb and 3j2\PQb categories, enriched in background processes with different compositions, are used to assess the background modelling.

{\tolerance=800
One top quark candidate is reconstructed per event in the 2j1\PQb signal category assuming \tchannel single top quark production. The procedure commences by first reconstructing the \PW~boson. The component of the neutrino candidate momentum along the beam direction $p_{z}$ is found by imposing a \PW~boson mass constraint (80.4\GeV) on the system formed by the charged lepton and $\pvmiss$, the latter being interpreted as the projection in the transverse plane of the four-momentum of the unknown neutrino, as in Ref.~\cite{Chatrchyan:2011vp}. The four-momentum of the top quark candidate (from which its mass, $\pt$, and rapidity are derived) is then calculated as the vector sum of the four-momenta of the charged lepton, the \PQb-tagged jet, and the neutrino candidate. The other (nontagged) jet is interpreted as originating from the spectator quark, which recoils against the \PW~boson.\par
}

\section{Multijet background estimation}
\label{sec:multijet}

Since the probability for a simulated multijet event to mimic the final state of the signal process is very small, it becomes impractical to simulate a sufficiently large number of events for this background. Therefore, the background from multijet events in the analysis phase space region is estimated in a two-step procedure based on data in a sideband region. First, templates of the $\mtw$ distribution from multijet events are obtained from data in a sideband region. Their normalisations are then estimated in a second step through a template-based ML fit to the events in the 2j1\PQb and 3j2\PQb categories, simultaneously with the number of signal events, as described in Section~\ref{sec:fit}. In this section, a dedicated ML fit is discussed that is performed on events in the 2j0\PQb category only for validating the procedure. The outcome of this ML fit is not used further in the measurement.

In the muon channel, the sideband region is defined by inverting the muon isolation requirement ($\muiso>20$\%), which results in a region dominated by multijet events. In the electron channel, the electron candidate is required to fail loose identification criteria, yielding a sideband region consisting not only of nonisolated electrons but also of electrons that fail the photon conversion criteria or are accompanied by large amounts of bremsstrahlung, thus reflecting a combination of various effects. The templates used in the ML fit are determined for this category by subtracting the contamination from other processes, estimated using simulation and which amounts to about 10~(5)\% in the muon (electron) channel, from the data.

\begin{figure*}[htb!]
\centering
\includegraphics[width=0.48\textwidth]{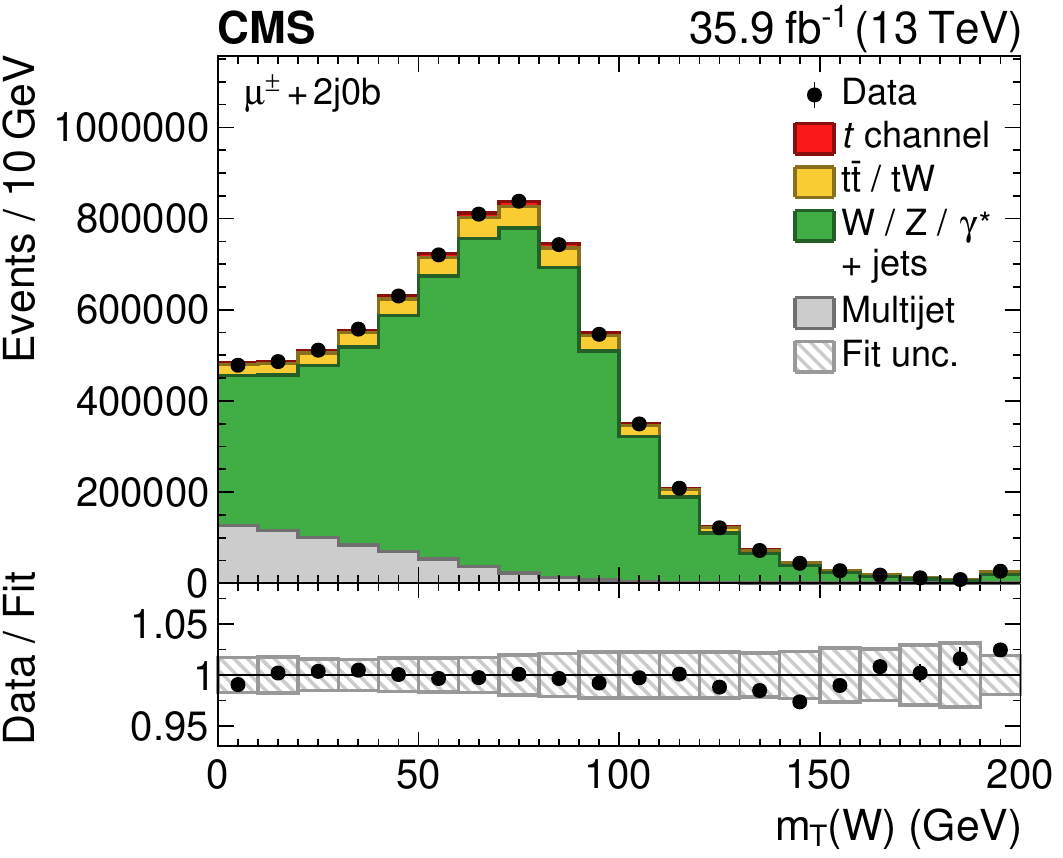}\hspace{0.03\textwidth}
\includegraphics[width=0.48\textwidth]{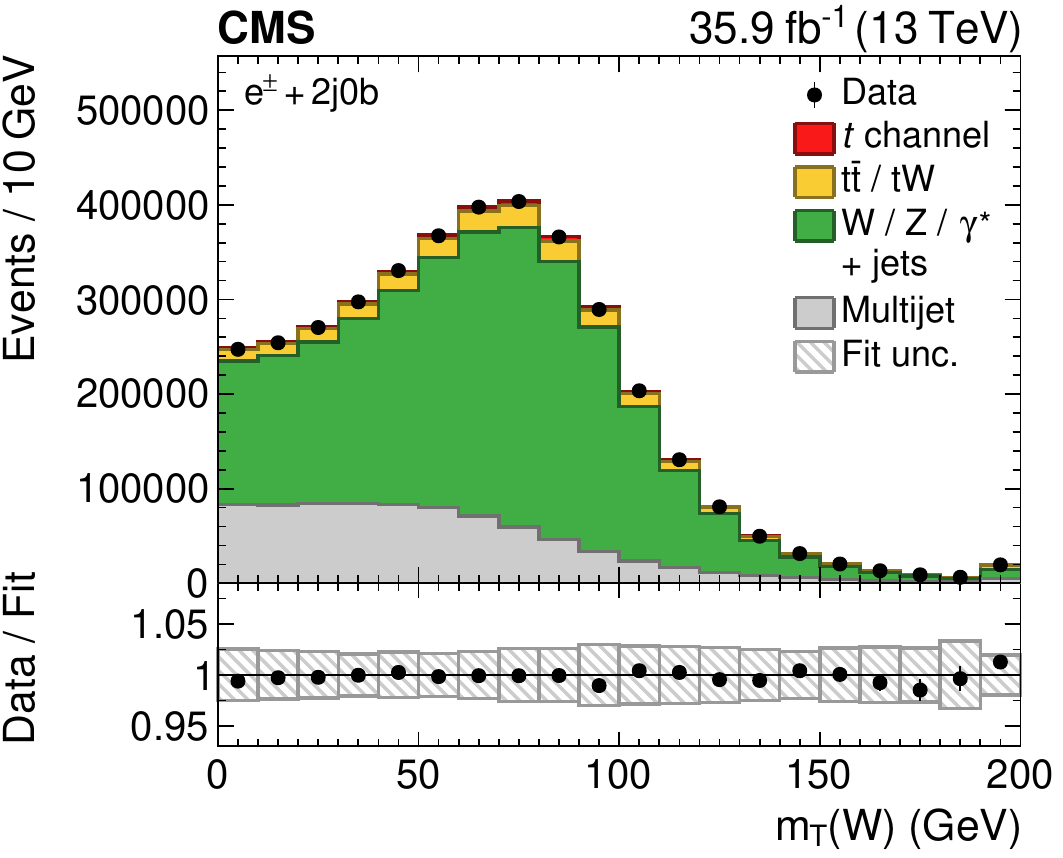}

\caption{\label{fig:multijet-mtw} Distributions of the transverse \PW~boson mass in the 2~jets, 0~\PQb~tag control category for the (left)~muon and (right)~electron channels after scaling the simulated and multijet templates to the result of a dedicated ML~fit performed on this category of events. The hatched band displays the fit uncertainty. The lower plots give the ratio of the data to the fit results. The right-most bins include the event overflows.}
\end{figure*}

The template shapes have been validated for various observables in the 2j0\PQb \wjets control category where the fraction of selected multijet events amounts to approximately 10 (20)\% for muon (electron) events, which is comparable to those in the signal category. The $\mtw$ distributions are shown in Fig.~\ref{fig:multijet-mtw} for the muon~(left) and electron~(right) channel after the multijet templates (extracted from data) and the templates of the processes with prompt leptons (extracted from the simulated events) have been normalised to the result of a dedicated ML fit using only events in the 2j0\PQb category. This dedicated fit encompasses only two components, which are the multijet template whose yield is unconstrained in the fit, and all other processes grouped together, with a constraint of ${\pm}30$\% on their combined yield using a log-normal prior. The fit is performed while simultaneously profiling the impact of experimental systematic uncertainties (as discussed in Section~\ref{sec:systematics}) affecting the yield and shape of the templates. After the fit, the derived multijet templates and the simulated samples in both channels are found to describe the distributions of data well, thus validating the procedure for estimating the contribution of multijet events from data. For the measurement, the normalisations of the multijet templates in the 2j1\PQb and 3j2\PQb categories are estimated using a different procedure, as described in Section~\ref{sec:fit}.

\section{Signal yield estimation}
\label{sec:fit}

The number of \tchannel single top quark events in data is determined from an ML fit using the distributions of $\mtw$ and of two boosted decision tree~(BDT) discriminants in the 2j1\PQb category, and the $\mtw$ distribution in the 3j2\PQb category. Simultaneously, the background yields and the impact of the experimental systematic uncertainties, modelled using nuisance parameters that influence yield and shape, are profiled.

The first BDT, labelled  \bdttch, has been trained separately on muon and electron events to discriminate \tchannel single top quark events from \ttbar, \wjets, and multijet events using corresponding samples of simulated events. The following five observables have been chosen as input:

\begin{itemize}
\item the absolute value of the pseudorapidity of the untagged jet, $\abs{\eta(\jprime)}$;
\item the reconstructed top quark mass, \mtop;
\item the transverse \PW~boson mass, \mtw;
\item the distance in $\eta\text{--}\phi$ space ($\Delta R$) between the \PQb-tagged and the untagged jet, \Rbj;
\item the absolute difference in pseudorapidity between the \PQb-tagged jet used to reconstruct the top quark and the selected lepton, \etabl.
\end{itemize}

\begin{figure*}[htb!]
\centering
\includegraphics[width=0.48\textwidth]{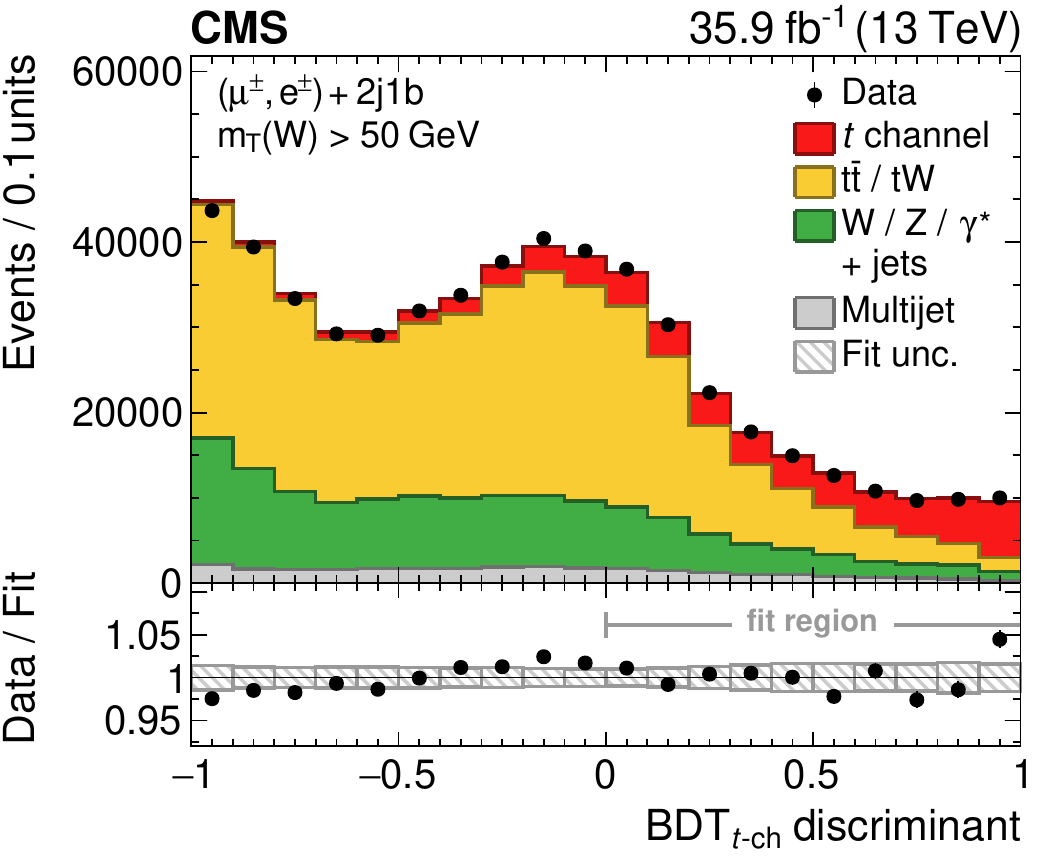}\hspace{0.03\textwidth}
\includegraphics[width=0.48\textwidth]{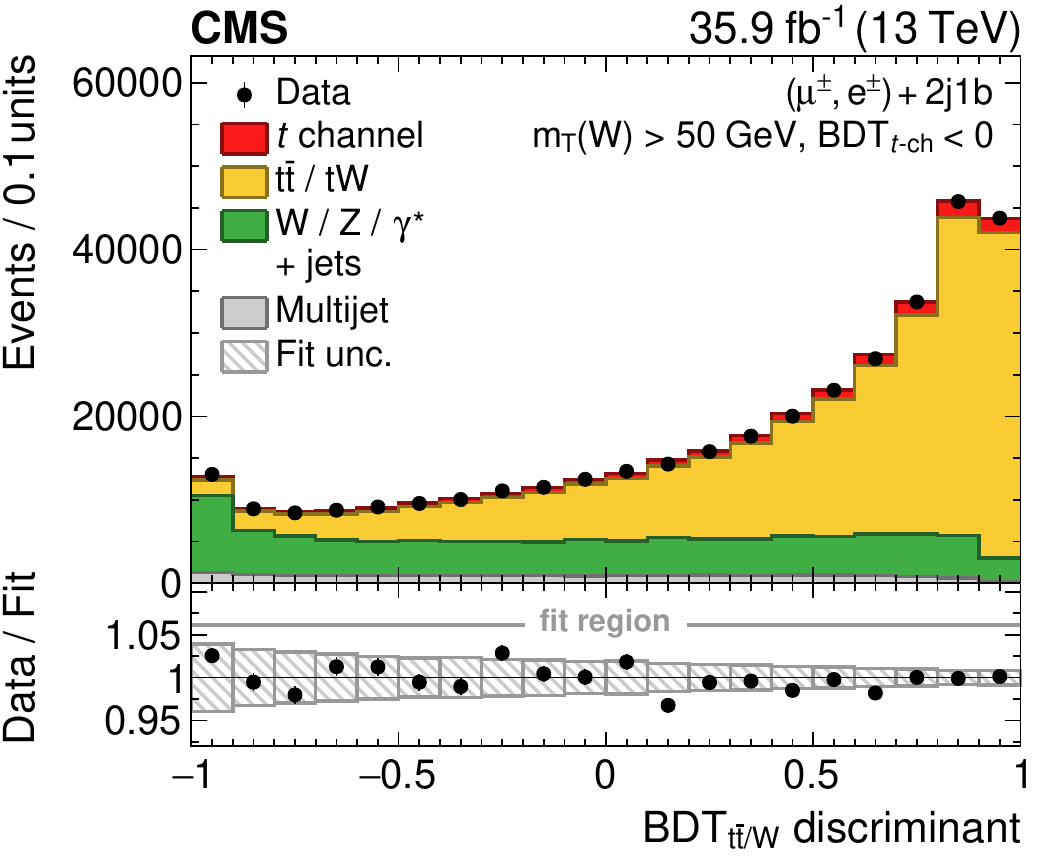}
\caption{\label{fig:fit-bdts} Distributions of the BDT discriminants in the 2~jets, 1~\PQb~tag category: (left)~\bdttch trained to separate signal from background events; (right)~\bdttw trained to separate \ttbar from \wjets events in a background-dominated category. Events in the muon and electron channels have been summed. The predictions have been scaled to the result of the inclusive ML~fit and the hatched band displays the fit uncertainty. The regions of the distributions used in the fits are indicated in the lower panels, which show the ratio of the data to the fit result.}
\end{figure*}

These have been selected based on their sensitivity for separating signal from background events, while exhibiting low correlations with the observables used to measure the differential cross sections. The resulting distribution of the \bdttch discriminant is presented in Fig.~\ref{fig:fit-bdts} (left).

The \bdttch discriminant shapes of the \wjets and \ttbar backgrounds are found to be very similar. To obtain sensitivity in the fit to both backgrounds individually, a second BDT, labelled \bdttw, has been trained separately on muon and electron events to classify events only for these two processes using the following six input observables: \mtop; \met; \Rbj; \etabl; the \PW~boson helicity angle, \cosw, defined as the angle between the lepton momentum and the negative of the top quark momentum in the \PW~boson rest frame~\cite{AguilarSaavedra:2010nx}; and the event shape $C$, defined using the momentum tensor
\begin{linenomath}
\begin{equation}
S^{ab}=\frac{\sum_{i}^{\text{jets},\Pell,\pvmiss}p_{i}^{a} p_{i}^{b}}{\sum_{i}^{\text{jets},\Pell,\pvmiss}\abs{\vec{p}_{i}}^2},
\end{equation}
\end{linenomath}
as $C=3(\lambda_1\lambda_2+\lambda_1\lambda_3+\lambda_2\lambda_3)$, where $\lambda_{1}$, $\lambda_{2}$, and $\lambda_{3}$ denote the eigenvalues of the momentum tensor $S^{ab}$ with $\lambda_1+\lambda_2+\lambda_3=1$. In the two most extreme cases, the event shape $C$ vanishes for perfectly back-to-back dijet events ($C=0$) and reaches its maximum ($C=1$) if the final-state momenta are distributed isotropically. For the measurement, the \bdttw discriminant is evaluated only in the phase space region defined by $\mtw>50$\GeV and $\bdttch<0$, which is found to be largely dominated by background events. Thus, the \bdttw input observables do not have to be selected explicitly such that they possess low correlation with the observables used to measure the differential cross sections. The resulting \bdttw discriminant distribution is displayed in Fig.~\ref{fig:fit-bdts} (right).

The ML fit is performed using the following four distributions from events in various categories:

\begin{itemize}
\item the \mtw distribution for events with $\mtw<50$\GeV in the 2j1\PQb category, which is particularly sensitive to the number of multijet events;
\item the \bdttw discriminant distribution for events with $\mtw>50$\GeV and $\bdttch<0$ in the 2j1\PQb category, which defines a region enriched in \ttbar and \wjets but depleted of signal and multijet events;
\item the \bdttch discriminant distribution for events with $\mtw>50$\GeV and $\bdttch>0$ in the 2j1\PQb category, which is enriched in signal events;
\item the \mtw distribution in the 3j2\PQb category, which provides additional sensitivity to the \ttbar yield, and thus further reduces the correlation between the estimated yields.
\end{itemize}

\begin{figure*}[htb!]
\centering
\includegraphics[width=0.48\textwidth]{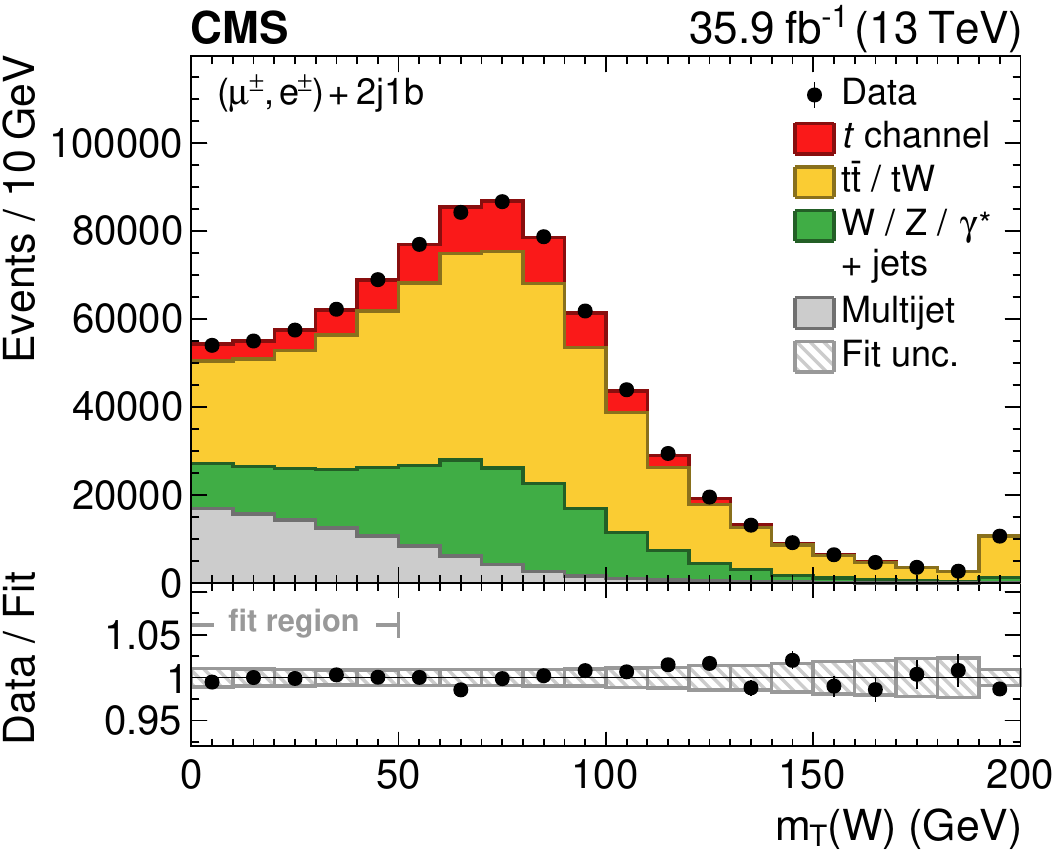}\hspace{0.03\textwidth}
\includegraphics[width=0.48\textwidth]{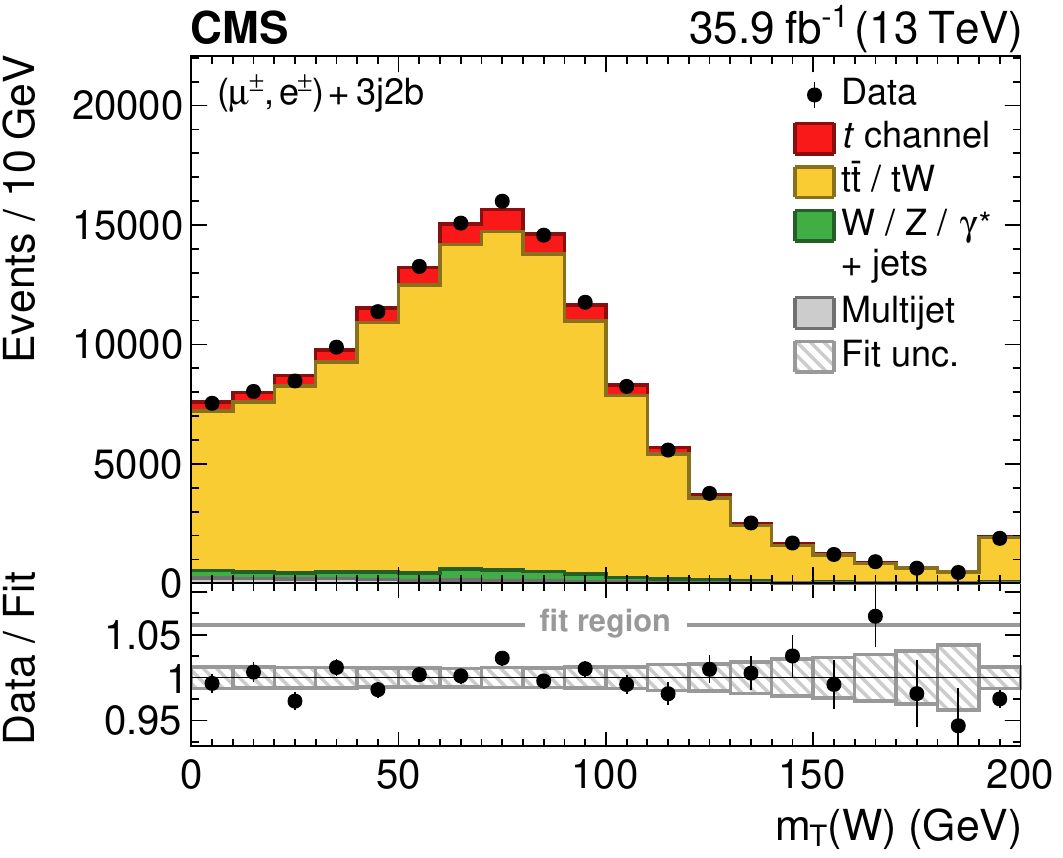}
\caption{\label{fig:fit-mtw} Distributions of the transverse \PW~boson mass for events in the (left)~2~jets, 1~\PQb~tag and (right)~3~jets, 2~\PQb~tags categories. Events in the muon and electron channels have been summed. The predictions have been scaled to the result of the inclusive ML~fit and the hatched band displays the fit uncertainty. The regions of the distributions used in the fits are indicated in the lower panels, which show the ratio of the data to the fit result. The right-most bins include the event overflows.}
\end{figure*}

The \mtw distributions in the 2j1\PQb and 3j2\PQb categories are shown in Fig.~\ref{fig:fit-mtw} on the left and right, respectively. In the fit, each distribution is split in two by separating events depending on the charge of the selected muon or electron in the event. This results in eight distributions per lepton channel and thus 16 distributions in the $\PGm/\Pe$ combined fit. A coarser equidistant binning of the distributions, as opposed to the one shown in Figs.~\ref{fig:fit-bdts} and~\ref{fig:fit-mtw}, is used in the ML fits to prevent cases where single bins are depleted of background estimates as follows: four bins are used for each of the \mtw and \bdttch distributions in the 2j1\PQb category; eight bins are used for the \bdttw distribution; and ten bins are used for the \mtw distribution in the 3j2\PQb category.

The yields of \tchannel single top quark and antiquark events are measured independently. Background events containing top quarks (\ttbar, \tw) are grouped together, and only their total yield is estimated. The top quark background yield is constrained using a log-normal prior with a width of ${\pm}10$\% to account for the uncertainty in the theoretical \ttbar and \tw production cross sections, and the uncertainty when two out of the four jets expected from semileptonic \ttbar production are not within the acceptance, as is the case in the 2j1\PQb category. The electroweak background processes, \wjets and \zjets, are grouped together as well, and an uncertainty of ${\pm}30$\% in their combined yield is applied using a log-normal prior constraint. This is motivated by the theoretical uncertainty in the modelling of the \PW{} and $\PZ/\cPgg^{*}$ production rates in association with two or more (heavy-flavour) jets~\cite{Kallweit:2015dum,Anger:2017glm}. The yields of multijet events are assumed to be independent per lepton type and event category. Their yields are constrained by a log-normal prior with a width of ${\pm}100$\% with respect to the template normalisations obtained from data in the sideband regions. In addition, an uncertainty in the predicted lepton charge ratio per background process, accounting for charge misreconstruction and uncertainties in the charge ratio~\cite{Kom:2010mv}, is taken into account using a Gaussian prior with a width of ${\pm}1$\% in the fit, for a total of 14 fit parameters. The impact of the finite number of simulated events on the templates is accounted for by employing the ``Barlow--Beeston-lite'' method~\cite{BARLOW1993219}.

Experimental systematic uncertainties, as detailed in Section~\ref{sec:systematics}, are profiled in the fit simultaneously with the yields and charge ratios. Each source is assigned a nuisance parameter according to which the shape and yield of the fit templates are modified.

The resulting event yields from a simultaneous fit to the data in the muon and electron channels are listed in Table~\ref{tab:fit-yields}. Overall, the distributions used in the fit, shown in Figs.~\ref{fig:fit-bdts} and~\ref{fig:fit-mtw}, are found to be well modelled by the samples of simulated events and the multijet templates from data after normalising them to the fit result.

\begin{table*}[ht!]
\centering
\topcaption{\label{tab:fit-yields}Measured and observed event yields in the 2j1\PQb category for each lepton channel and charge. The uncertainties in the yields are the combination of statistical and experimental systematic uncertainties.}
\begin{tabular}{ @{}r r@{$\,\pm\,$}l r@{$\,\pm\,$}l r@{$\,\pm\,$}l r@{$\,\pm\,$}l@{} }
\hline
Process &  \multicolumn{2}{c}{\PGmp} &  \multicolumn{2}{c }{\PGmm}           & \multicolumn{2}{c}{\Pep} &  \multicolumn{2}{c}{\Pem}  \\
\hline
                       \wzjets     &     72\,000 & 6\,800   &     62\,800 & 5\,600   &     33\,400 & 3\,200   &     30\,700 & 2\,800    \\
                    \ttbar/\tw     &    142\,400 & 2\,400   &    143\,400 & 2\,500   &     84\,500 & 1\,400   &     84\,800 & 1\,500    \\
                      Multijet     &     35\,150 &  550   &     35\,710 &  760       &     13\,500 & 1\,000   &     12\,700 & 1\,000    \\
         $t$~channel (top quark)   &     34\,400 & 1\,500   &         10 &    3      &     17\,720 &  820   &        27 &    2    \\
     $t$~channel (top antiquark)   &        13 &    2   &     21\,600 & 1\,600       &        25 &    3   &     11\,460 &  880   \\[\cmsTabSkip]
                         Total     &    284\,100 & 5\,800   &    263\,700 & 4\,600   &    149\,300 & 2\,400   &    139\,700 & 2\,200    \\
                          Data     &  \multicolumn{2}{c}{283\,391}   &  \multicolumn{2}{c}{260\,044} &  \multicolumn{2}{c}{148\,418}   &  \multicolumn{2}{c}{138\,781}    \\

\hline
\end{tabular}
\end{table*}

For each differential cross section measurement, the observable of interest is divided into intervals, discussed in Section~\ref{sec:unfolding}, and a fit is performed in which the signal and background yields can vary independently in each of the intervals. The likelihood $L$ to be maximised in such fits can be expressed as
\begin{linenomath}
\ifthenelse{\boolean{cms@external}}
{
\begin{align}
&\ln\left(L(\vec{\beta},\vec{\nu},\vec{R})\right)=\nonumber\\
&\hphantom{=}-\sum_{k}^\text{dist}\sum_{j}^\text{int}\sum_{i}^\text{bins}\left(d_{kji}\ln{p_{kji}(\vec{\beta_{j}},\vec{\nu},\vec{R})-p_{kji}(\vec{\beta_{j}},\vec{\nu},\vec{R})}\right)\nonumber\\
&\hphantom{=}+\text{constraints},
\end{align}
}
{
\begin{equation}
\ln\left(L(\vec{\beta},\vec{\nu},\vec{R})\right)=-\sum_{k}^\text{dist}\sum_{j}^\text{int}\sum_{i}^\text{bins}\left(d_{kji}\ln{p_{kji}(\vec{\beta_{j}},\vec{\nu},\vec{R})-p_{kji}(\vec{\beta_{j}},\vec{\nu},\vec{R})}\right)+\text{constraints},
\end{equation}
}
\end{linenomath}
where $d$ denotes the number of observed events and $p$ is the estimated yield. The summation over $k$ denotes the 16 distributions (``dist''), $j$ denotes the interval (``int'') in the observable (e.g. for the top quark \pt: 0--50\GeV, 50--80\GeV, 80--120\GeV, 120--180\GeV, and 180--300\GeV), and $i$ denotes a bin in one of the 16 distributions per interval. The prediction $\vec{p}_{kj}$, which includes all bins $i$ for distribution $k$ and interval $j$, is given by
\begin{linenomath}
\ifthenelse{\boolean{cms@external}}
{
\begin{align}
\vec{p}_{kj}(\vec{\beta}_{j},\vec{\nu},\vec{R})&=\beta_{\PQt,j}\vec{T}_{\PQt,kj}^{\tch}(\vec{\nu})+\beta_{\PAQt,j}\vec{T}_{\PAQt,kj}^{\tch}(\vec{\nu})\nonumber\\
&\hphantom{=}+\beta_{\ttbar/\tw,j}\vec{T}_{kj}^{\ttbar/\tw}(R_{j},\vec{\nu})\nonumber\\
&\hphantom{=}+\beta_{\wzjets,j}\vec{T}_{kj}^{\wzjets}(R_{j},\vec{\nu})\nonumber\\
&\hphantom{=}+\beta_{\text{multijet},j}(\Pell,r)\vec{T}_{kj}^{\text{multijet}}(R_{j}(\Pell,r),\vec{\nu}),
\end{align}
}
{
\begin{align}
\vec{p}_{kj}(\vec{\beta}_{j},\vec{\nu},\vec{R})&=\beta_{\PQt,j}\vec{T}_{\PQt,kj}^{\tch}(\vec{\nu})+\beta_{\PAQt,j}\vec{T}_{\PAQt,kj}^{\tch}(\vec{\nu})\nonumber\\
&\hphantom{=}+\beta_{\ttbar/\tw,j}\vec{T}_{kj}^{\ttbar/\tw}(R_{j},\vec{\nu})+\beta_{\wzjets,j}\vec{T}_{kj}^{\wzjets}(R_{j},\vec{\nu})\nonumber\\
&\hphantom{=}+\beta_{\text{multijet},j}(\Pell,r)\vec{T}_{kj}^{\text{multijet}}(R_{j}(\Pell,r),\vec{\nu}),
\end{align}
}
\end{linenomath}
where $\vec{\nu}$ are the nuisance parameters, $R$ the charge ratios of each background process, and $\beta$ the normalisations of the templates $\vec{T}$, which are independent per lepton flavour $\Pell$ and category $r\in$\{2j1\PQb, 3j2{\PQb}\} for the multijet templates. The profiling of systematic uncertainties leads to a correlation between the \tchannel top quark and antiquark yields in the same interval of about 20--30\%. These correlations are propagated to the differential cross sections for each top quark charge, and are accounted for when calculating their sum and ratio.

Since the kinematic selection of electron events is restricted to $\pt>35$\GeV and $\abs{\eta}<1.48$, which is tighter than for muon events~($\pt>26$\GeV, $\abs{\eta}<2.4$), the signal yields in the lowest interval of the lepton $\pt$ and in the highest two intervals of the lepton rapidity spectra are estimated from the muon channel alone in the combined $\PGm/\Pe$ fit.

\section{Validation of signal and background modelling}
\label{sec:validation}

The distributions of the observables that are unfolded are validated by comparing the predictions to the data in a background-dominated as well as in a signal-enriched region before unfolding. Both regions are defined for events in the 2j1\PQb category that also satisfy $\mtw>50$\GeV to suppress the contribution from multijet production. The modelling of the \ttbar/\tw and \wzjets backgrounds is validated in a background-dominated region obtained from events having $\bdttch<0$. To validate the modelling of the \tchannel process, events are instead required to pass $\bdttch>0.7$, resulting in a sample enriched in signal events. These two regions and their selections are only defined and applied for validation purposes, and not used for measuring the differential cross sections for which the individual fit results are used in the unfolding instead.

\begin{figure*}[hp!]
\centering
\includegraphics[width=0.48\textwidth]{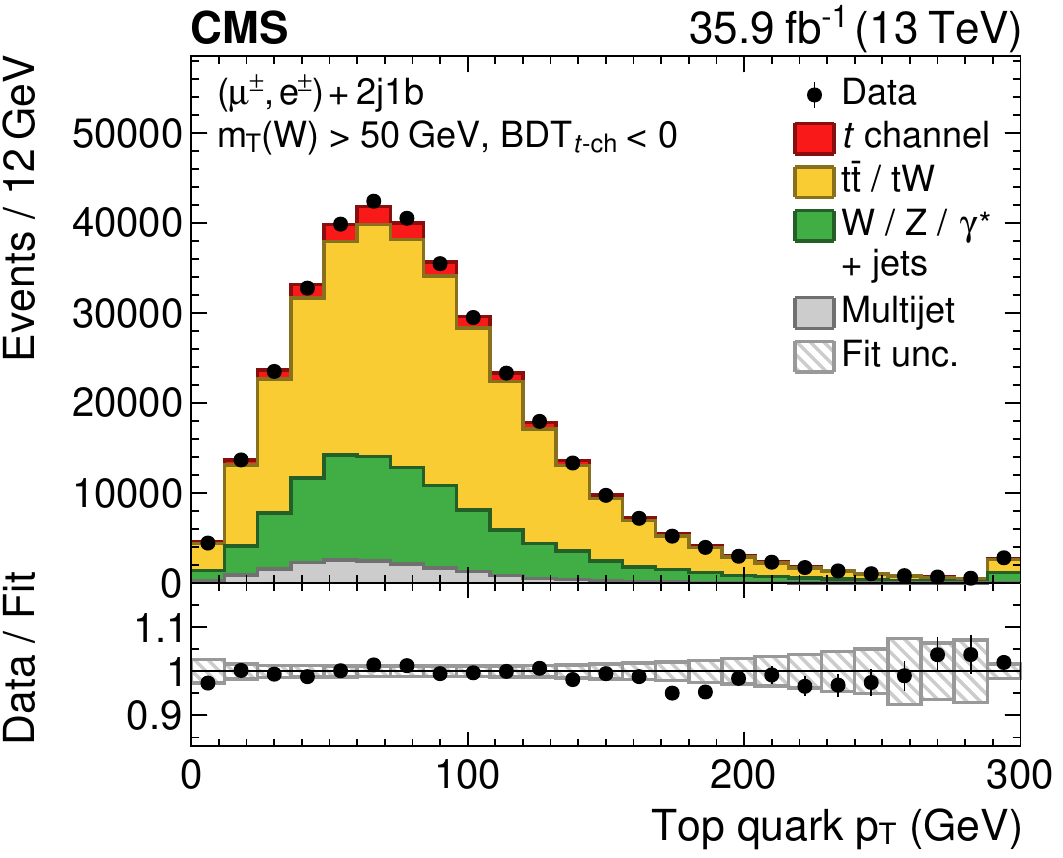}\hspace{0.03\textwidth}
\includegraphics[width=0.48\textwidth]{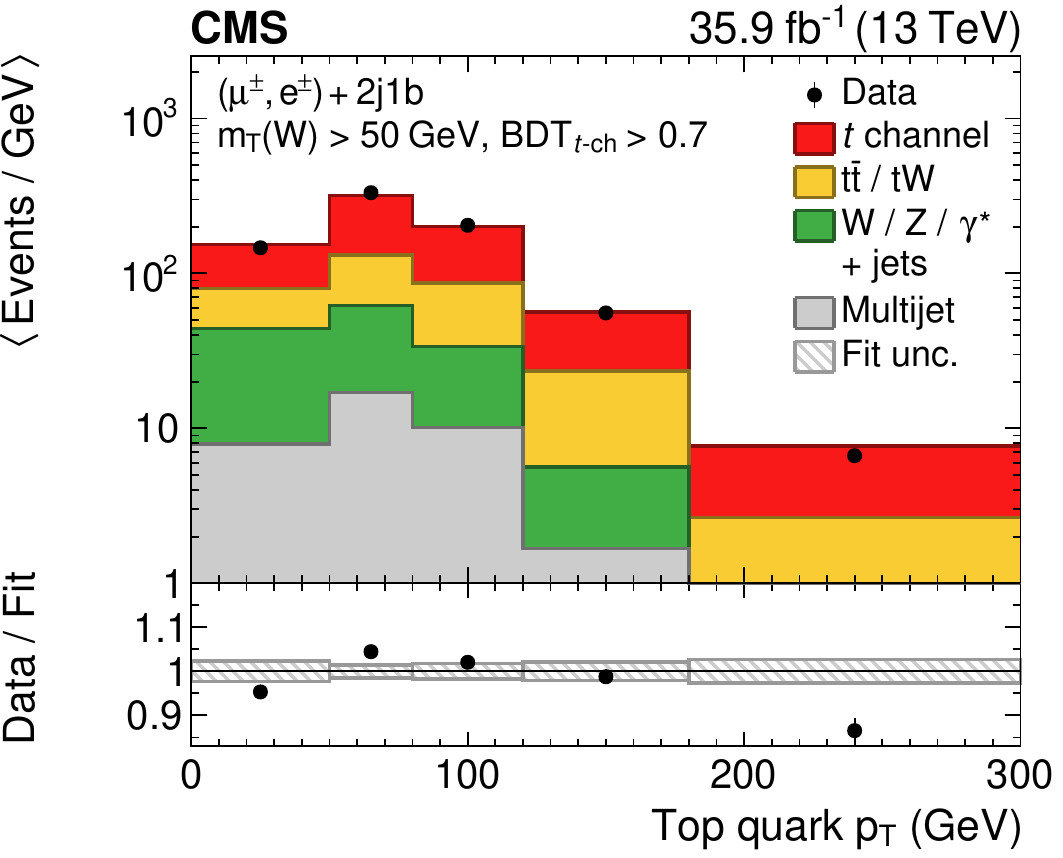}\\[0.015\textheight]
\includegraphics[width=0.48\textwidth]{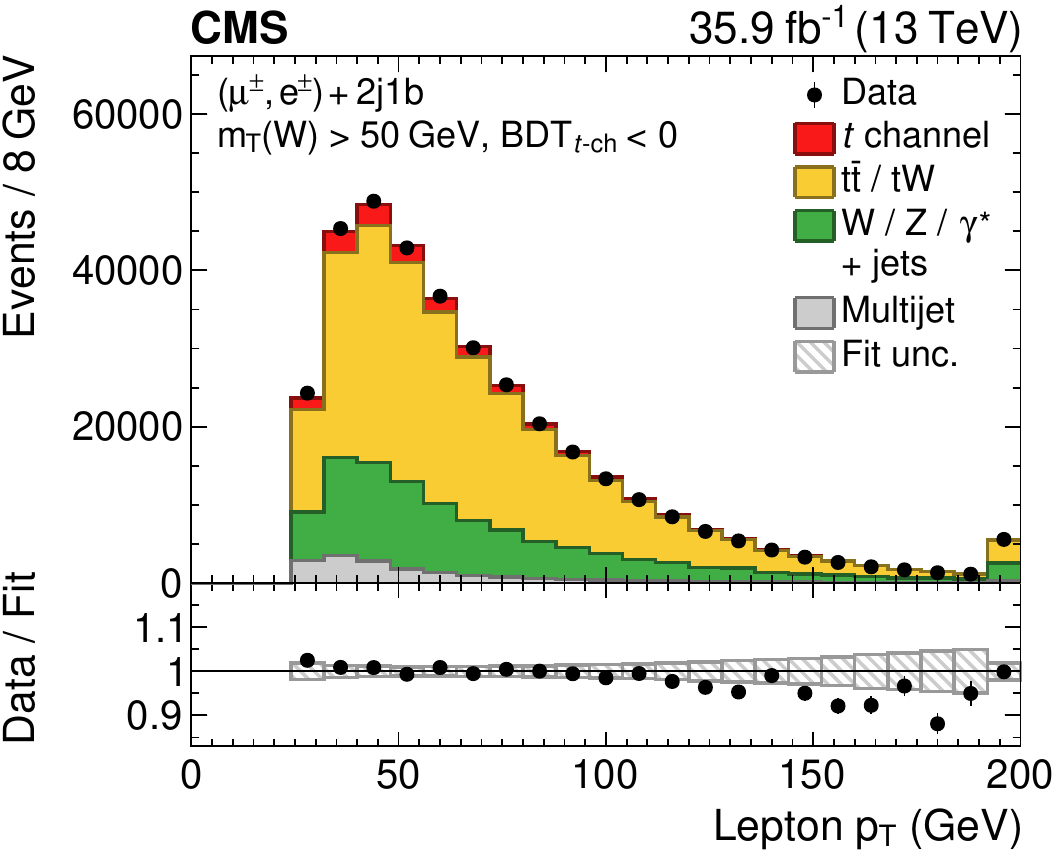}\hspace{0.03\textwidth}
\includegraphics[width=0.48\textwidth]{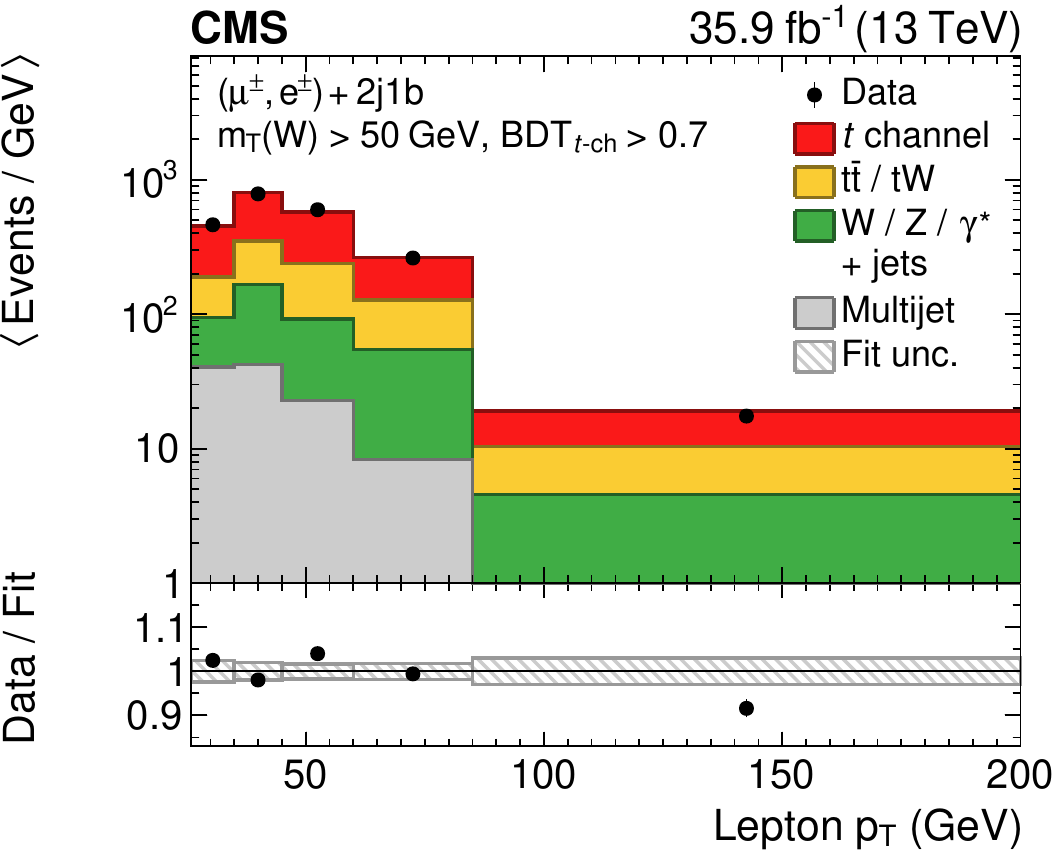}\\[0.015\textheight]
\includegraphics[width=0.48\textwidth]{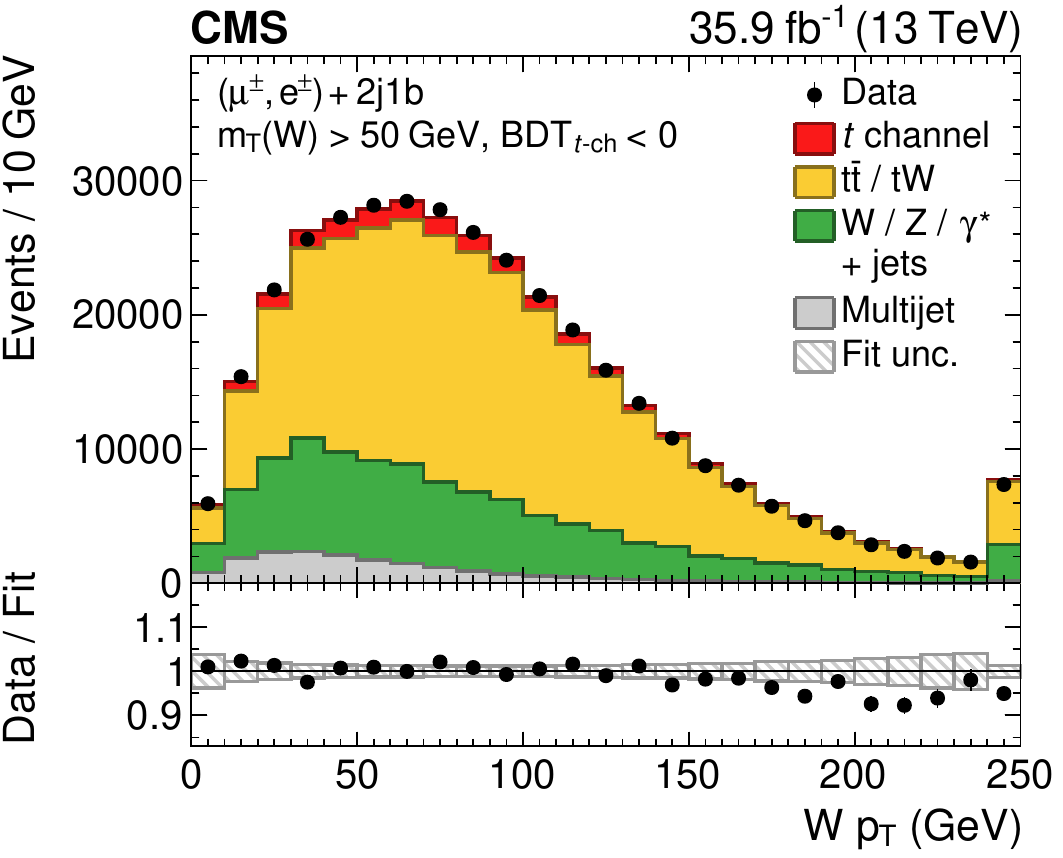}\hspace{0.03\textwidth}
\includegraphics[width=0.48\textwidth]{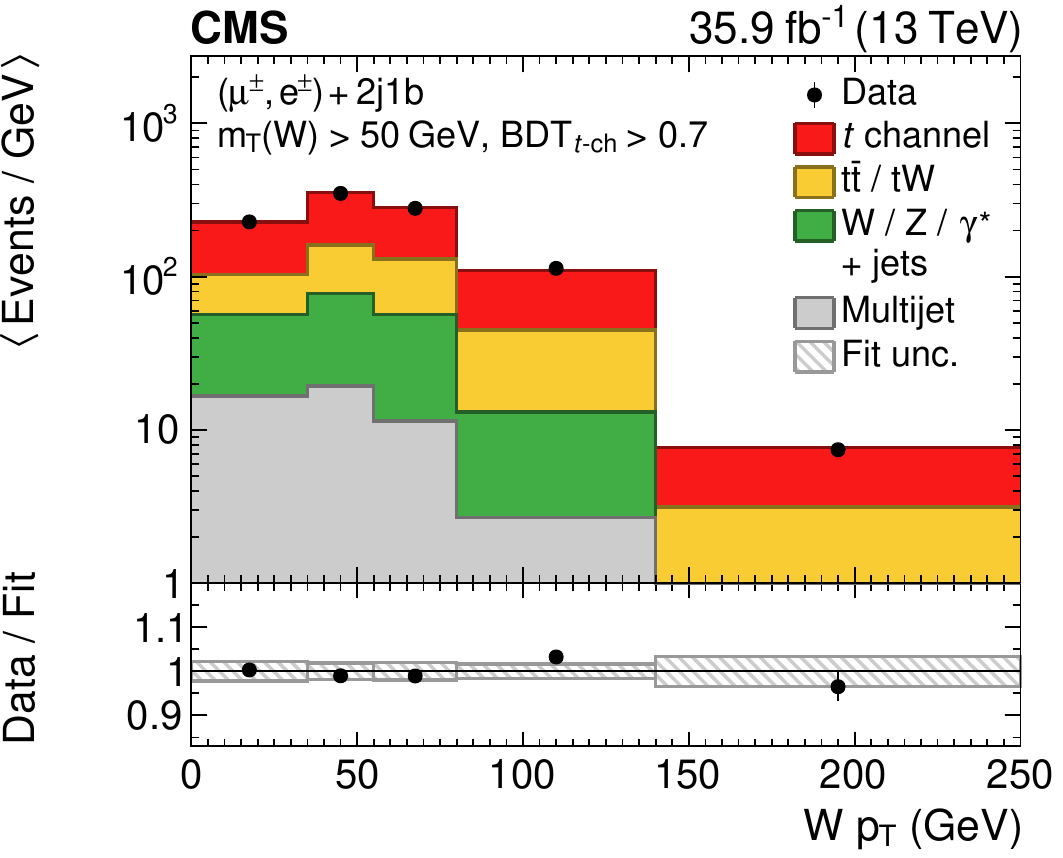}
\caption{\label{fig:validation-dists1} Distributions of the observables in a (left column)~background-dominated and a (right column)~signal-enriched region for events passing the 2~jets, 1~\PQb~tag selection: (upper row)~top quark \pt; (middle row)~charged lepton \pt; (lower row)~\PW~boson \pt. Events in the muon and electron channels have been summed. The predictions have been scaled to the result of the inclusive ML~fit and the hatched band displays the fit uncertainty. The plots on the left give the number of events per bin, while those on the right show the number of events per bin divided by the bin width. The lower panel in each plot gives the ratio of the data to the fit results. The right-most bins include the event overflows.}
\end{figure*}

\begin{figure*}[hp!]
\centering
\includegraphics[width=0.48\textwidth]{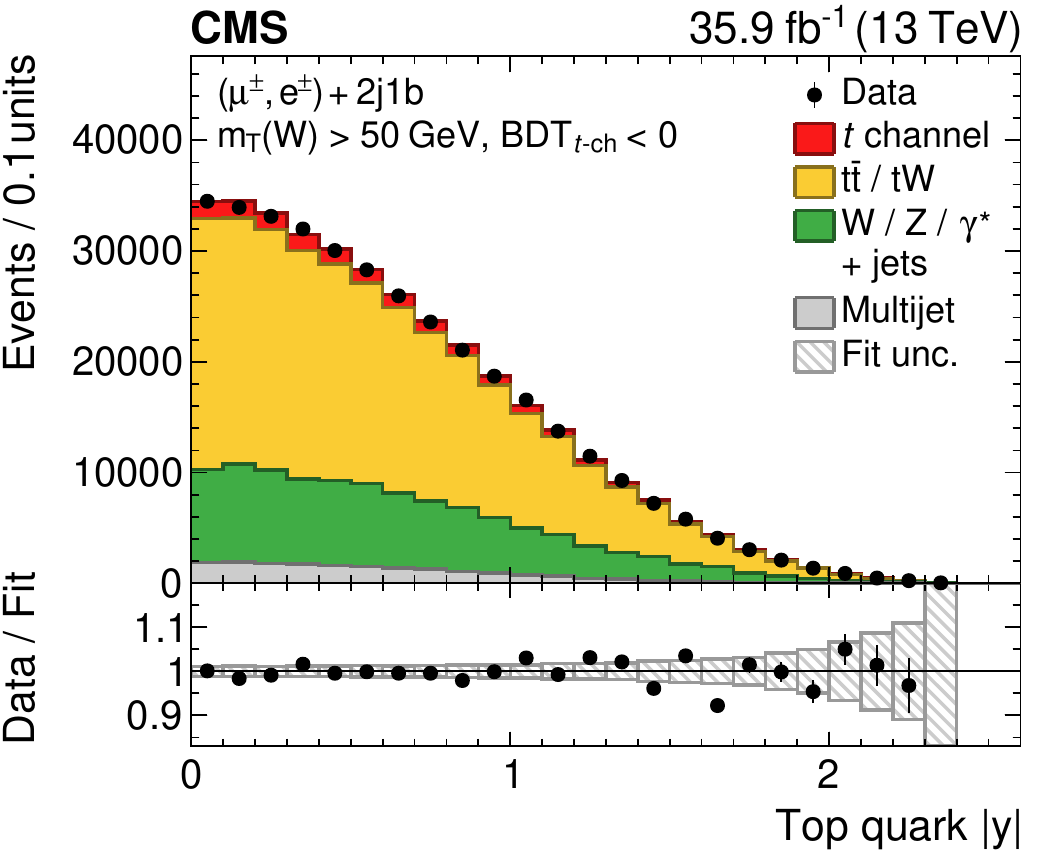}\hspace{0.03\textwidth}
\includegraphics[width=0.48\textwidth]{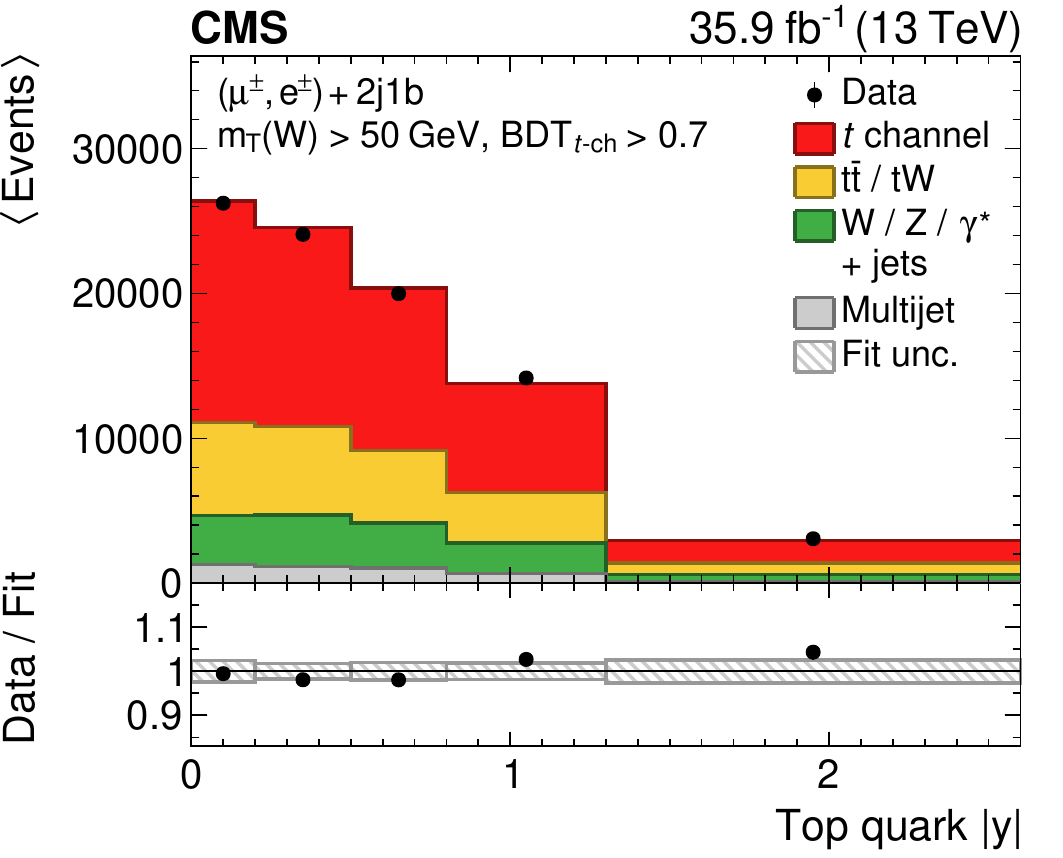}\\[0.015\textheight]
\includegraphics[width=0.48\textwidth]{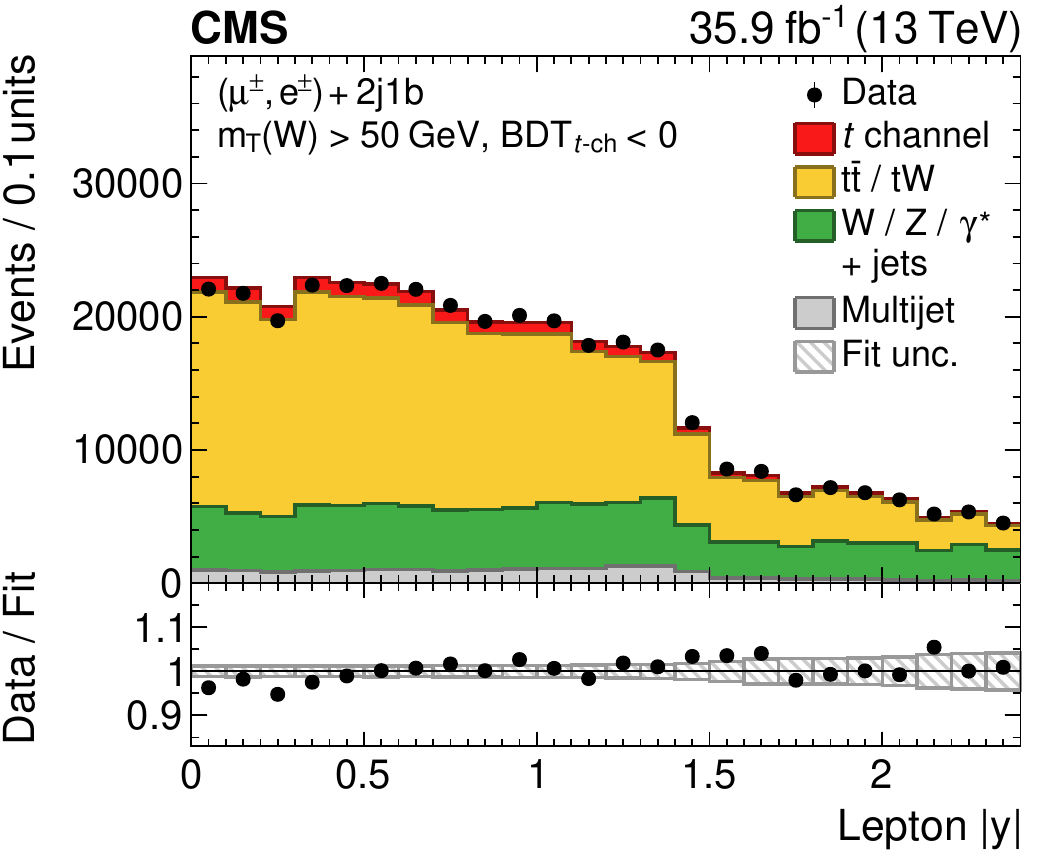}\hspace{0.03\textwidth}
\includegraphics[width=0.48\textwidth]{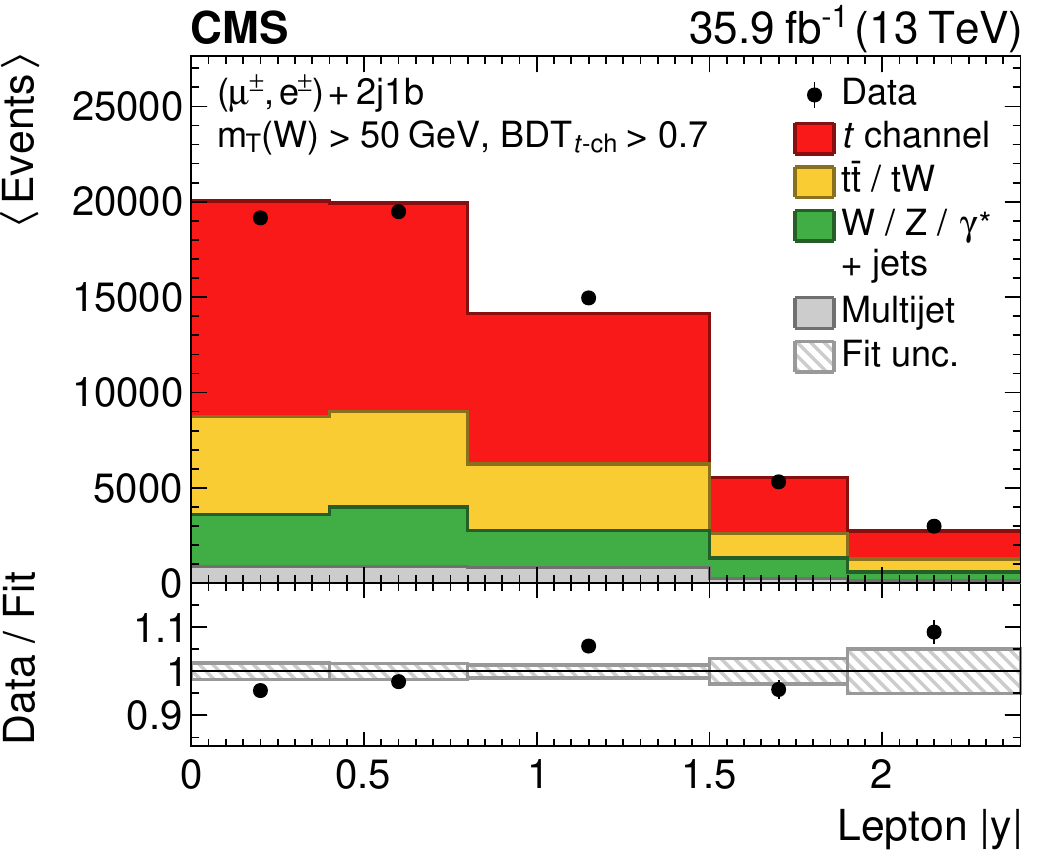}\\[0.015\textheight]
\includegraphics[width=0.48\textwidth]{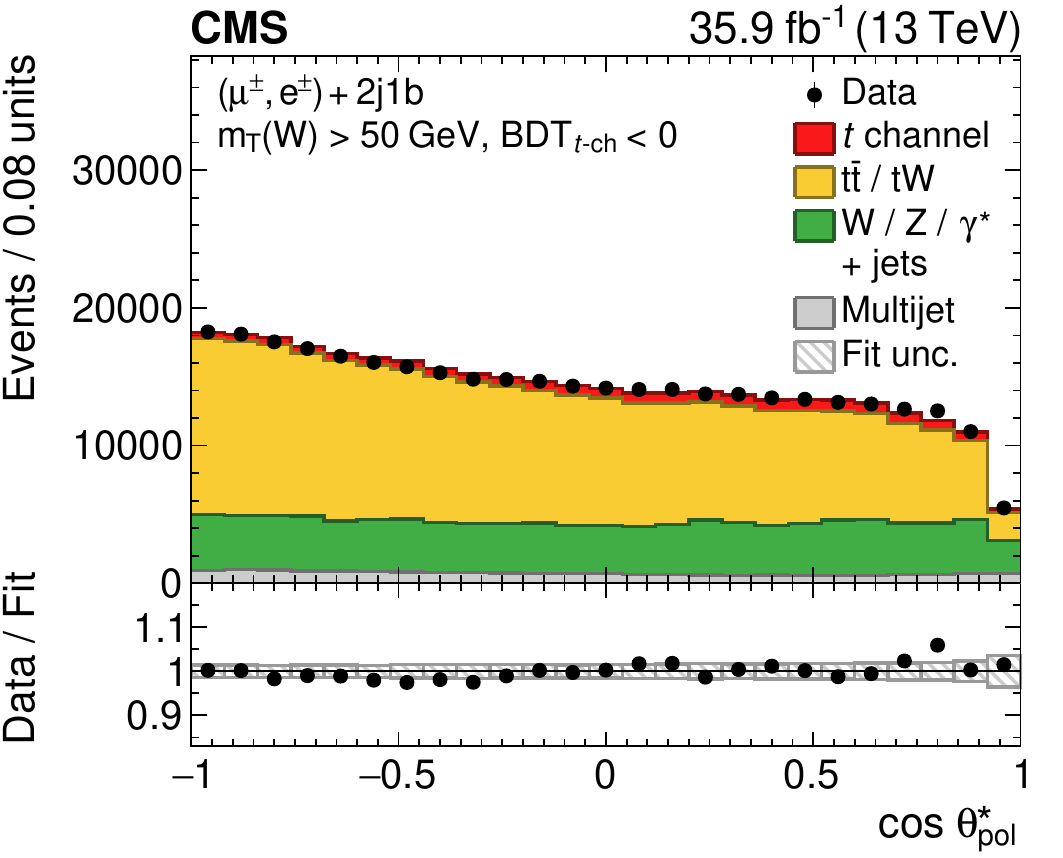}\hspace{0.03\textwidth}
\includegraphics[width=0.48\textwidth]{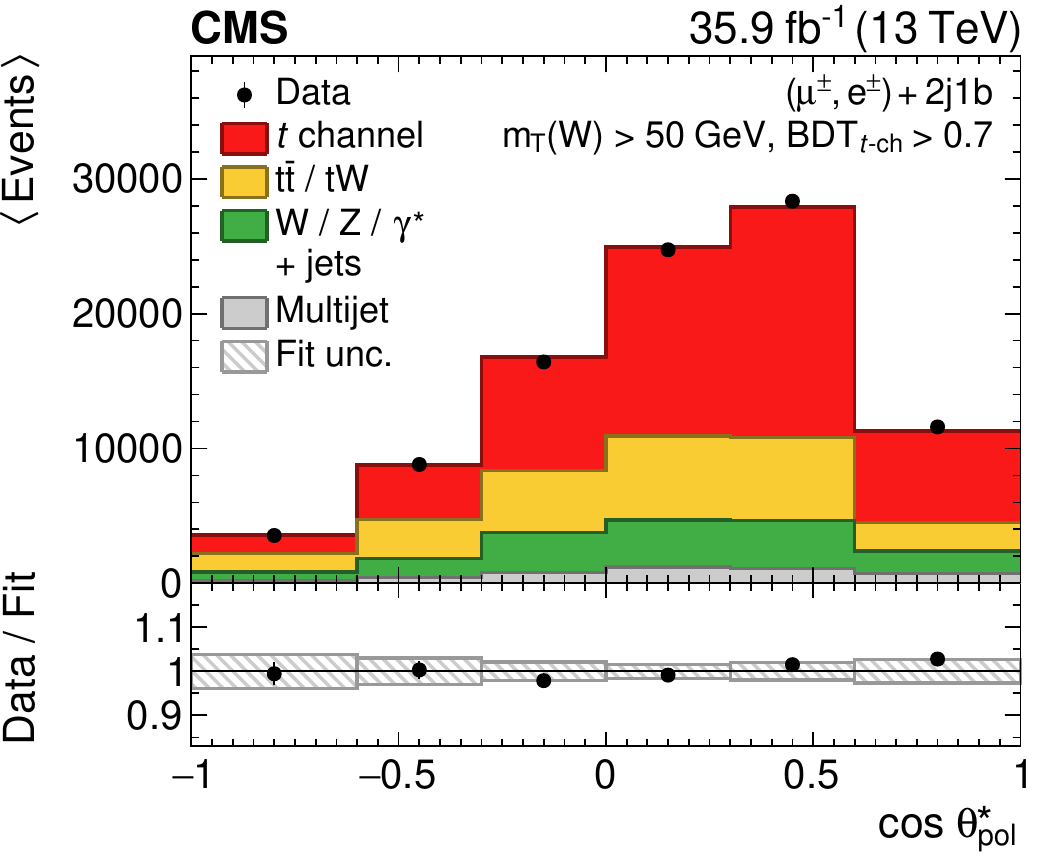}
\caption{\label{fig:validation-dists2} Distributions of the observables in a (left column)~background-dominated and a (right column)~signal-enriched region for events passing the 2~jets, 1~\PQb~tag selection: (upper row)~top quark rapidity; (middle row)~charged lepton rapidity; (lower row)~cosine of the top quark polarisation angle. Events in the muon and electron channels have been summed. The predictions have been scaled to the result of the inclusive ML~fit and the hatched band displays the fit uncertainty. The plots on the left give the number of events per bin, while those on the right show the number of events per bin divided by the bin width. The lower panel in each plot gives the ratio of the data to the fit results.}
\end{figure*}

The resulting distributions in both regions for all six observables that are unfolded are shown in Figs.~\ref{fig:validation-dists1} and~\ref{fig:validation-dists2} after the predictions have been scaled to the inclusive fit result. Overall good agreement between the data and the fit result is observed in the background-dominated region, thus validating the modelling of the \ttbar/\tw and \wzjets backgrounds. In the signal region, reasonable agreement is also observed.

\section{Unfolding}
\label{sec:unfolding}

The distributions from reconstructed events are affected by the detector resolution, selection efficiencies, and kinematic reconstruction, which lead to distortions with respect to the corresponding distributions at the parton or particle levels. The size of these effects varies with the event kinematics. In order to correct for these effects and determine the parton- and particle-level distributions, an unfolding method is applied to the reconstructed distributions. In this analysis, the \textsc{tunfold} algorithm~\cite{Schmitt:2012kp} is chosen, which treats unfolding as a minimisation problem of the function
\begin{linenomath}
\ifthenelse{\boolean{cms@external}}
{
\begin{align}
\chi^2&=\left(\vec{y}-\boldsymbol{R}\boldsymbol{\epsilon}\vec{x}\right)^\mathrm{T}\boldsymbol{V}_{y}^{-1}\left(\vec{y}-\boldsymbol{R}\boldsymbol{\epsilon}\vec{x}\right)\nonumber\\
&\hphantom{=}+\underbrace{\tau^2\left\lVert\boldsymbol{L}(\vec{x}-\vec{x}_{0})\right\rVert^2}_\text{regularisation}+\lambda\sum_{i}\left(\vec{y}-\boldsymbol{R}\boldsymbol{\epsilon}\vec{x}\right)_{i},
\end{align}
}
{
\begin{equation}
\chi^2=\left(\vec{y}-\boldsymbol{R}\boldsymbol{\epsilon}\vec{x}\right)^\mathrm{T}\boldsymbol{V}_{y}^{-1}\left(\vec{y}-\boldsymbol{R}\boldsymbol{\epsilon}\vec{x}\right)+\underbrace{\tau^2\left\lVert\boldsymbol{L}(\vec{x}-\vec{x}_{0})\right\rVert^2}_\text{regularisation}+\lambda\sum_{i}\left(\vec{y}-\boldsymbol{R}\boldsymbol{\epsilon}\vec{x}\right)_{i},
\end{equation}
}
\end{linenomath}
where $\vec{y}$ denotes the measured yields in data, $\boldsymbol{V}_{y}$ is the covariance matrix of the measured yields, and $\vec{x}$ is the corresponding differential cross section at parton or particle level. The matrices $\boldsymbol{R}$ and $\boldsymbol{\epsilon}$ denote the transition probability and selection efficiencies, respectively, both estimated from simulation. The signal yields and covariances are estimated through ML fits using the \mtw, \bdttw, and \bdttch distributions, as detailed in Section~\ref{sec:fit}.

A penalty term, based on the curvature of the unfolded spectrum~\cite{Tikhnov1,Blobel:2002pu} encoded in the matrix $\boldsymbol{L}$, is added in the minimisation to suppress oscillating solutions originating from amplified statistical fluctuations. This ``regularisation'' procedure has a strength $\tau$ that is chosen to minimise the global correlation between the unfolded bins. The ``bias vector'' $\vec{x}_{0}$ is set to the expected spectrum from simulation. Pseudo-experiments using simulated data are performed to verify that the unfolding method estimates the uncertainties correctly, while keeping the regularisation bias at a minimum. No regularisation is applied when unfolding the lepton \pt and rapidity spectra since the migrations between bins are found to be negligible. The overall normalisation of the unfolded spectrum is determined by performing a simultaneous minimisation with respect to the Lagrange multiplier $\lambda$.

The parton-level top quark in simulation is defined as the generated on-shell top quark after quantum electrodynamic (QED) and QCD radiation, taking into account the intrinsic transverse momentum of initial-state partons. Events are required to contain either a muon or an electron from the top quark decay chain. This also includes muons or electrons from intermediately produced $\PGt$ leptons. In such events, the \PW~boson is chosen to be the direct daughter of the top quark. The spectator quark is selected from among the light quarks after QED and QCD radiation that are not products of the top quark decay. In case of ambiguities arising from initial-state radiation, the spectator quark that minimises the $\pt$ of the combined spectator quark and top quark system is chosen.

The top quark at the particle level (called ``pseudo top quark'') is defined in simulated events by performing an event reconstruction based on the set of stable simulated particles after hadronisation~\cite{Collaboration:2267573}.
In the context of this study, all particles with a lifetime of more than 30\unit{ps} are considered stable. So-called ``dressed'' muons and electrons are constructed by accounting for the additional momenta carried by photons within a cone of $\Delta R<0.1$ around the corresponding prompt lepton that do not originate from hadronisation products. The $\pvmiss$ is defined as the summed momentum of all prompt neutrinos in the event. Jets at the particle level are clustered from all stable particles excluding prompt muons, prompt electrons, prompt photons, and all neutrinos using the anti-\kt algorithm with a distance parameter of $R=0.4$. From these objects, a pseudo top quark is reconstructed by first solving for the unknown neutrino $p_{z}$ momentum, which is identical to the top quark reconstruction procedure applied to data, as described in Section~\ref{sec:selection}. Events containing a single dressed muon or electron with $\pt>26$\GeV and $\abs{\eta}<2.4$, together with two jets with $\pt>40$\GeV and $\abs{\eta}<4.7$, are considered at the particle level. Jets that are closer than $\Delta R=0.4$ to the selected dressed muon or electron are ignored. The jet that yields a top quark mass closest to 172.5\GeV is assumed to come from the top quark decay, while the other jet is taken as the spectator jet.

The size of the binning intervals are chosen to minimise the migrations between the reconstructed bins while retaining sensitivity to the shapes of the distributions. The stability (purity) is defined as the probability that the parton- or particle-level (reconstructed) values of an observable within a certain range also have their reconstructed (parton-/particle-level) counterparts in the same range. Both quantities are found to be greater than or equal to 50\% in most bins of all distributions, with the exception of a few bins at the parton level where purity and stability drop to 40\%, and the first two bins of the polarisation angle distribution at the parton level where both quantities drop to about 25\%. The stability and purity values are about 10\% larger for the particle-level distributions than for the parton-level ones. The acceptance times efficiency for selecting \tchannel single top quark events at the detector level is found to be 2--8 (20--30)\% for muon events and 1--5 (10--20)\% for electron events with respect to the parton (particle) level across the unfolding bins.

\section{Systematic uncertainties}
\label{sec:systematics}

{\tolerance=800
The measurements are affected by various sources of systematic uncertainty. For each systematic variation, new templates and response matrices are derived. Systematic variations can create correlations between the \tchannel top quark and antiquark yields since both yields are estimated simultaneously from data through an ML fit, as described in Section~\ref{sec:fit}.\par
}

The following experimental systematic uncertainties are profiled in the ML fit.

\begin{itemize}
\item Background composition: As described in Section~\ref{sec:fit}, the \zjets and \wjets processes and the \ttbar and \tw processes are separately grouped together in the ML fit. The ratios of the \zjets to the \wjets yields and the \ttbar to the \tw yields are assigned a ${\pm}20$\% uncertainty. This covers the uncertainty in the small  \zjets and \tw yields, for which the analysis has little sensitivity.

{\tolerance=800
\item Multijet shape estimation: The multijet event distributions are estimated from data by inversion of the muon isolation criterion or the electron identification criteria. The uncertainty in the shape of these distributions is estimated by varying the
criteria. The requirement on the muon isolation parameter in the sideband region is modified from $\muiso>20$\% to either $20 < \muiso < 40$\% or $\muiso > 40$\%, and the electron isolation parameter to either $\eiso < 30$\% or $\eiso > 5.88$\%, while inverting the identification criteria. Another variation is done by requiring electrons in the sideband region to explicitly pass or fail the photon conversion criterion, which is also part of the electron identification requirement.\par
}

{\tolerance=9999
\item Efficiency of \cPqb~tagging and misidentification:
The scale factors used to reweight the \cPqb~tagging and misidentification efficiencies in simulation to the ones estimated from data are varied within their uncertainties based on the true flavour of the selected jets~\cite{Sirunyan:2017ezt}.\par
}

{\tolerance=800
\item Jet energy scale and resolution:
The jet energy scale and resolution corrections are varied within their uncertainties~\cite{Chatrchyan:2011ds}. The shifts induced in the jet momenta are propagated to $\pvmiss$ as well.\par
}

\item Unclustered energy:
The contributions to \met of PF candidates that have not been clustered into jets are varied within their respective energy resolutions~\cite{Khachatryan:2014gga}.

\item Pileup: The simulated distribution of pileup interactions is modified by shifting the total inelastic $\Pp\Pp$ cross section by ${\pm}5$\%~\cite{Sirunyan:2018nqx}.

\item Lepton efficiencies: The scale factors that account for differences in the lepton selection and reconstruction efficiencies between data and simulation are varied within their uncertainties~\cite{Sirunyan:2018fpa,Khachatryan:2015hwa}.

\end{itemize}

The systematic uncertainties in the theoretical modelling of the simulated samples are estimated by using new templates and response matrices in the ML fit and unfolding for each variation. For each uncertainty source, the maximum difference of the up/down variations with the result using the nominal templates and response matrix is taken as the estimated uncertainty per bin. These are added in quadrature to the experimental uncertainty per bin.

The following sources of theoretical uncertainty have been evaluated.

{\tolerance=1200
\begin{itemize}

\item Modelling of top quark \pt in \ttbar events: Differential cross section measurements of \ttbar production by CMS~\cite{Khachatryan:2015oqa,Khachatryan:2016mnb} have shown that the \pt spectrum of top quarks in \ttbar events is significantly softer than predicted by NLO simulations. To correct for this effect, simulated \ttbar events are reweighted according to the scale factors derived from measurements at 13\TeV~\cite{Khachatryan:2016mnb}. The difference in the predictions when using the default \ttbar simulation sample is taken as an additional uncertainty.

{\tolerance=8000
\item Top quark mass: The nominal top quark mass of 172.5\GeV is modified by ${\pm}0.5$\GeV in the simulation~\cite{Khachatryan:2015hba}. The difference with respect to the nominal simulation results is taken as the corresponding uncertainty.\par
}

\item Parton distribution functions: The effect of the uncertainty in the PDFs is estimated by reweighting the simulated events using the recommended variations in the NNPDF3.0 NLO set, including a variation of $\alpS$~\cite{Ball:2014uwa}. The reweighting is performed using precomputed weights stored in the event record by the matrix element generator~\cite{Kalogeropoulos:2018cke}.

\item Renormalisation/factorisation scales: A reweighting procedure similar to that used for the PDFs is carried out on simulated \tchannel, \wjets, and \ttbar simulated events to estimate the effect of the uncertainties in the renormalisation and factorisation scales. The weights correspond to independent variations by factors of 0.5 and 2 in the scales with respect to their nominal values. The envelope of all possible combinations of up-varied/down-varied scales with the exception of the extreme up/down combinations is considered as an uncertainty. This uncertainty is evaluated independently for the \tchannel, \wjets, and \ttbar simulated event samples.

\item Parton shower: The uncertainties in the parton shower simulation are evaluated by comparing the nominal samples to dedicated samples with varied shower parameters. For \tchannel single top quark production, the differences with respect to samples with a varied factorisation scale by a factor of 0.5 or 2 or with a varied \POWHEG $h_\text{damp}$ parameter are taken as two independent uncertainties. For the simulated \ttbar samples, the variation of the factorisation scale in both initial- and final-state radiation, and the $h_\text{damp}$ parameter are evaluated as three independent uncertainties.

\item Underlying event tune: The impact of uncertainties arising from the CUETP8M2T4 underlying event tune \cite{CMS-PAS-TOP-16-021} used in the simulation of \ttbar events is evaluated using dedicated samples with the tune varied within its uncertainties.

\item Colour reconnection: The default model of colour reconnection in \PYTHIA is based on multiple-particle interactions (MPI) with early resonance decays switched off. An uncertainty in the choice of this model is taken into account by repeating the measurement using three alternative models of colour reconnection in the simulation of \tchannel single top quark and \ttbar production: the MPI-based scheme with early resonance decays switched on, a gluon-move scheme~\cite{Argyropoulos:2014zoa}, and a QCD-inspired scheme~\cite{Christiansen:2015yqa}.

\item Fragmentation model: The fragmentation of \PQb quarks, modelled by the Bowler-Lund function~\cite{Bowler1981}, is varied within its uncertainties for \tchannel single top quark and \ttbar production. Additionally, the impact when using the Peterson model~\cite{Peterson} for \PQb quark fragmentation instead is assessed.

\end{itemize}
\par}

In addition, an uncertainty of ${\pm}2.5$\% in the measurement of the integrated luminosity of the data set~\cite{CMS-PAS-LUM-17-001} is taken into account by scaling the evaluated covariance matrix per observable accordingly.

\section{Results}
\label{sec:results}

\begin{figure*}[phtb!]
\centering
\includegraphics[width=0.48\textwidth]{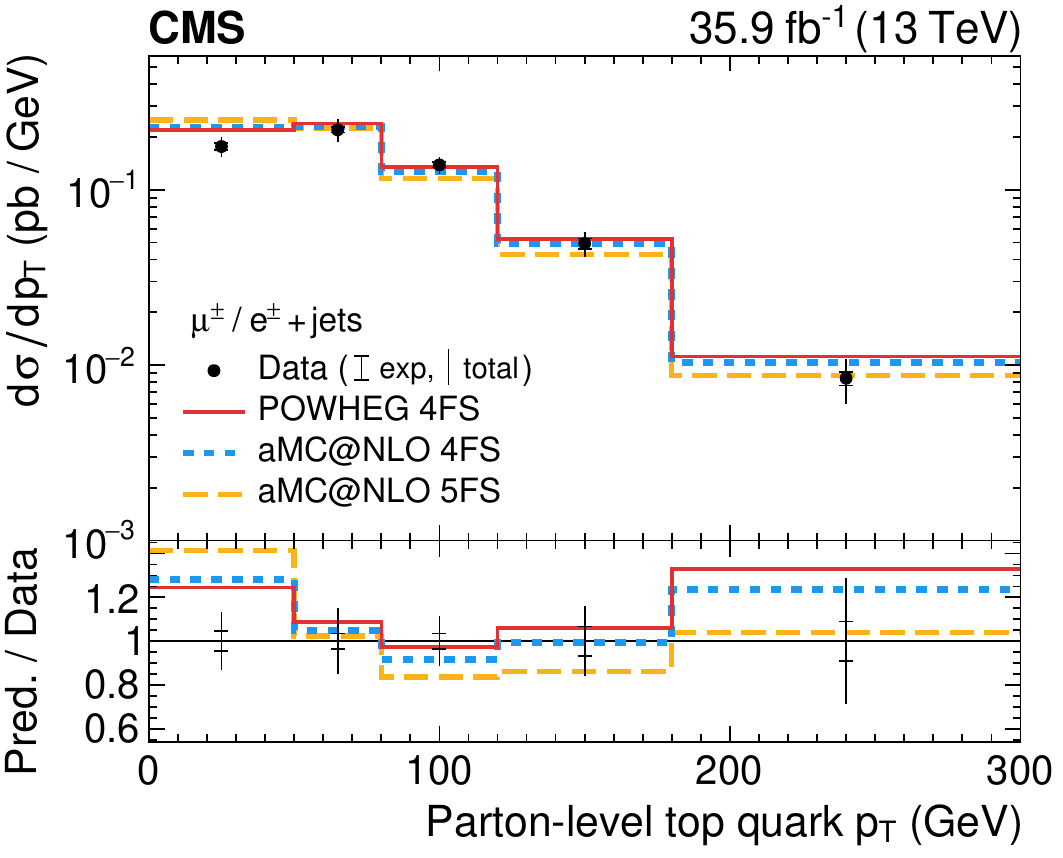}\hspace{0.03\textwidth}
\includegraphics[width=0.48\textwidth]{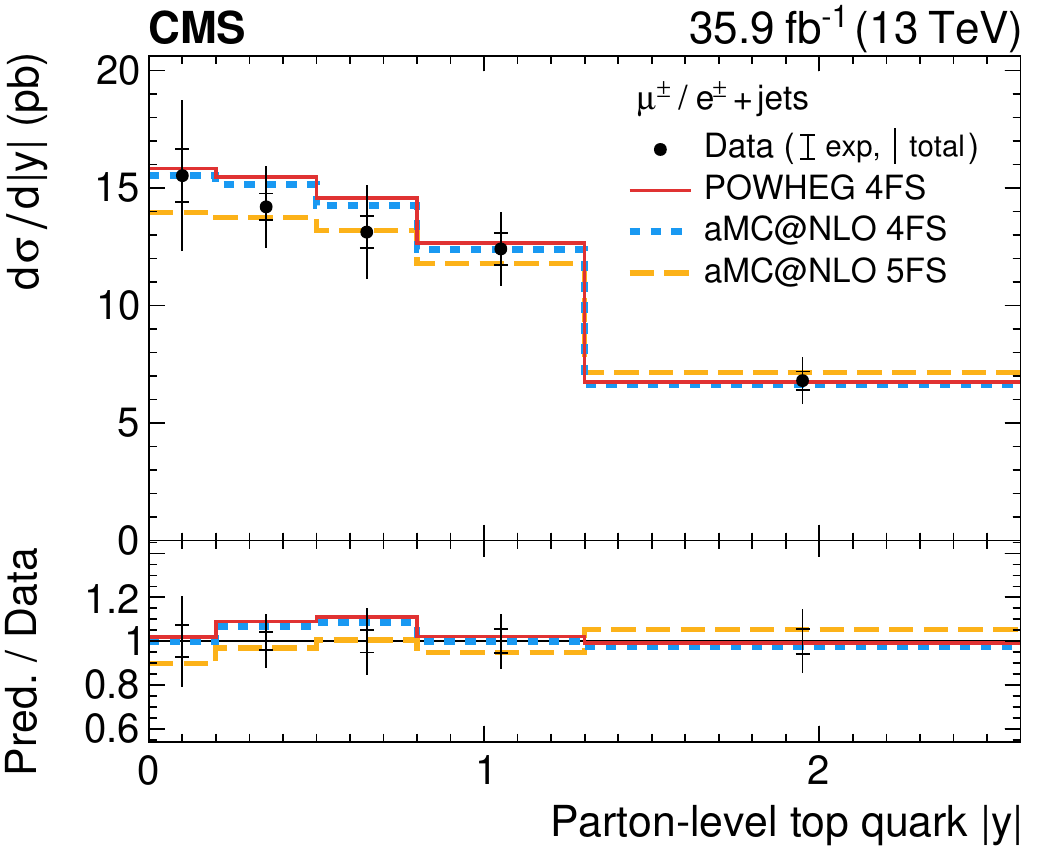}\\[0.015\textheight]
\includegraphics[width=0.48\textwidth]{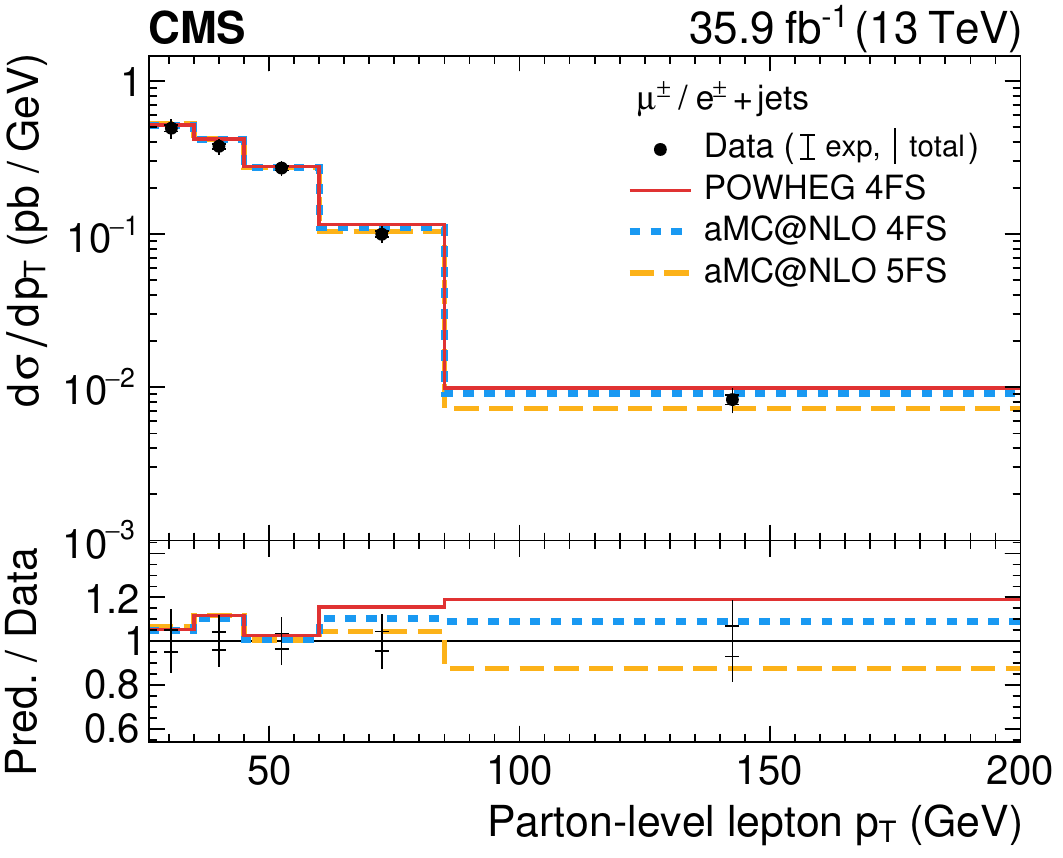}\hspace{0.03\textwidth}
\includegraphics[width=0.48\textwidth]{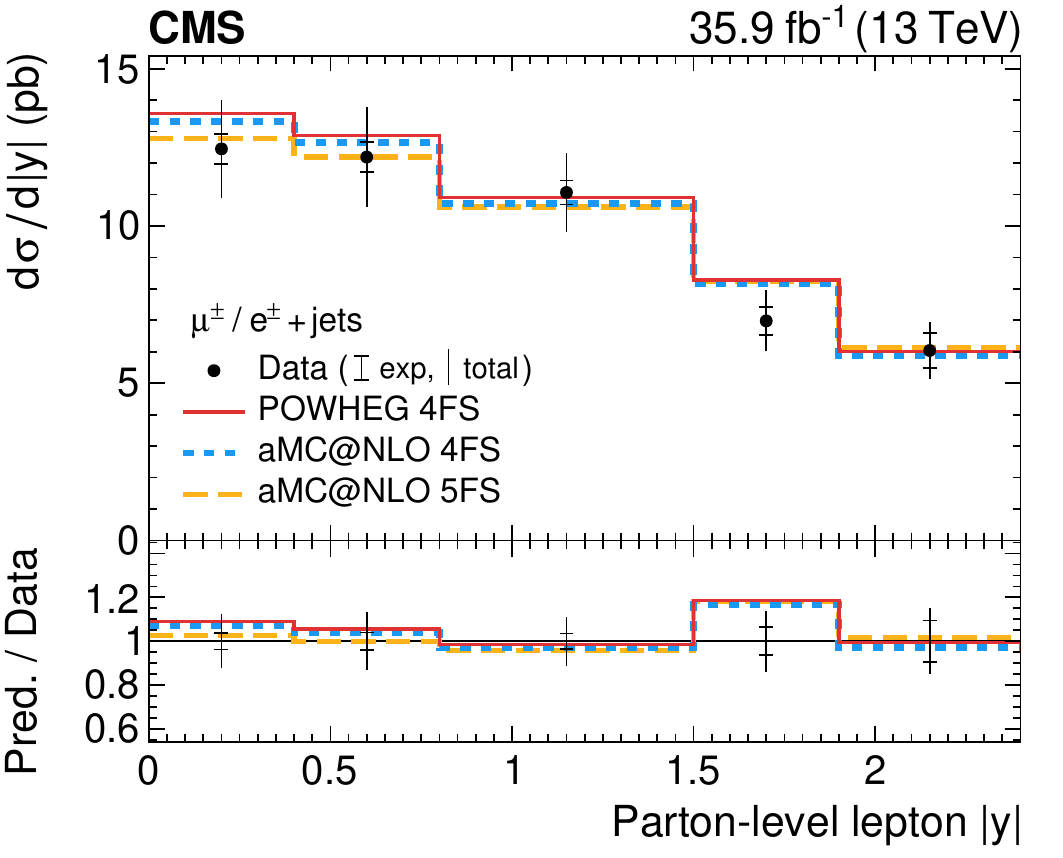}\\[0.015\textheight]
\includegraphics[width=0.48\textwidth]{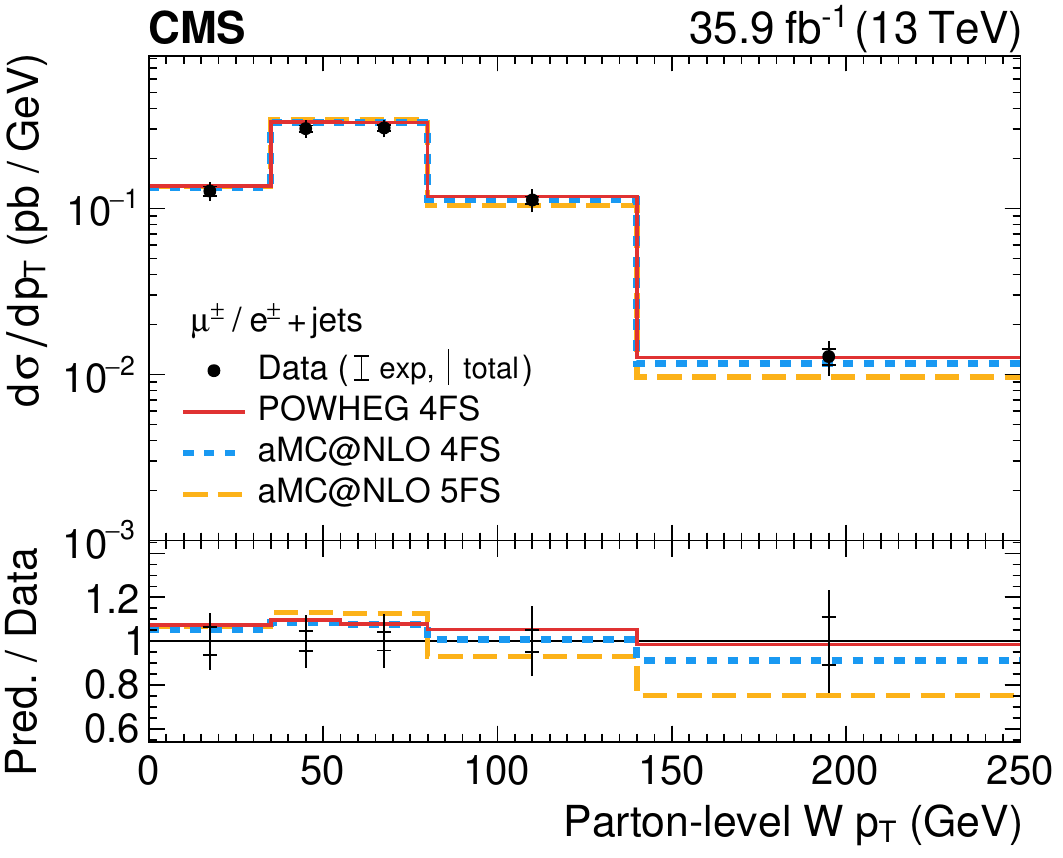}\hspace{0.03\textwidth}
\includegraphics[width=0.48\textwidth]{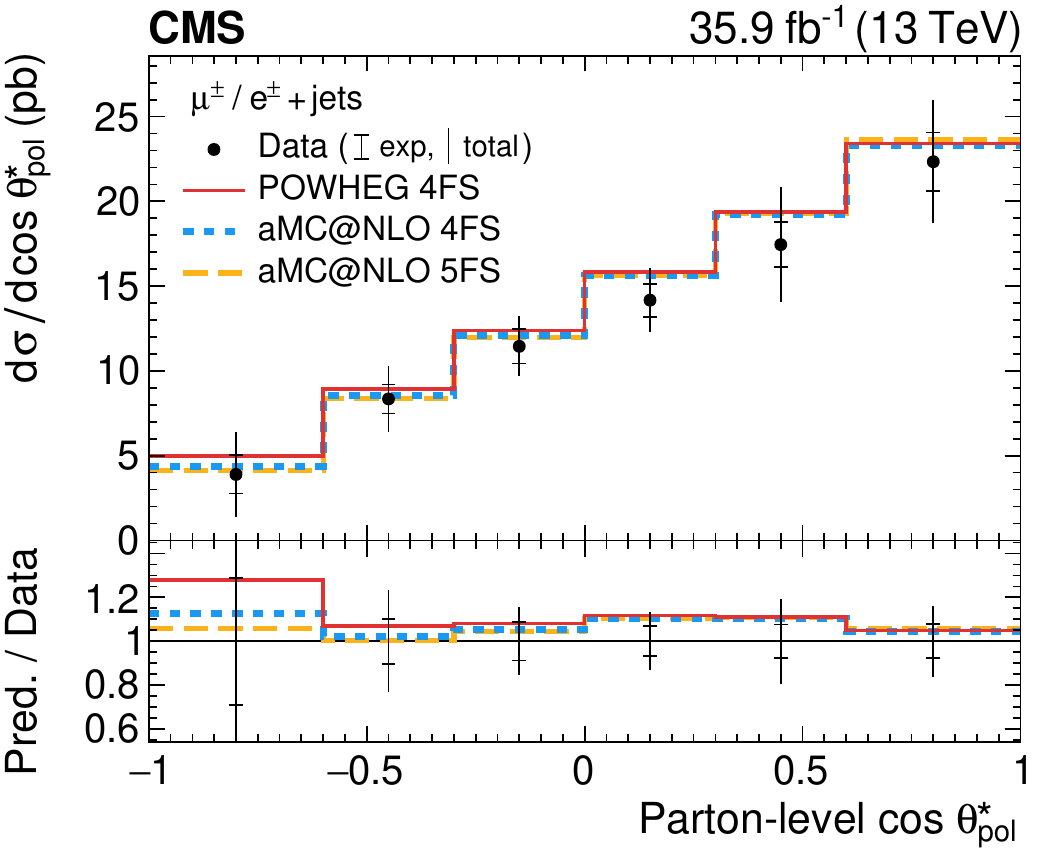}
\caption{\label{fig:result-parton-sum}Differential cross sections for the sum of \tchannel single top quark and antiquark production at the parton level: (upper row)~top quark \pt and rapidity; (middle row)~charged lepton \pt and rapidity; (lower left)~\PW~boson \pt; (lower right)~cosine of the top quark polarisation angle. The total uncertainty is indicated by the vertical lines, while horizontal bars indicate the statistical and experimental uncertainties, which have been profiled in the ML fit, and thus exclude the uncertainties in the theoretical modelling and the luminosity. Three different predictions from event generators are shown by the solid, dashed, and dotted lines. The lower panels show the ratios of the predictions to the data.}
\end{figure*}

\begin{figure*}[phtb!]
\centering
\includegraphics[width=0.48\textwidth]{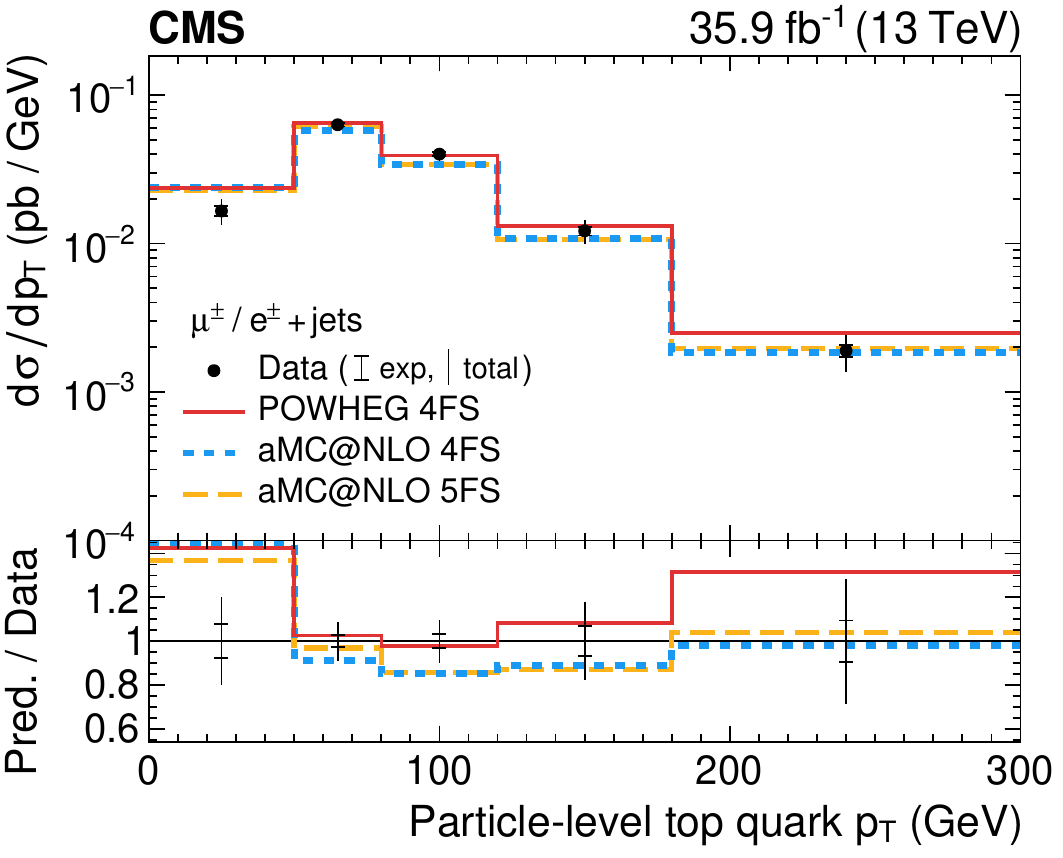}\hspace{0.03\textwidth}
\includegraphics[width=0.48\textwidth]{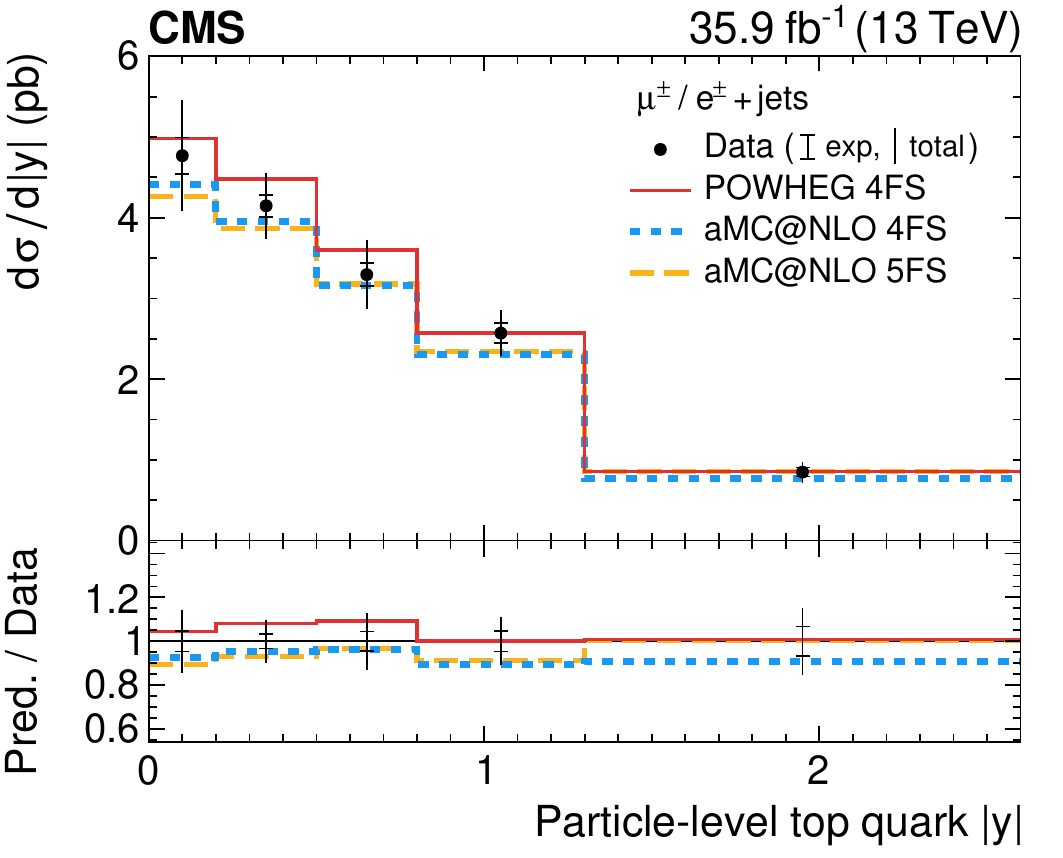}\\[0.015\textheight]
\includegraphics[width=0.48\textwidth]{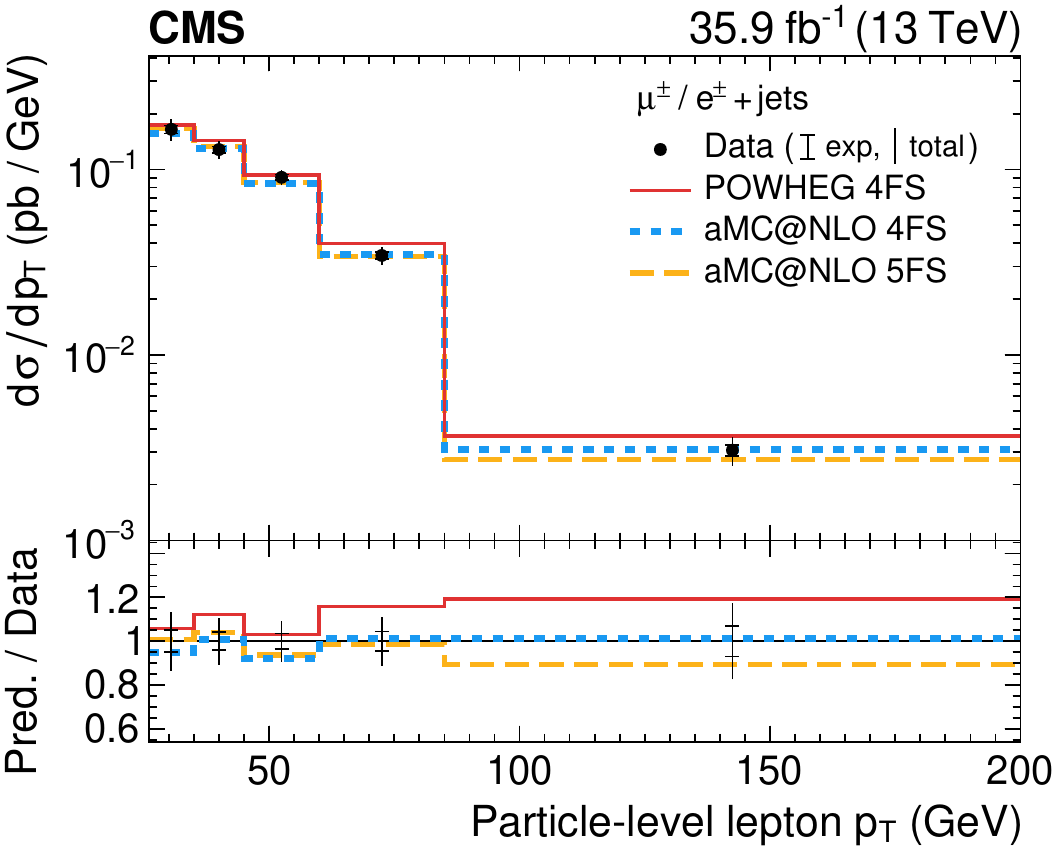}\hspace{0.03\textwidth}
\includegraphics[width=0.48\textwidth]{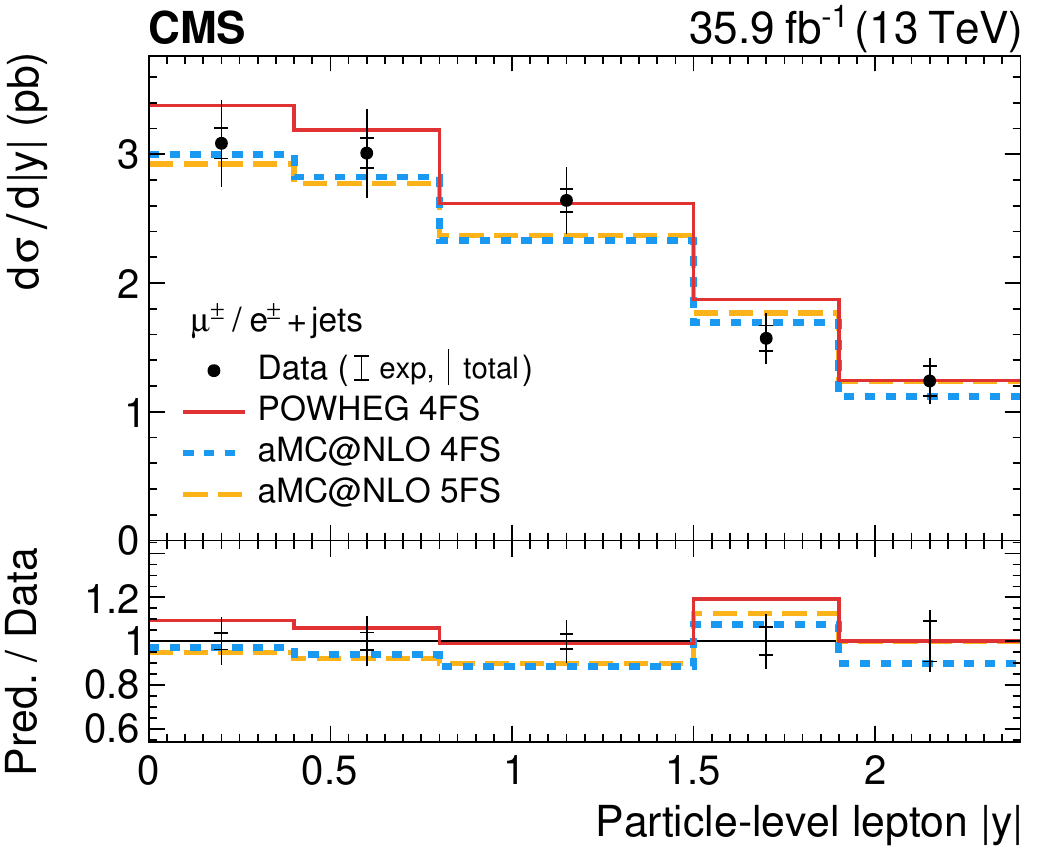}\\[0.015\textheight]
\includegraphics[width=0.48\textwidth]{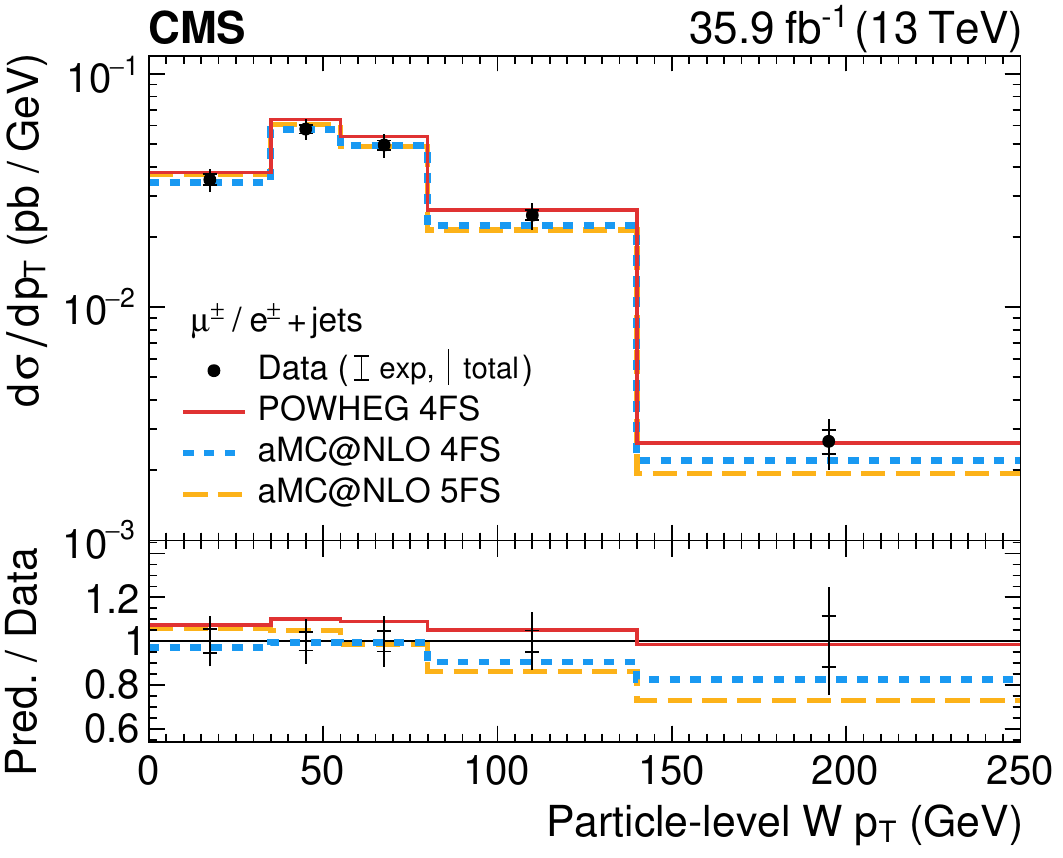}\hspace{0.03\textwidth}
\includegraphics[width=0.48\textwidth]{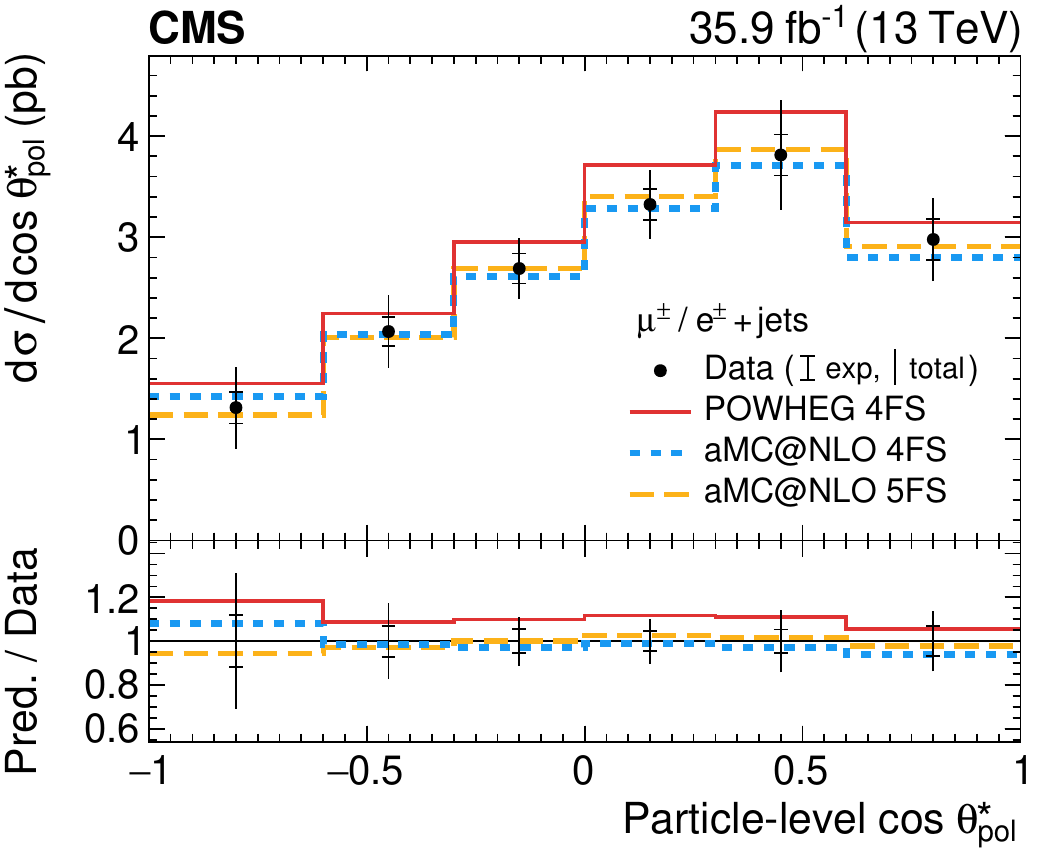}
\caption{\label{fig:result-particle-sum}Differential cross sections for the sum of \tchannel single top quark and antiquark production at the particle level: (upper row)~top quark \pt and rapidity; (middle row)~charged lepton \pt and rapidity; (lower left)~\PW~boson \pt; (lower right)~cosine of the top quark polarisation angle. The total uncertainty is indicated by the vertical lines, while horizontal bars indicate the statistical and experimental uncertainties, which have been profiled in the ML fit, and thus exclude the uncertainties in the theoretical modelling and the luminosity. Three different predictions from event generators are shown by the solid, dashed, and dotted lines. The lower panels show the ratios of the predictions to the data.}
\end{figure*}

\begin{figure*}[phtb!]
\centering
\includegraphics[width=0.48\textwidth]{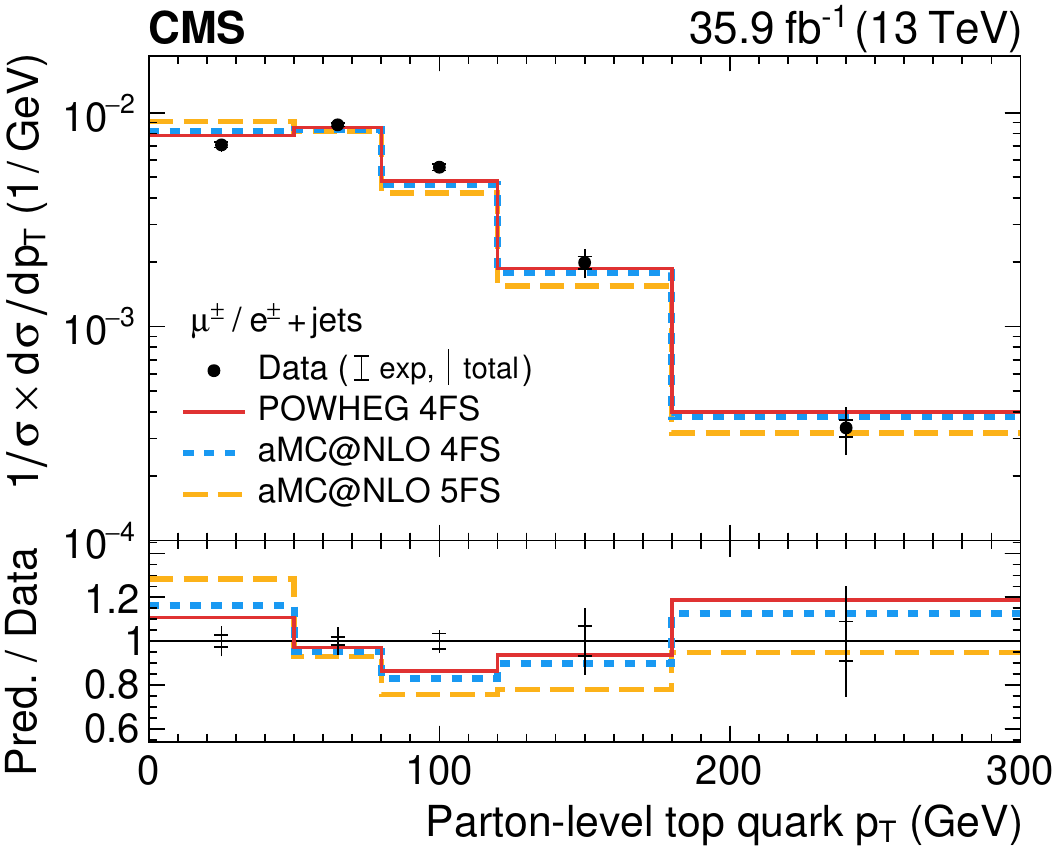}\hspace{0.03\textwidth}
\includegraphics[width=0.48\textwidth]{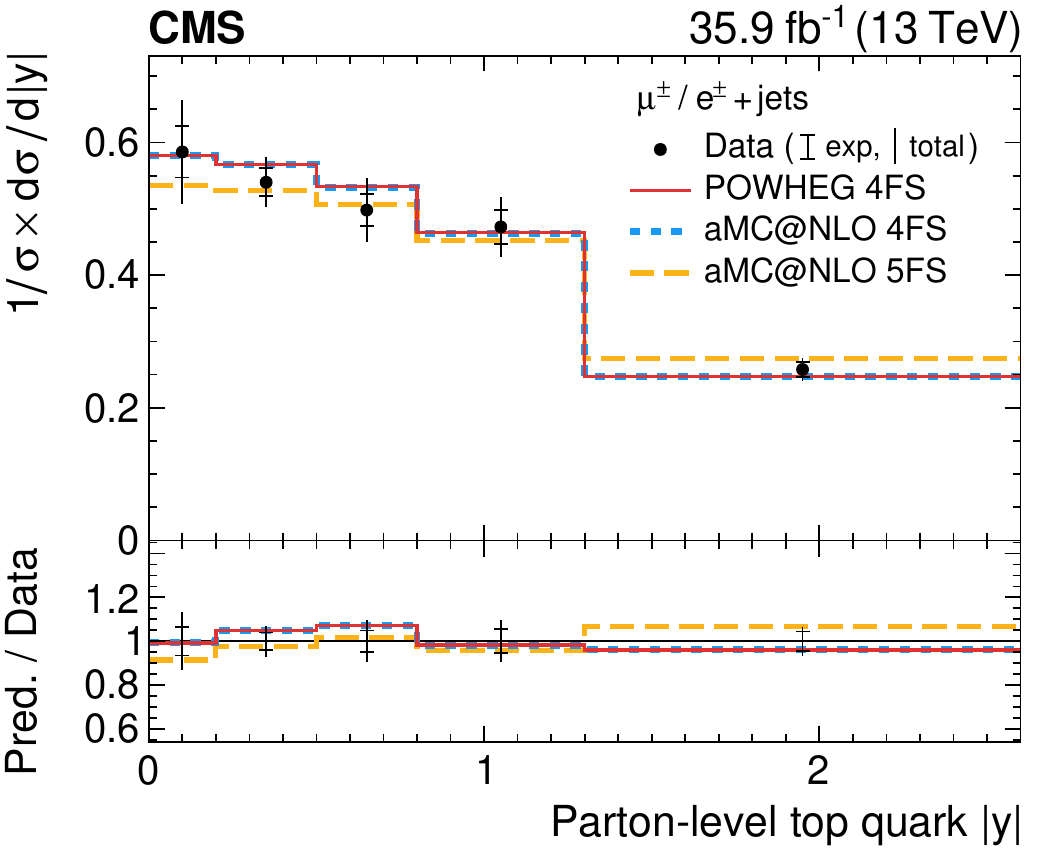}\\[0.015\textheight]
\includegraphics[width=0.48\textwidth]{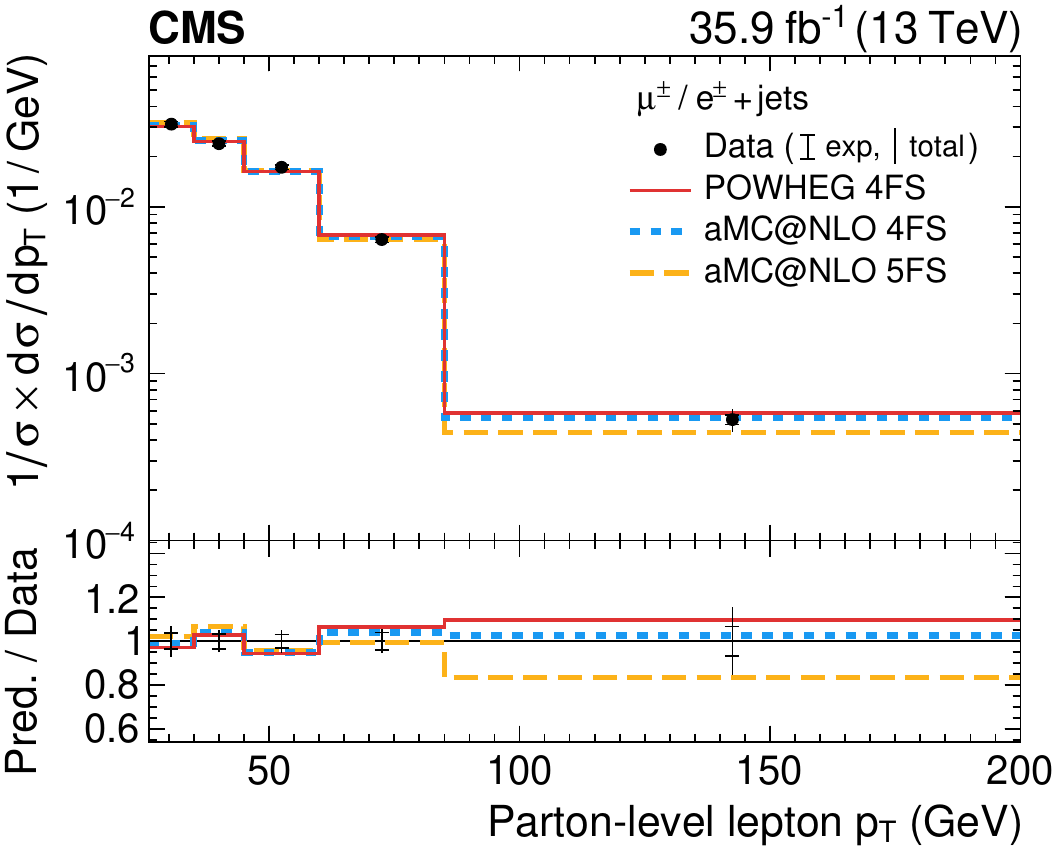}\hspace{0.03\textwidth}
\includegraphics[width=0.48\textwidth]{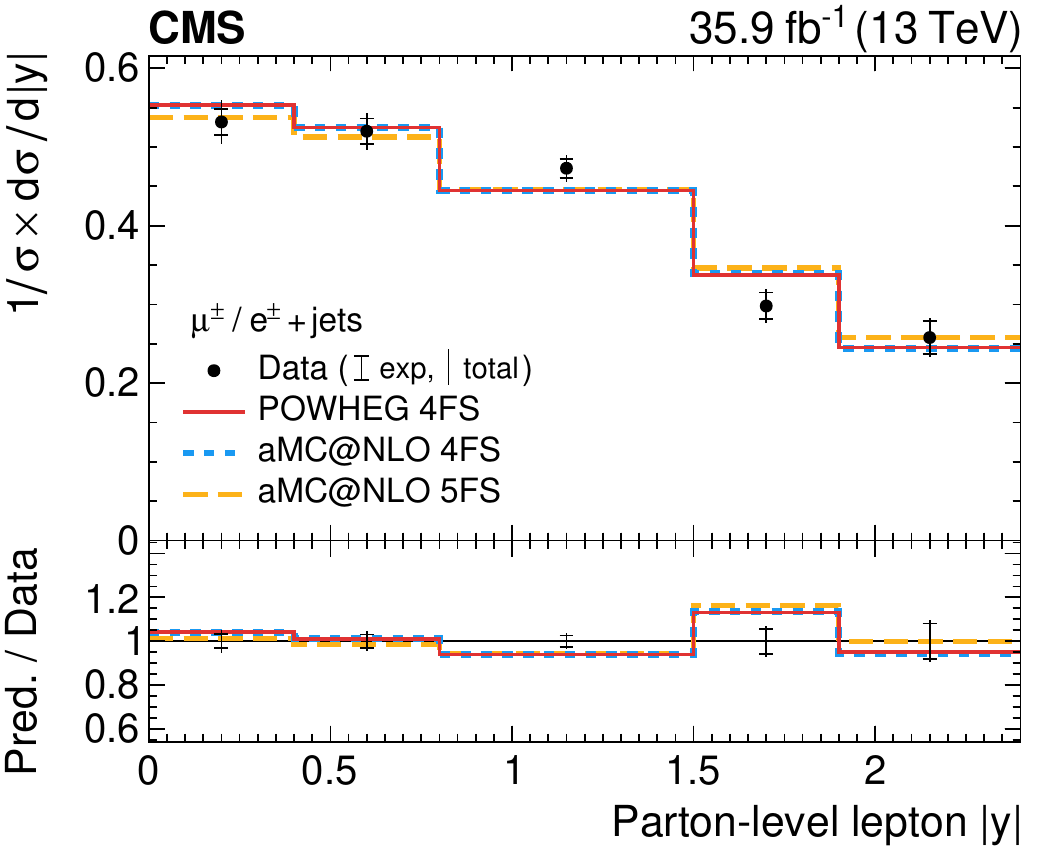}\\[0.015\textheight]
\includegraphics[width=0.48\textwidth]{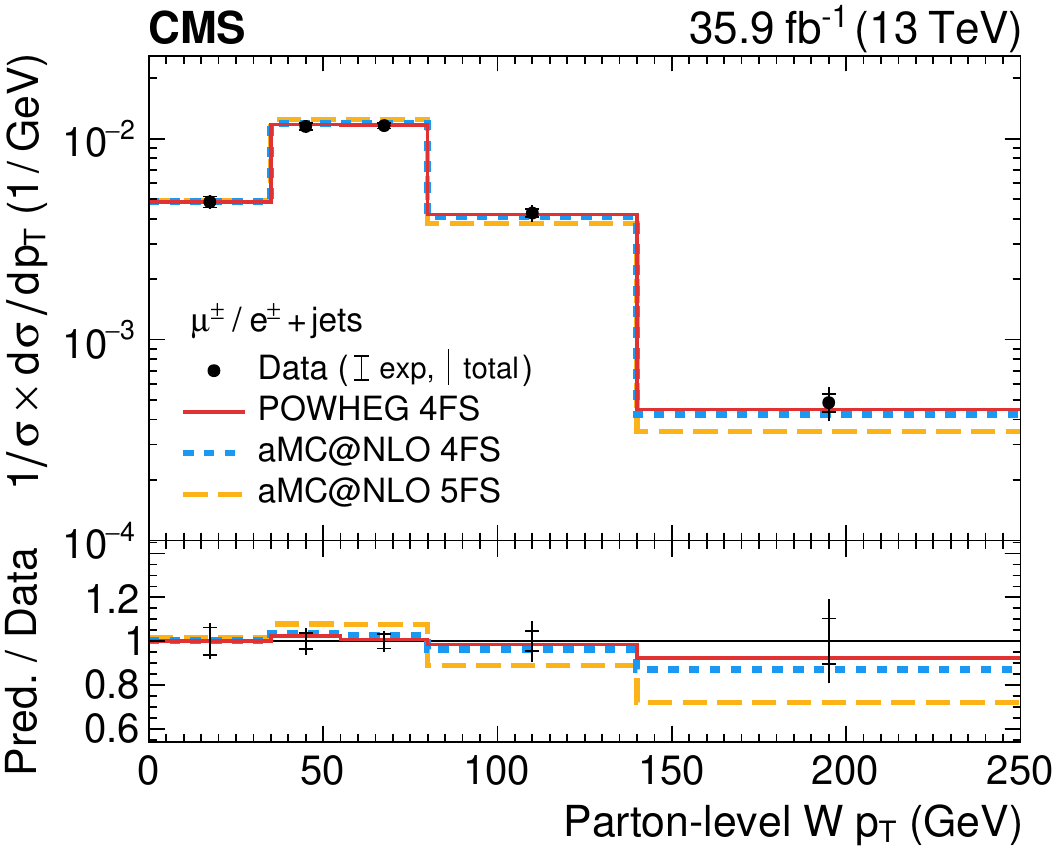}\hspace{0.03\textwidth}
\includegraphics[width=0.48\textwidth]{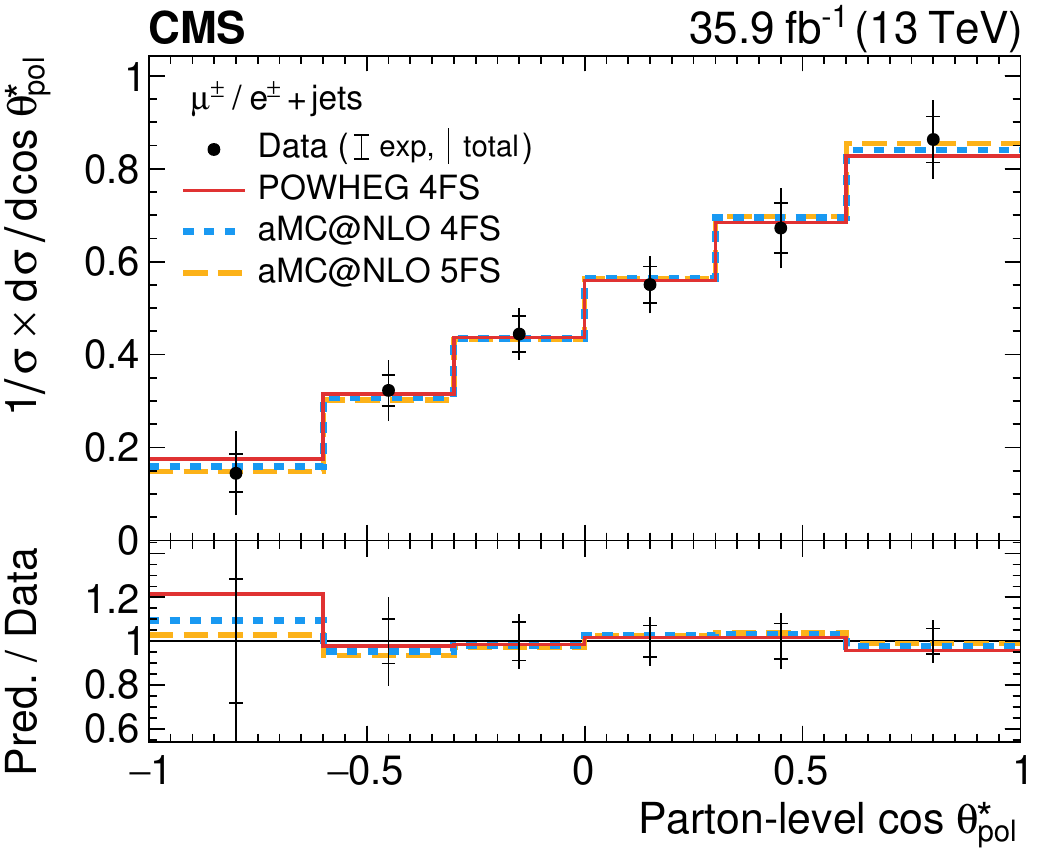}
\caption{\label{fig:result-parton-sumnorm}Normalised differential cross sections for the sum of \tchannel single top quark and antiquark production at the parton level: (upper row)~top quark \pt and rapidity; (middle row)~charged lepton \pt and rapidity; (lower left)~\PW~boson \pt; (lower right)~cosine of the top quark polarisation angle. The total uncertainty is indicated by the vertical lines, while horizontal bars indicate the statistical and experimental uncertainties, which have been profiled in the ML fit, and thus exclude the uncertainties in the theoretical modelling. Three different predictions from event generators are shown by the solid, dashed, and dotted lines. The lower panels show the ratios of the predictions to the data.}
\end{figure*}

\begin{figure*}[phtb!]
\centering
\includegraphics[width=0.48\textwidth]{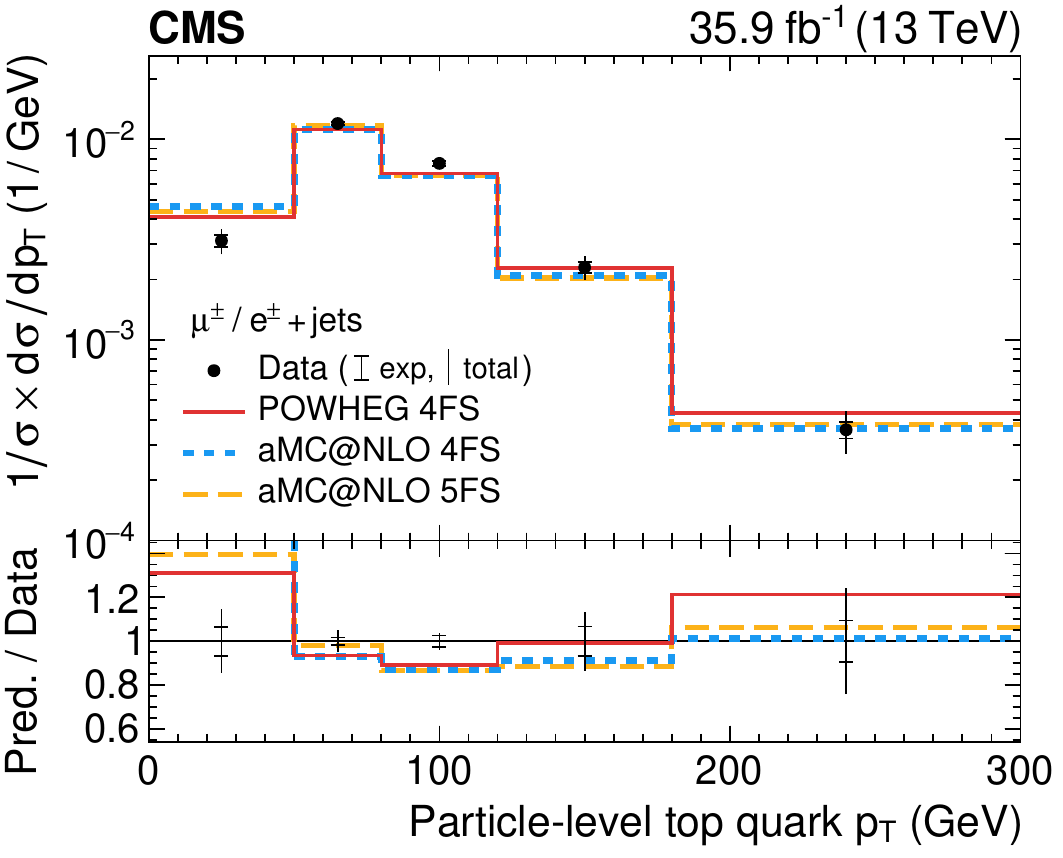}\hspace{0.03\textwidth}
\includegraphics[width=0.48\textwidth]{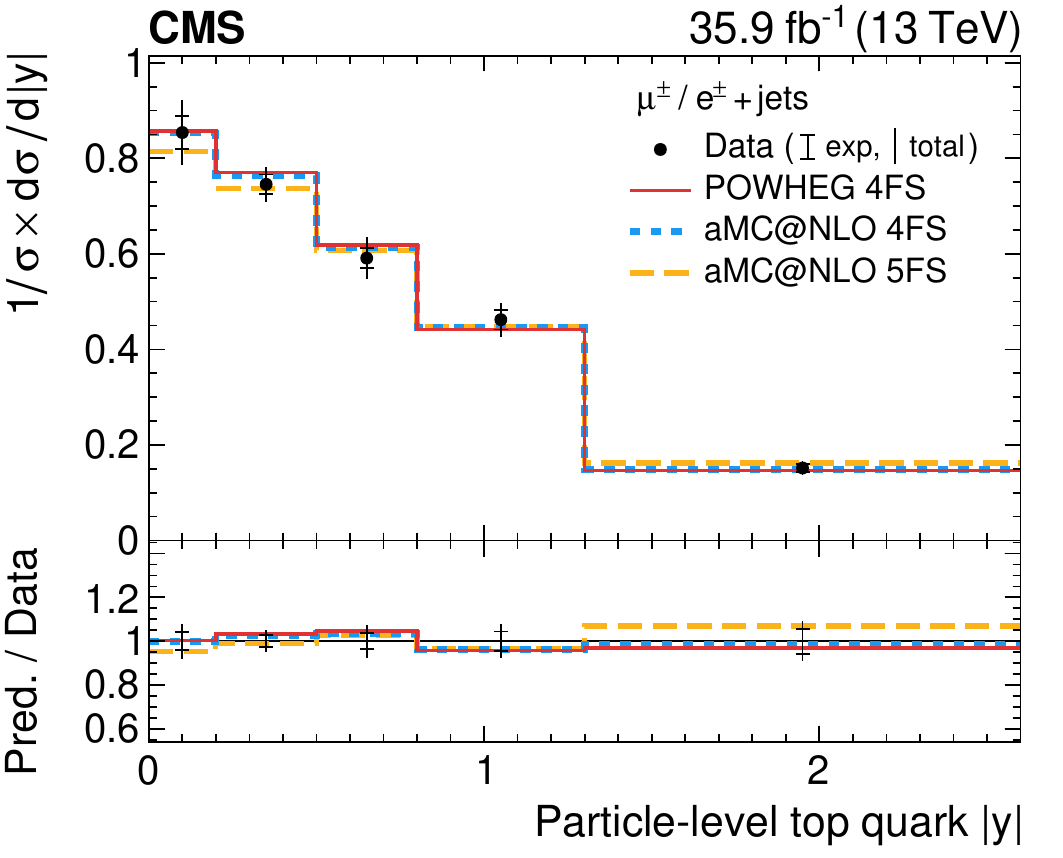}\\[0.015\textheight]
\includegraphics[width=0.48\textwidth]{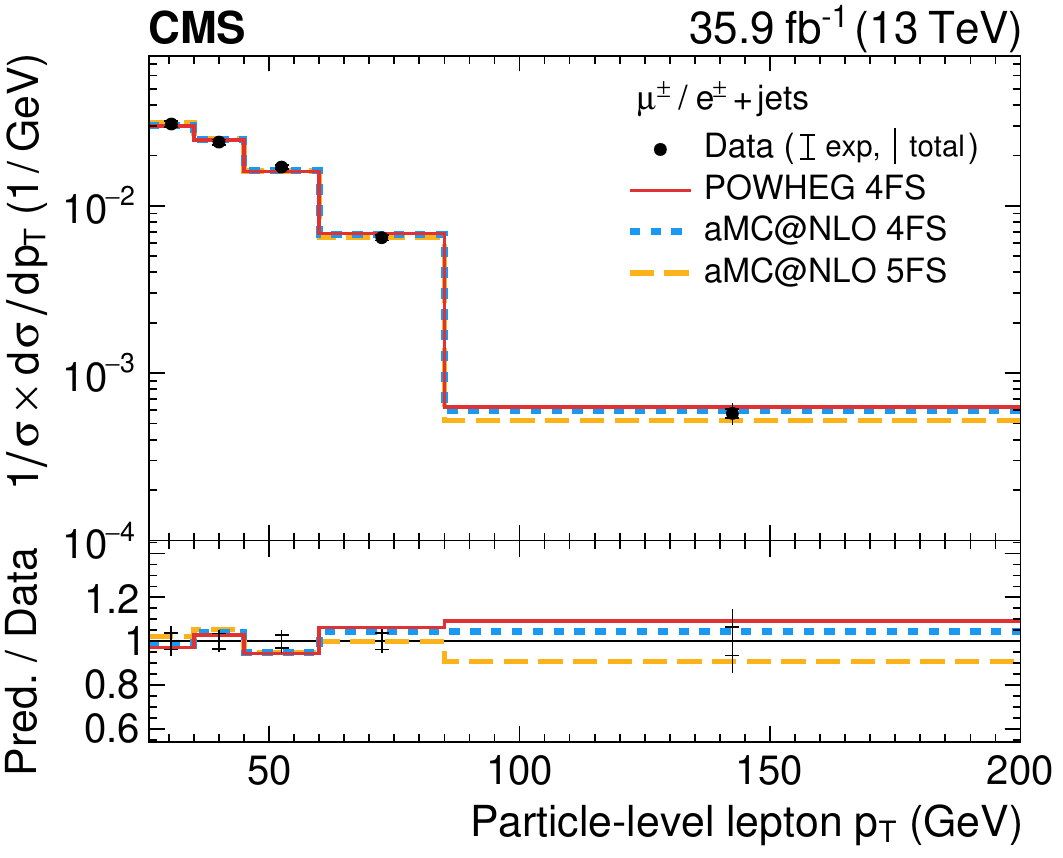}\hspace{0.03\textwidth}
\includegraphics[width=0.48\textwidth]{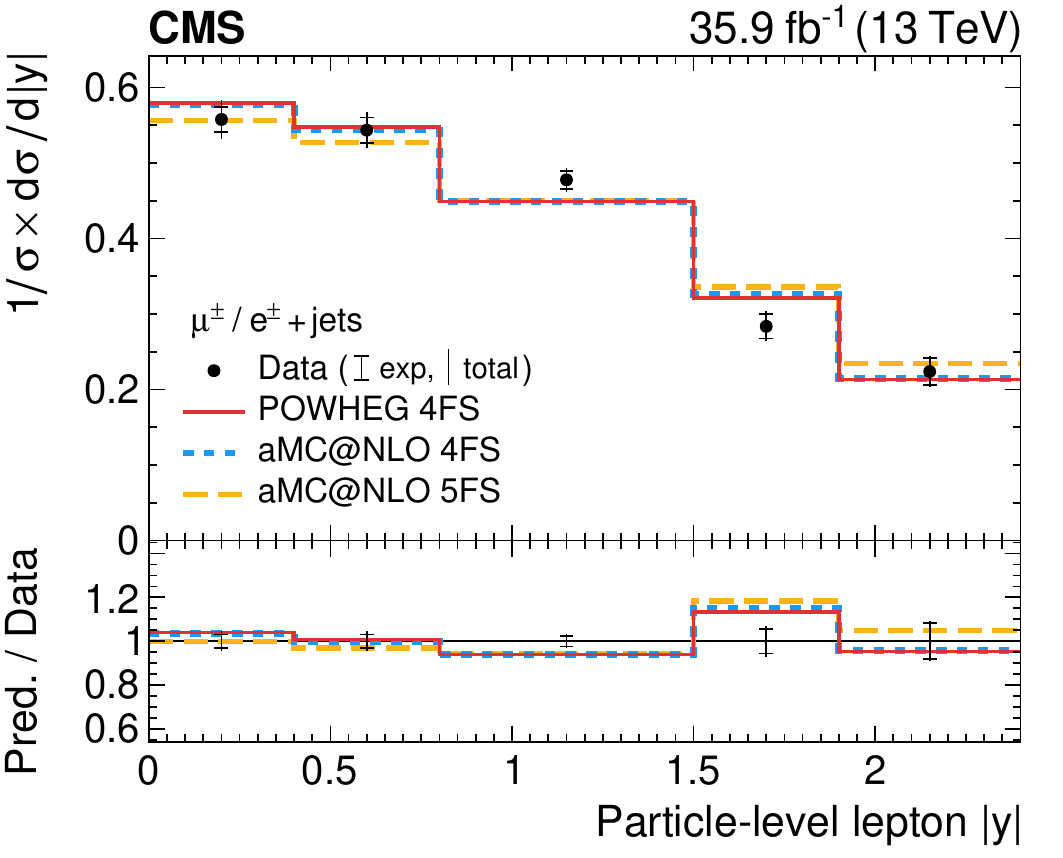}\\[0.015\textheight]
\includegraphics[width=0.48\textwidth]{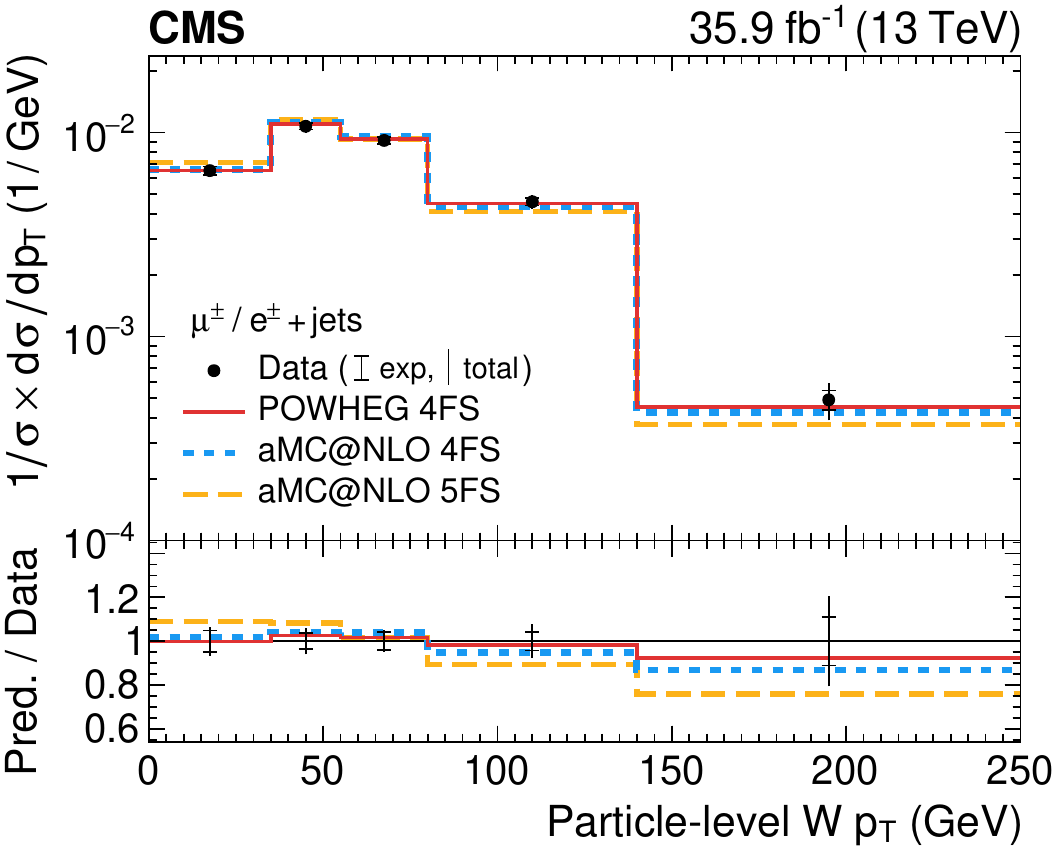}\hspace{0.03\textwidth}
\includegraphics[width=0.48\textwidth]{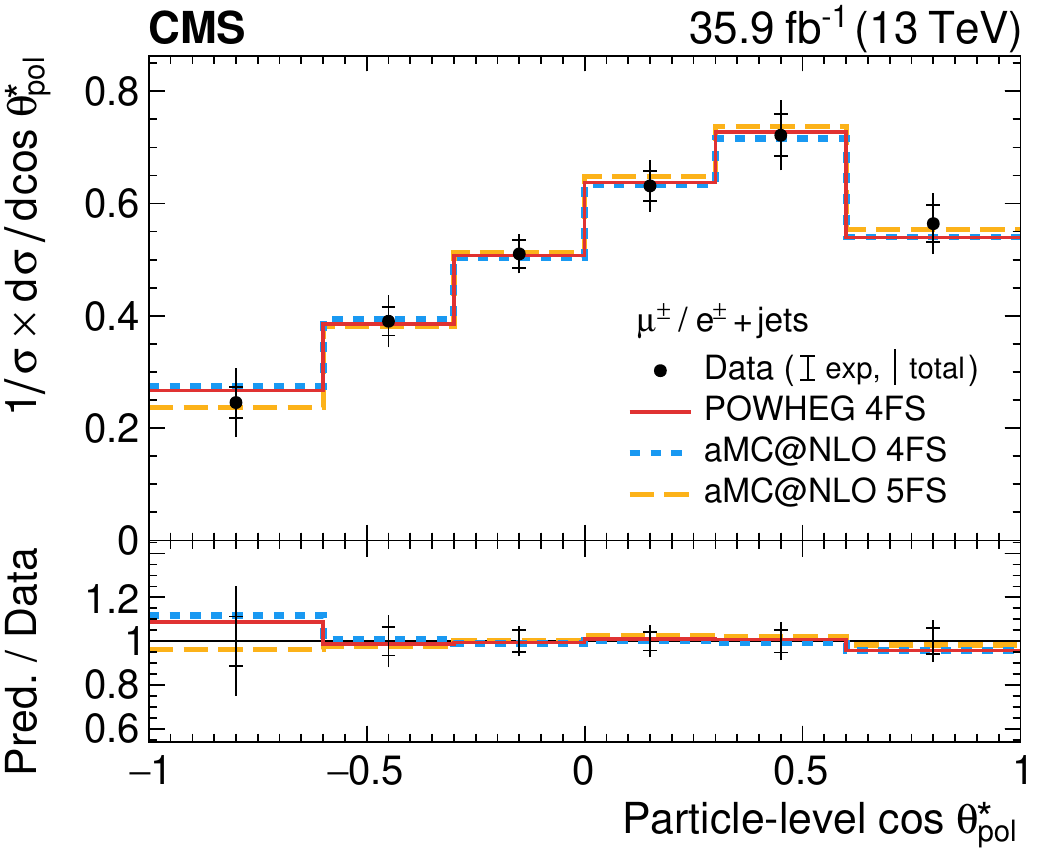}
\caption{\label{fig:result-particle-sumnorm}Normalised differential cross sections for the sum of \tchannel single top quark and antiquark production at the particle level: (upper row)~top quark \pt and rapidity; (middle row)~charged lepton \pt and rapidity; (lower left)~\PW~boson \pt; (lower right)~cosine of the top quark polarisation angle. The total uncertainty is indicated by the vertical lines, while horizontal bars indicate the statistical and experimental uncertainties, which have been profiled in the ML fit, and thus exclude the uncertainties in the theoretical modelling. Three different predictions from event generators are shown by the solid, dashed, and dotted lines. The lower panels show the ratios of the predictions to the data.}
\end{figure*}

Differential cross sections of \tchannel single top quark production as a function of the top quark $\pt$, rapidity, and polarisation angle, the $\pt$ and rapidity of the charged lepton (muon or electron) that originates from the top quark decay, and the $\pt$ of the \PW~boson from the top quark decay are presented in Figs.~\ref{fig:result-parton-sum} and~\ref{fig:result-particle-sum} at the parton and particle levels, respectively. The normalised differential cross sections of the same observables at the parton and particle levels are provided in Figs.~\ref{fig:result-parton-sumnorm} and~\ref{fig:result-particle-sumnorm}. The total uncertainty is indicated by the vertical lines, while horizontal bars indicate the statistical and experimental uncertainties, which have been profiled in the ML fit, and thus exclude the uncertainties in the theoretical modelling and the luminosity.
The differential cross sections refer to \tchannel single top quark production where the top quark decays semileptonically (into either muon or electron) including events where the charged lepton stems from an intermediate $\PGt$ lepton decay. The results are compared to the predictions by the \POWHEG generator interfaced with \PYTHIA in the 4FS and the \MGvATNLO generator interfaced with \PYTHIA in the 4FS and 5FS.

An overall good agreement of the results with the predictions from the 4FS is observed, except for a slight deviation at low top quark $\pt$. The predictions from the 5FS for the top quark and \PW~boson \pt distributions do not agree as well with the data.

\begin{figure*}[phtb!]
\centering
\includegraphics[width=0.48\textwidth]{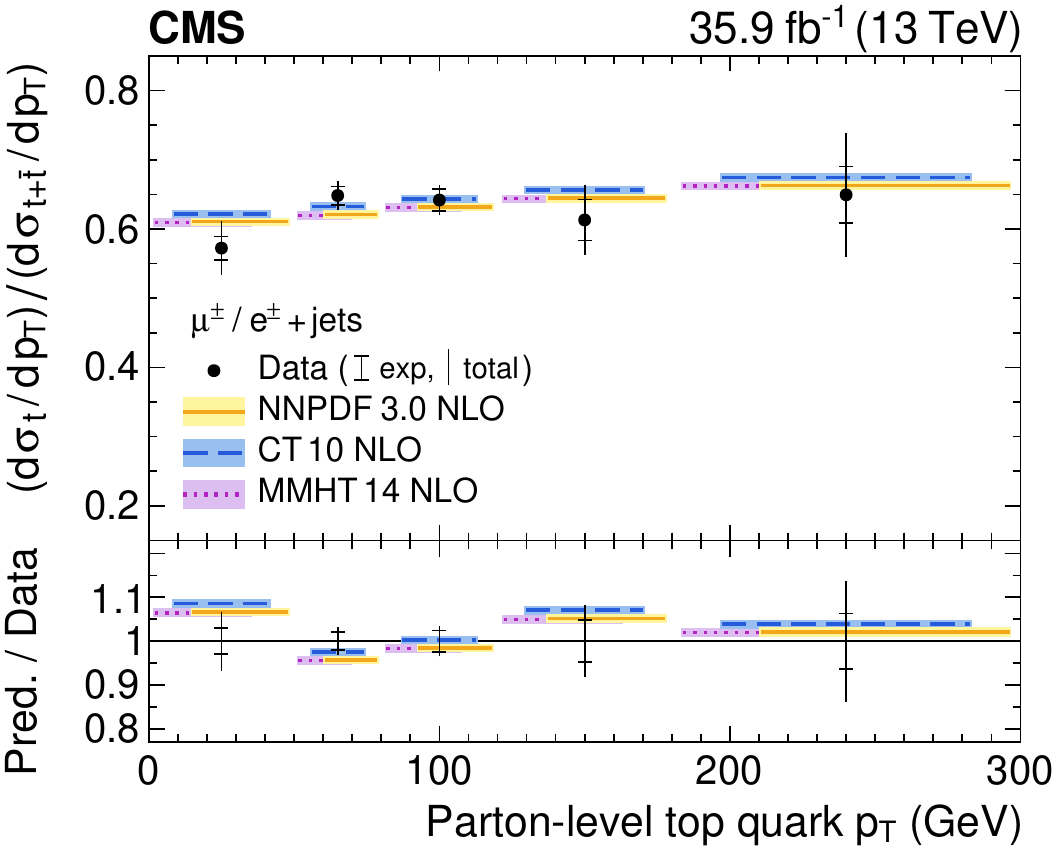}\hspace{0.03\textwidth}
\includegraphics[width=0.48\textwidth]{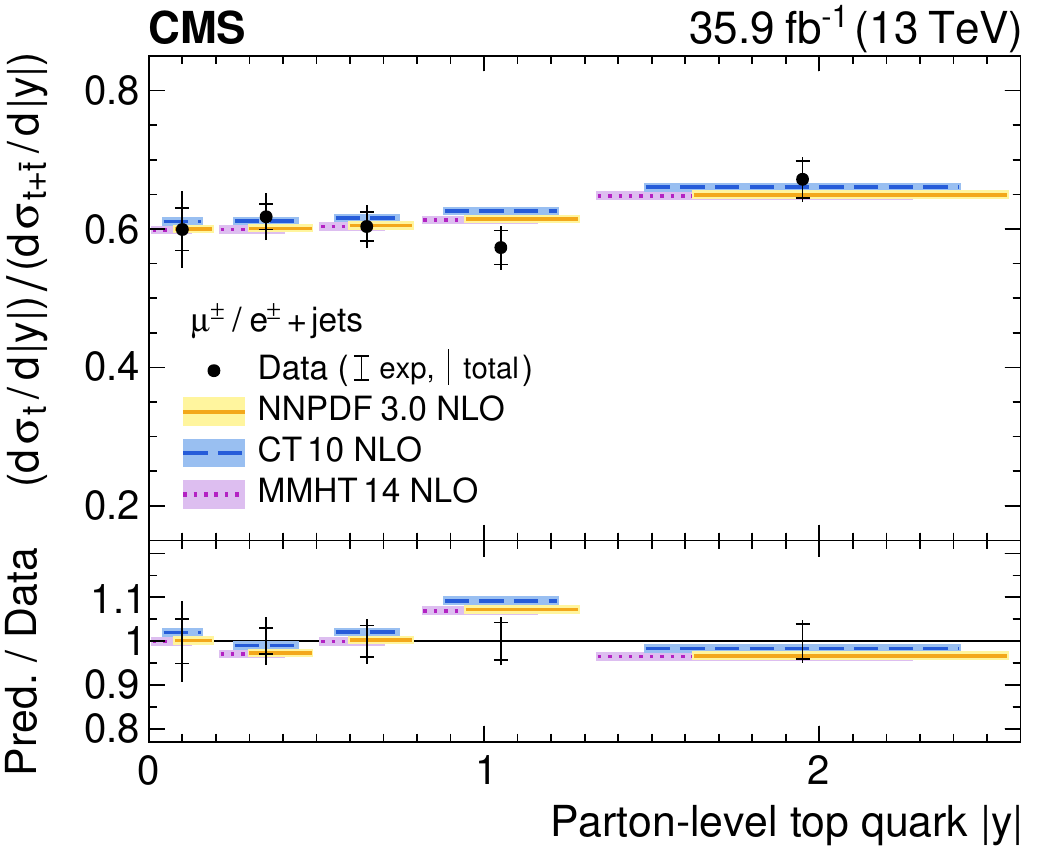}\\[0.015\textheight]
\includegraphics[width=0.48\textwidth]{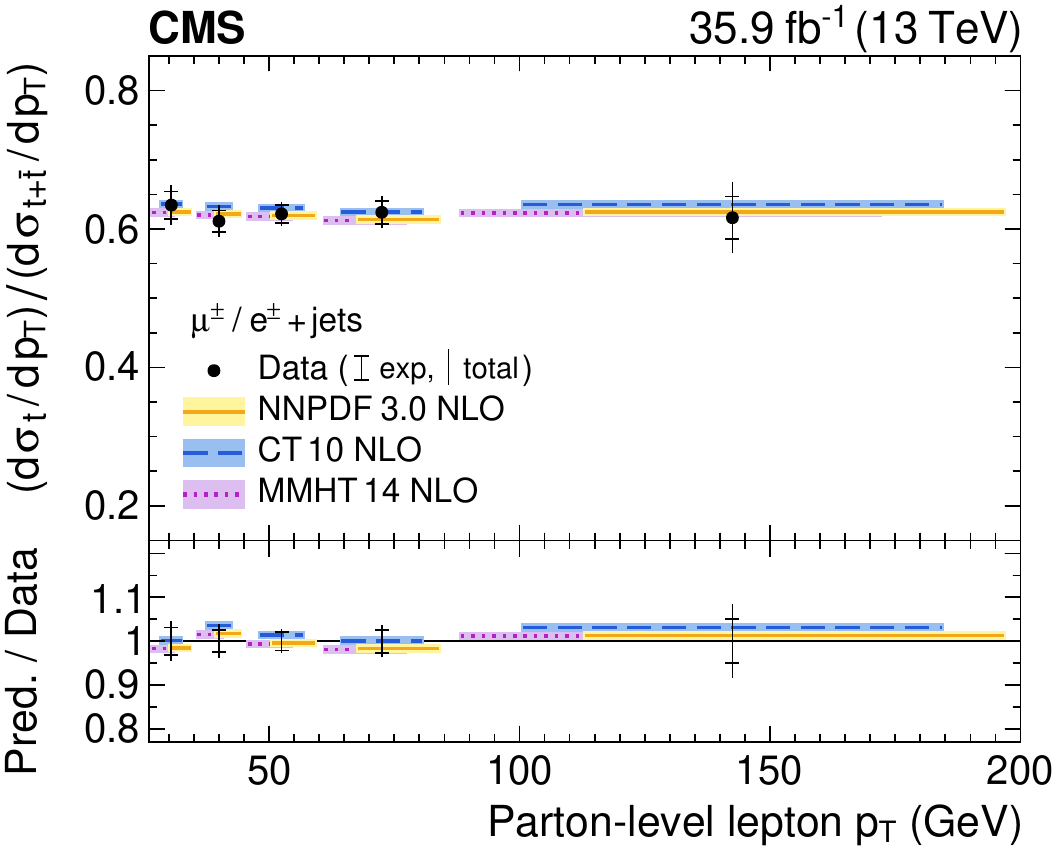}\hspace{0.03\textwidth}
\includegraphics[width=0.48\textwidth]{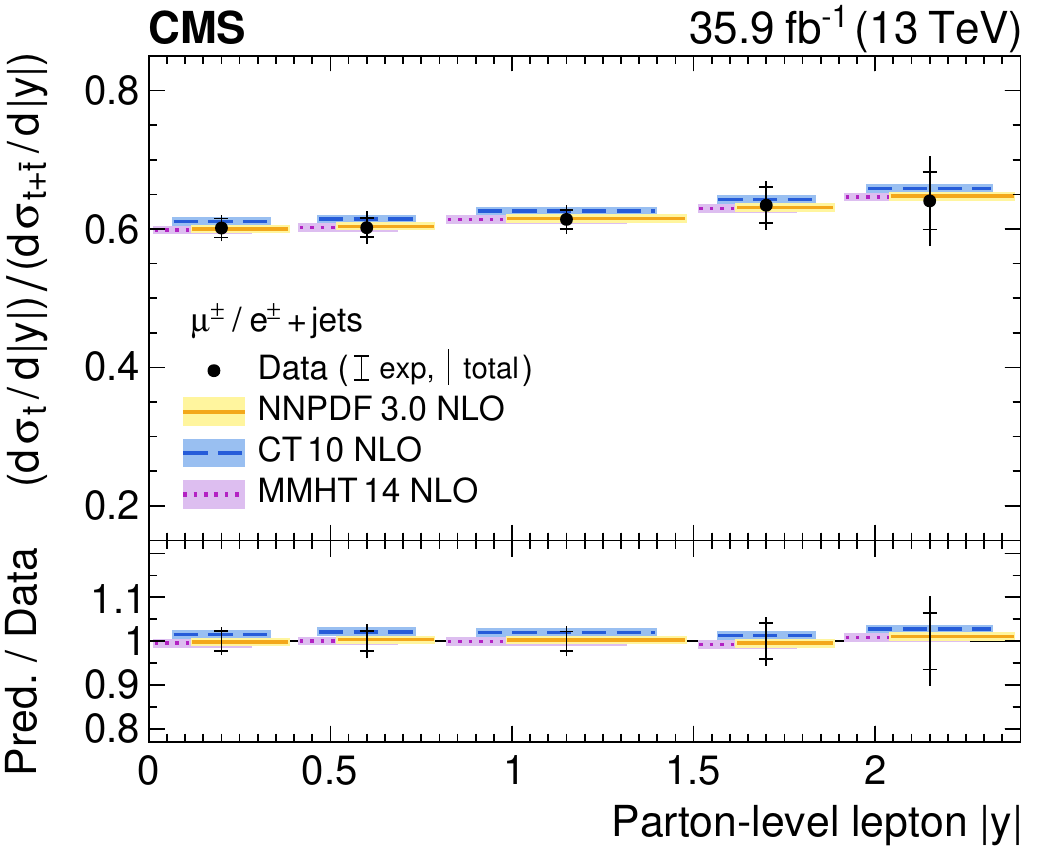}\\[0.015\textheight]
\includegraphics[width=0.48\textwidth]{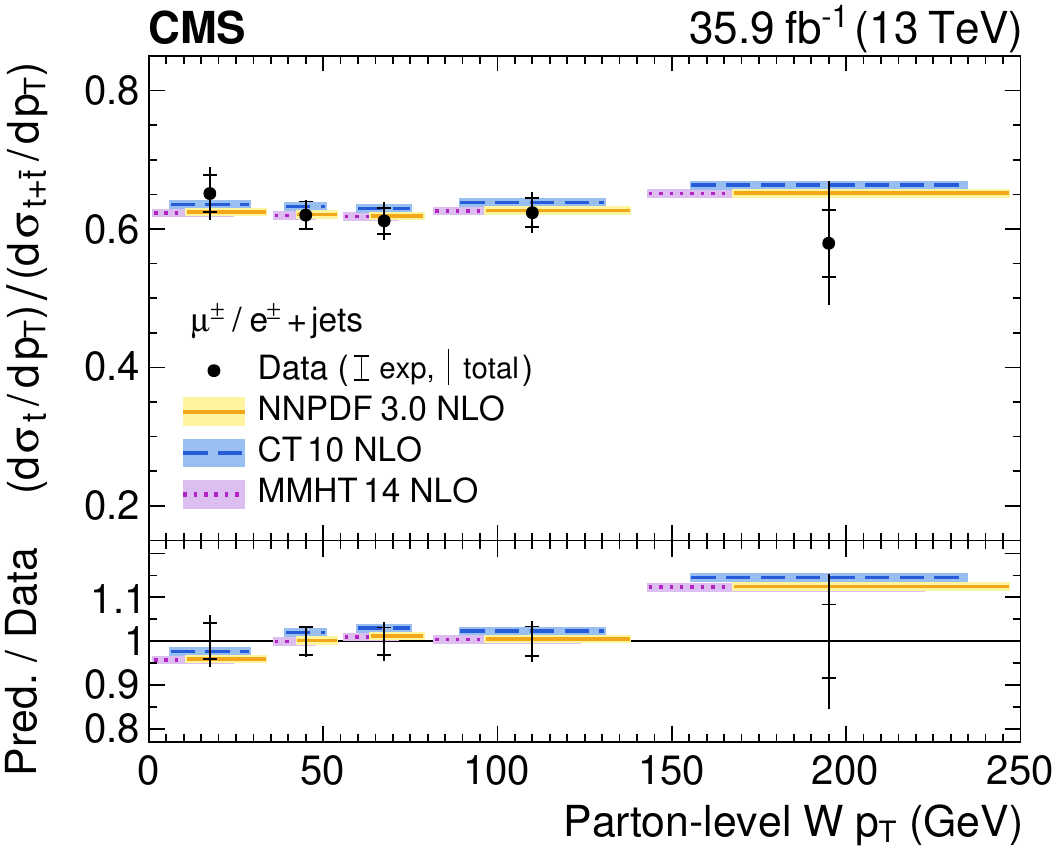}
\caption{\label{fig:result-parton-ratio}Ratio of the top quark to the sum of the top quark and antiquark \tchannel differential cross section at the parton level: (upper row)~top quark \pt and rapidity; (middle row)~charged lepton \pt and rapidity; (lower row)~\PW~boson \pt. The total uncertainty is indicated by the vertical lines, while horizontal bars indicate the statistical and experimental uncertainties, which have been profiled in the ML fit, and thus exclude the uncertainties in the theoretical modelling. Predictions from three different PDF sets are shown by the solid, dashed, and dotted lines. The lower panels show the ratios of the predictions to the data.}
\end{figure*}

\begin{figure*}[phtb!]
\centering
\includegraphics[width=0.48\textwidth]{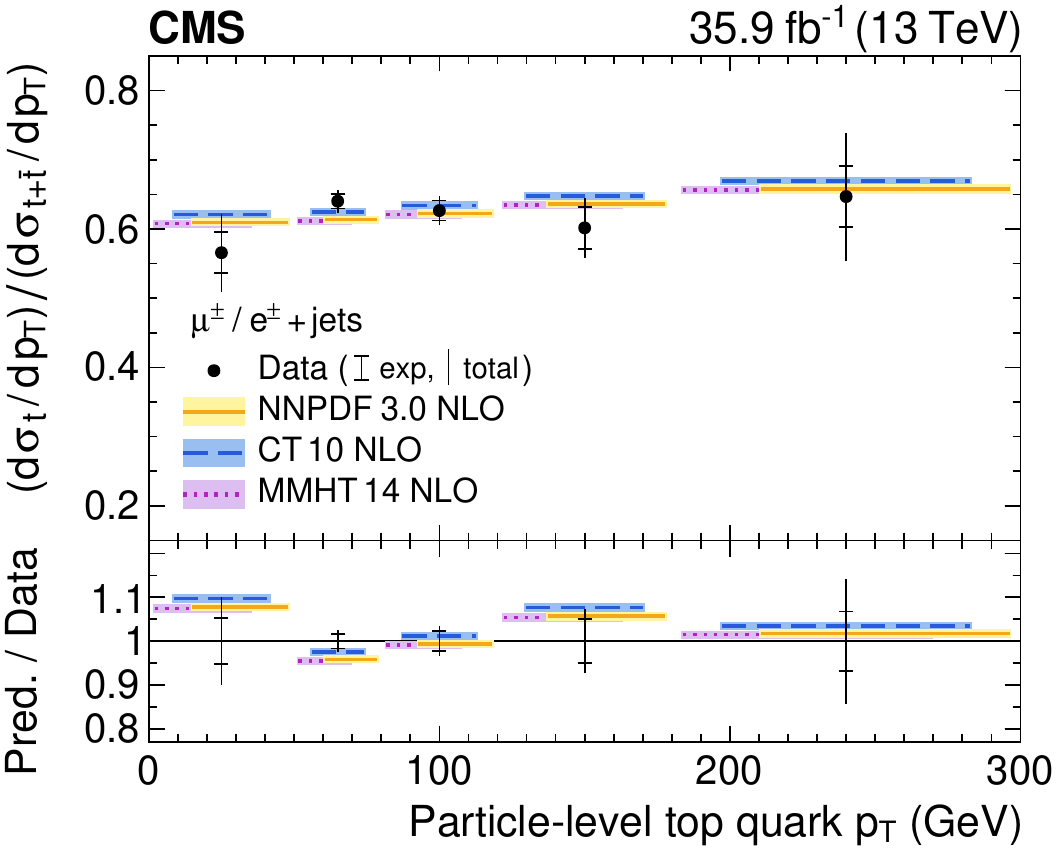}\hspace{0.03\textwidth}
\includegraphics[width=0.48\textwidth]{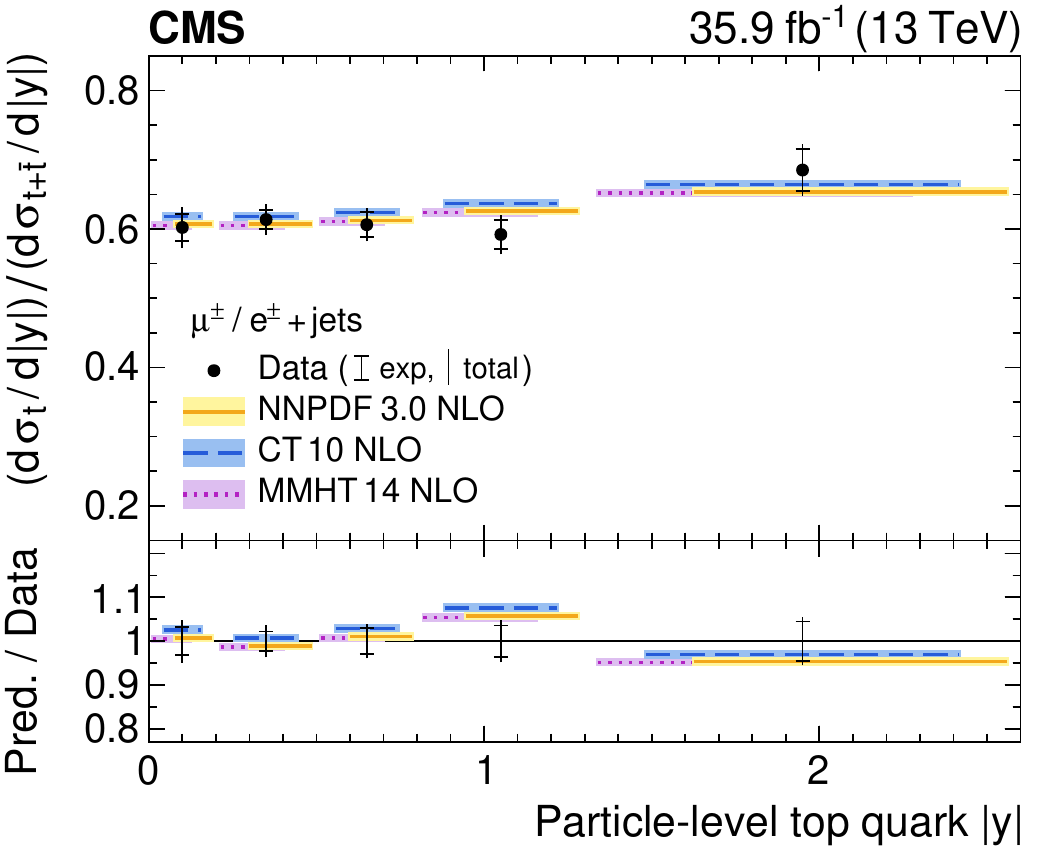}\\[0.015\textheight]
\includegraphics[width=0.48\textwidth]{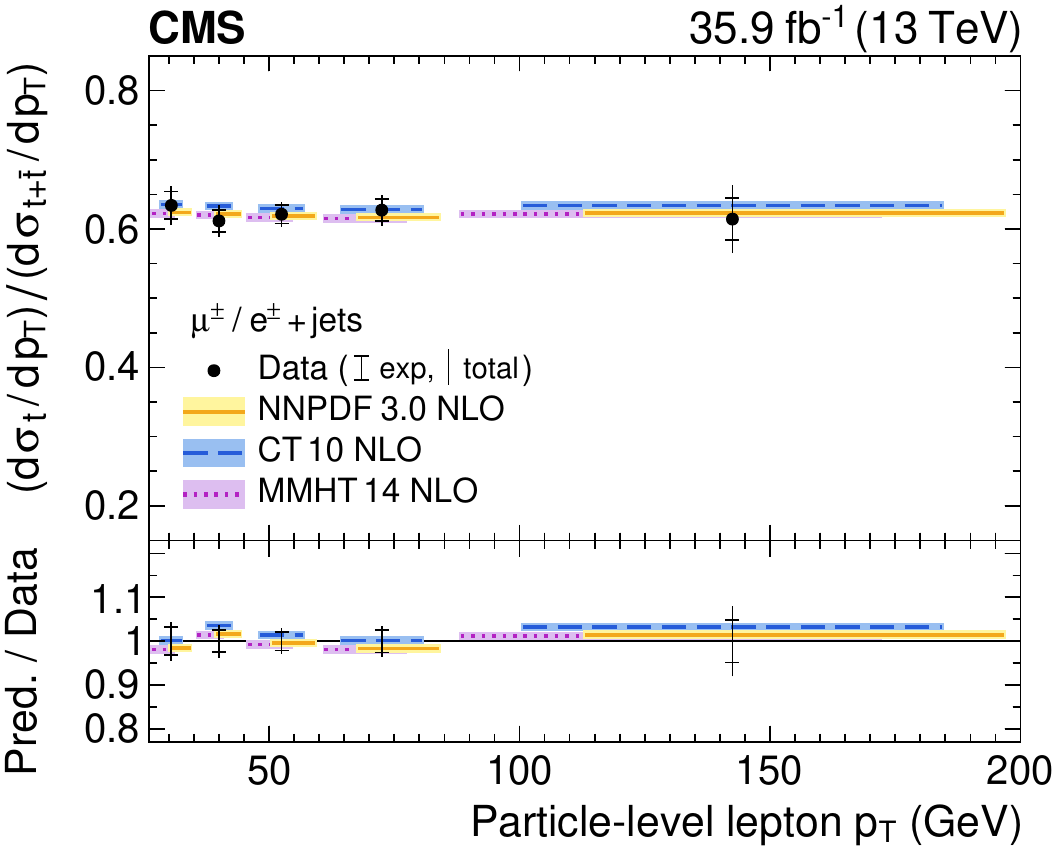}\hspace{0.03\textwidth}
\includegraphics[width=0.48\textwidth]{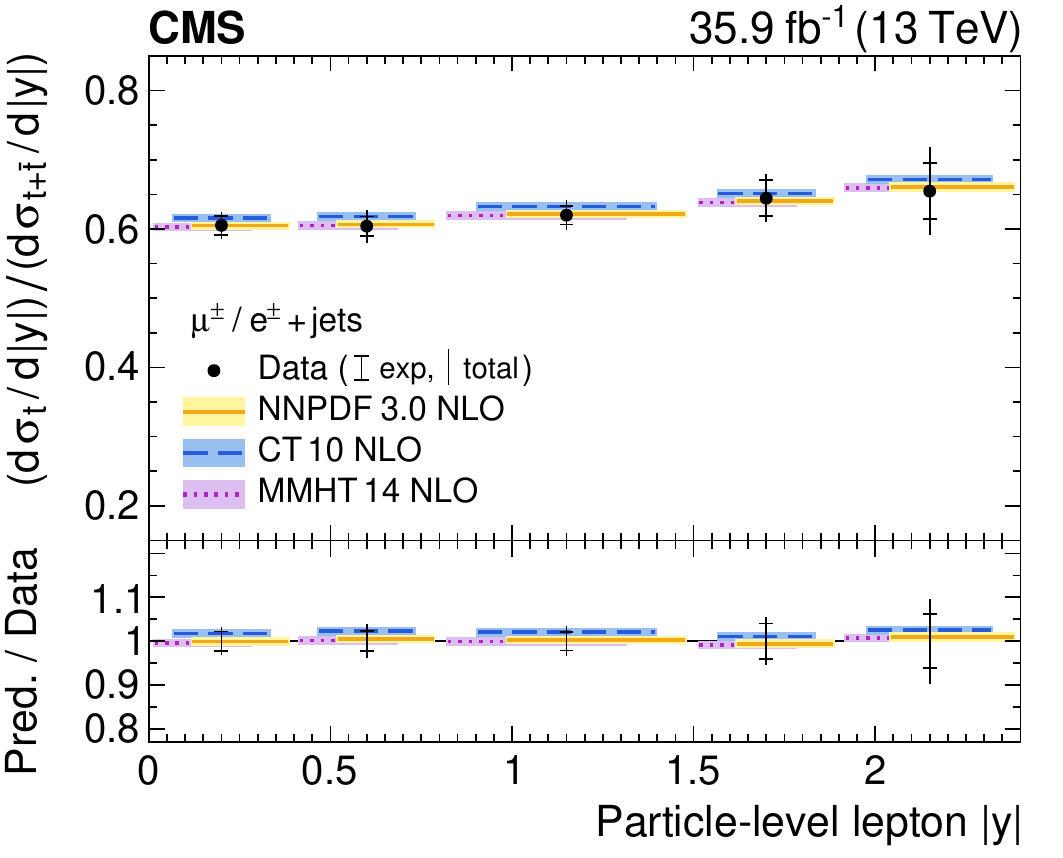}\\[0.015\textheight]
\includegraphics[width=0.48\textwidth]{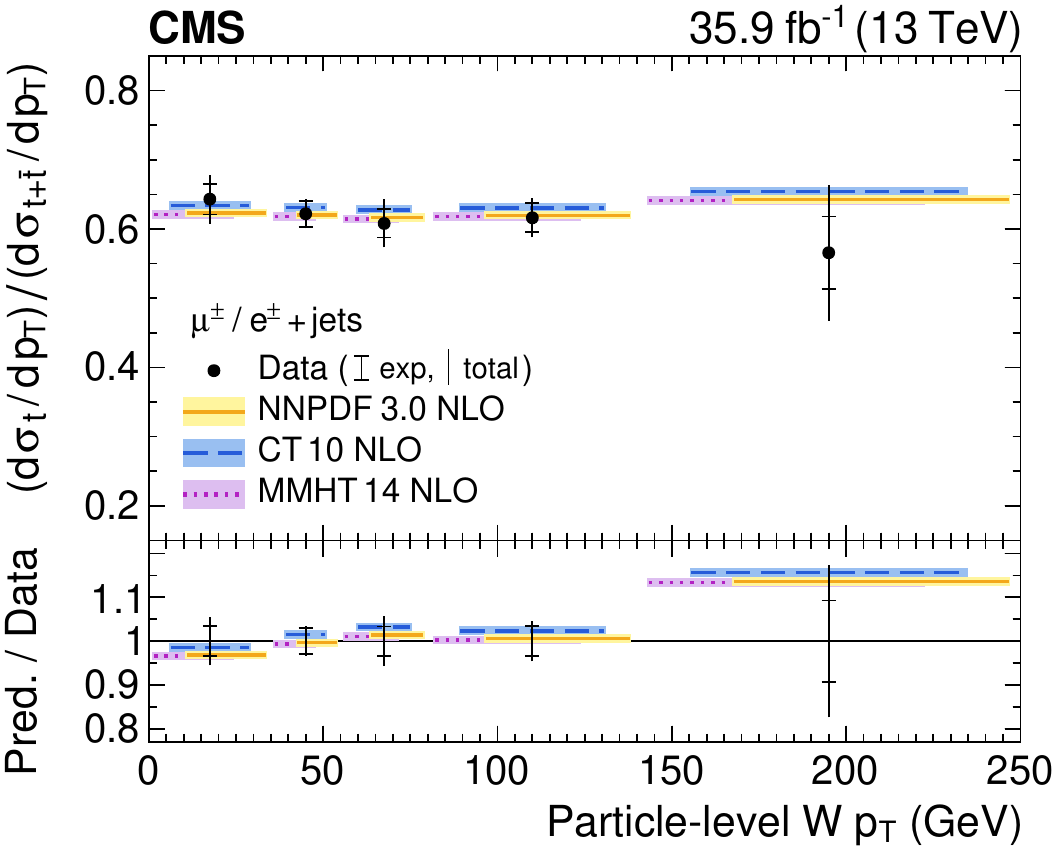}
\caption{\label{fig:result-particle-ratio}Ratio of the top quark to the sum of the top quark and antiquark \tchannel differential cross section at the particle level: (upper row)~top quark \pt and rapidity; (middle row)~charged lepton \pt and rapidity; (lower row)~\PW~boson \pt. The total uncertainty is indicated by the vertical lines, while horizontal bars indicate the statistical and experimental uncertainties, which have been profiled in the ML fit, and thus exclude the uncertainties in the theoretical modelling. Predictions from three different PDF sets are shown by the solid, dashed, and dotted lines. The lower panels show the ratios of the predictions to the data.}
\end{figure*}

Differential ratios of the top quark production rates to the sum of the top quark and antiquark rates as a function of the top quark $\pt$ and rapidity, the $\pt$ and rapidity of the charged lepton, and the \PW~boson $\pt$ are presented in Figs.~\ref{fig:result-parton-ratio} and~\ref{fig:result-particle-ratio} at the parton and particle levels, respectively. It is found that the standard definition of the charge ratio in the literature, \ie $\sigmat/\sigmatbar$, can yield large variances when the precision in certain intervals of the differential cross section for the top antiquark is low. Therefore, the charge ratio is defined as $\sigmat/\sigmatsum$ in this paper. The ratios have been calculated from the measured cross sections at the parton and particle levels, while accounting for correlations between the top quark and antiquark spectra, as detailed in Sections~\ref{sec:fit} and~\ref{sec:systematics}. The resulting charge ratios are compared to the predictions by the NNPDF3.0 NLO, MMHT14 NLO~\cite{Harland-Lang:2014zoa}, and CT10 NLO PDF sets, which have been calculated using the \POWHEG signal sample---generated in the 4FS and interfaced with \PYTHIA. The uncertainty bands shown in Figs.~\ref{fig:result-parton-ratio} and \ref{fig:result-particle-ratio} represent the total uncertainty from varying the corresponding PDF eigenvectors and $\alpS$. Within the uncertainties, the measured charge ratios are in good agreement with the predictions from all three PDF sets.

The spin asymmetry, sensitive to the top quark polarisation, is determined from the differential cross section as a function of the polarisation angle at the parton level (Fig.~\ref{fig:result-parton-sum}, lower right). A linear $\chi^{2}$-based fit, assuming the expected functional dependence given in Eq.~(\ref{eq:intro-pol}), is used to take the correlations between the unfolded bins into account. The measured spin asymmetry in the muon and electron channel and their combination is given in Table~\ref{tab:app-asym-sys}.

\begin{table*}[h!]
\centering
\topcaption{\label{tab:app-asym-sys} The measured spin asymmetry in the muon and electron channel and their combination. A breakdown of the systematic uncertainties is also provided. Minor systematic uncertainties (lepton efficiencies, pileup, and unclustered energy) have been grouped into the ``Others'' category.}
\begin{tabular}{@{}l r r@{.}l r@{.}l r@{.}l@{}}
\hline
&& \multicolumn{2}{c}{$A_{\PGm}$} & \multicolumn{2}{c}{$A_{\Pe}$} & \multicolumn{2}{c@{}}{$A_{\PGm\text{+}\Pe}$} \\
&                          Central values &           0&403 &           0&446 &           0&440 \\
\hline
\multirow{8}{*}{\rotatebox[origin=c]{90}{Profiled uncertainties}}
&                             Statistical &    ${\pm}$0&029 &    ${\pm}$0&038 &    ${\pm}$0&024 \\
&                \ttbar/\tw normalisation &    ${\pm}$0&010 &    ${\pm}$0&007 &    ${\pm}$0&007 \\
&                   \wzjets normalisation &    ${\pm}$0&012 &    ${\pm}$0&011 &    ${\pm}$0&012 \\
&                  Multijet normalisation &      ${<}$0&001 &      ${<}$0&001 &    ${\pm}$0&003 \\
&                          Multijet shape &      ${<}$0&001 &    ${\pm}$0&006 &      ${<}$0&001 \\
&             Jet energy scale/resolution &    ${\pm}$0&008 &      ${<}$0&001 &      ${<}$0&001 \\
&\PQb tagging efficiencies/misidentification &      ${<}$0&001 &    ${\pm}$0&009 &    ${\pm}$0&004 \\
&                                  Others &      ${<}$0&001 &    ${\pm}$0&003 &    ${\pm}$0&005 \\
\hline
\multirow{11}{*}{\rotatebox[origin=c]{90}{Theoretical uncertainties}}
&                          Top quark mass &    ${\pm}$0&033 &    ${\pm}$0&063 &    ${\pm}$0&044 \\
&                             PDF+$\alpS$ &    ${\pm}$0&011 &    ${\pm}$0&009 &    ${\pm}$0&011 \\
&        $t$ channel renorm./fact. scales &    ${\pm}$0&013 &    ${\pm}$0&018 &    ${\pm}$0&020 \\
&               $t$ channel parton shower &    ${\pm}$0&030 &    ${\pm}$0&008 &    ${\pm}$0&014 \\
&             \ttbar renorm./fact. scales &    ${\pm}$0&008 &    ${\pm}$0&019 &    ${\pm}$0&017 \\
&                    \ttbar parton shower &    ${\pm}$0&031 &    ${\pm}$0&037 &    ${\pm}$0&033 \\
&            \ttbar underlying event tune &      ${<}$0&001 &    ${\pm}$0&014 &    ${\pm}$0&014 \\
&                  \ttbar \pt reweighting &      ${<}$0&001 &    ${\pm}$0&010 &    ${\pm}$0&009 \\
&             \wjets renorm./fact. scales &      ${<}$0&001 &    ${\pm}$0&019 &    ${\pm}$0&014 \\
&                      Color reconnection &    ${\pm}$0&036 &    ${\pm}$0&056 &    ${\pm}$0&031 \\
&                     Fragmentation model &    ${\pm}$0&011 &    ${\pm}$0&011 &    ${\pm}$0&011 \\
\hline
\multicolumn{2}{@{}l}{Profiled uncertainties only} &  ${\pm}$0&041 &            ${\pm}$0&047 &            ${\pm}$0&031 \\
\multicolumn{2}{@{}l}{(statistical+experimental)} \\[\cmsTabSkip]
\multicolumn{2}{@{}l}{Total uncertainties} &   ${\pm}$ 0&071 &   ${\pm}$0&099 &   ${\pm}$0&070 \\
\hline
\end{tabular}
\end{table*}

The measured asymmetries are in good agreement with the predicted SM value of 0.436, found using \POWHEG at NLO, with a negligible uncertainty. Good agreement is also found with a corresponding measurement by the ATLAS Collaboration at $\sqrt{s}=8$\TeV~\cite{Aaboud:2017aqp}. This measurement is found to be more precise than a previous analysis of the spin asymmetry at $\sqrt{s}=8$\TeV by the CMS Collaboration~\cite{Khachatryan:2015dzz}. In particular, the deviation found therein, corresponding to 2.0 standard deviations, is not seen.

\section{Summary}
\label{sec:summary}

Differential cross sections for \tchannel single top quark and antiquark production in proton-proton collisions at $\sqrt{s}=13$\TeV have been measured by the CMS experiment at the LHC using a sample of proton-proton collision data, corresponding to an integrated luminosity of 35.9\fbinv. The cross sections are determined as a function of the top quark transverse momentum ($\pt$), rapidity, and polarisation angle, the charged lepton $\pt$ and rapidity, and the $\pt$ of the \PW~boson from the top quark decay. In addition, the charge ratio has been measured as a function of the top quark, charged lepton, and \PW~boson kinematic observables. Events containing one muon or electron and two or three jets are used. The single top quark and antiquark yields are determined through maximum-likelihood fits to the data distributions. The differential cross sections are then obtained at the parton and particle levels by unfolding the measured signal yields.

The results are compared to various next-to-leading-order predictions, and found to be in good agreement. Furthermore, the top quark spin asymmetry, which is sensitive to the top quark polarisation, has been measured using the differential cross section as a function of the top quark polarisation angle at the parton level. The resulting value of $0.440 \pm 0.070$ is in good agreement with the standard model prediction.

{\tolerance=800
These results demonstrate a good understanding of the underlying electroweak production mechanism of single top quarks at $\sqrt{s}=13$\TeV and in particular of the electroweak vector$-$axial-vector coupling predicting highly polarized top quarks. Lastly, the differential charge ratios, sensitive to the ratio of the up to down quark content of the proton, are found to be consistent with the predictions by various sets of parton distribution functions.\par
}

\begin{acknowledgments}
We congratulate our colleagues in the CERN accelerator departments for the excellent performance of the LHC and thank the technical and administrative staffs at CERN and at other CMS institutes for their contributions to the success of the CMS effort. In addition, we gratefully acknowledge the computing centres and personnel of the Worldwide LHC Computing Grid for delivering so effectively the computing infrastructure essential to our analyses. Finally, we acknowledge the enduring support for the construction and operation of the LHC and the CMS detector provided by the following funding agencies: BMBWF and FWF (Austria); FNRS and FWO (Belgium); CNPq, CAPES, FAPERJ, FAPERGS, and FAPESP (Brazil); MES (Bulgaria); CERN; CAS, MoST, and NSFC (China); COLCIENCIAS (Colombia); MSES and CSF (Croatia); RPF (Cyprus); SENESCYT (Ecuador); MoER, ERC IUT, PUT and ERDF (Estonia); Academy of Finland, MEC, and HIP (Finland); CEA and CNRS/IN2P3 (France); BMBF, DFG, and HGF (Germany); GSRT (Greece); NKFIA (Hungary); DAE and DST (India); IPM (Iran); SFI (Ireland); INFN (Italy); MSIP and NRF (Republic of Korea); MES (Latvia); LAS (Lithuania); MOE and UM (Malaysia); BUAP, CINVESTAV, CONACYT, LNS, SEP, and UASLP-FAI (Mexico); MOS (Montenegro); MBIE (New Zealand); PAEC (Pakistan); MSHE and NSC (Poland); FCT (Portugal); JINR (Dubna); MON, RosAtom, RAS, RFBR, and NRC KI (Russia); MESTD (Serbia); SEIDI, CPAN, PCTI, and FEDER (Spain); MOSTR (Sri Lanka); Swiss Funding Agencies (Switzerland); MST (Taipei); ThEPCenter, IPST, STAR, and NSTDA (Thailand); TUBITAK and TAEK (Turkey); NASU and SFFR (Ukraine); STFC (United Kingdom); DOE and NSF (USA).

\hyphenation{Rachada-pisek} Individuals have received support from the Marie-Curie programme and the European Research Council and Horizon 2020 Grant, contract Nos.\ 675440, 752730, and 765710 (European Union); the Leventis Foundation; the A.P.\ Sloan Foundation; the Alexander von Humboldt Foundation; the Belgian Federal Science Policy Office; the Fonds pour la Formation \`a la Recherche dans l'Industrie et dans l'Agriculture (FRIA-Belgium); the Agentschap voor Innovatie door Wetenschap en Technologie (IWT-Belgium); the F.R.S.-FNRS and FWO (Belgium) under the ``Excellence of Science -- EOS" -- be.h project n.\ 30820817; the Beijing Municipal Science \& Technology Commission, No. Z181100004218003; the Ministry of Education, Youth and Sports (MEYS) of the Czech Republic; the Lend\"ulet (``Momentum") Programme and the J\'anos Bolyai Research Scholarship of the Hungarian Academy of Sciences, the New National Excellence Program \'UNKP, the NKFIA research grants 123842, 123959, 124845, 124850, 125105, 128713, 128786, and 129058 (Hungary); the Council of Science and Industrial Research, India; the HOMING PLUS programme of the Foundation for Polish Science, cofinanced from European Union, Regional Development Fund, the Mobility Plus programme of the Ministry of Science and Higher Education, the National Science Center (Poland), contracts Harmonia 2014/14/M/ST2/00428, Opus 2014/13/B/ST2/02543, 2014/15/B/ST2/03998, and 2015/19/B/ST2/02861, Sonata-bis 2012/07/E/ST2/01406; the National Priorities Research Program by Qatar National Research Fund; the Ministry of Science and Education, grant no. 3.2989.2017 (Russia); the Programa Estatal de Fomento de la Investigaci{\'o}n Cient{\'i}fica y T{\'e}cnica de Excelencia Mar\'{\i}a de Maeztu, grant MDM-2015-0509 and the Programa Severo Ochoa del Principado de Asturias; the Thalis and Aristeia programmes cofinanced by EU-ESF and the Greek NSRF; the Rachadapisek Sompot Fund for Postdoctoral Fellowship, Chulalongkorn University and the Chulalongkorn Academic into Its 2nd Century Project Advancement Project (Thailand); the Welch Foundation, contract C-1845; and the Weston Havens Foundation (USA). \end{acknowledgments}

\bibliography{auto_generated}

\cleardoublepage \appendix\section{The CMS Collaboration \label{app:collab}}\begin{sloppypar}\hyphenpenalty=5000\widowpenalty=500\clubpenalty=5000\input{TOP-17-023-authorlist.tex}\end{sloppypar}
\end{document}

%% file: TOP-17-023-authorlist.tex
\vskip\cmsinstskip
\textbf{Yerevan Physics Institute, Yerevan, Armenia}\\*[0pt]
A.M.~Sirunyan$^{\textrm{\dag}}$, A.~Tumasyan
\vskip\cmsinstskip
\textbf{Institut f\"{u}r Hochenergiephysik, Wien, Austria}\\*[0pt]
W.~Adam, F.~Ambrogi, T.~Bergauer, J.~Brandstetter, M.~Dragicevic, J.~Er\"{o}, A.~Escalante~Del~Valle, M.~Flechl, R.~Fr\"{u}hwirth\cmsAuthorMark{1}, M.~Jeitler\cmsAuthorMark{1}, N.~Krammer, I.~Kr\"{a}tschmer, D.~Liko, T.~Madlener, I.~Mikulec, N.~Rad, J.~Schieck\cmsAuthorMark{1}, R.~Sch\"{o}fbeck, M.~Spanring, D.~Spitzbart, W.~Waltenberger, C.-E.~Wulz\cmsAuthorMark{1}, M.~Zarucki
% ...
\vskip\cmsinstskip
\dag: Deceased\\
1:  Also at Vienna University of Technology, Vienna, Austria\\
% ...
